\documentclass[acmsmall,screen]{acmart}

\usepackage{multirow,makecell}
\usepackage{tcolorbox}
\usepackage{color,xcolor}
\usepackage{listings,amsfonts}
\usepackage{caption}
\usepackage{subcaption}
\usepackage{threeparttable}
\usepackage{bbding}
\usepackage{graphicx}
\usepackage{fancyhdr}

\definecolor{dkgreen}{rgb}{0,0.6,0}
\definecolor{gray}{rgb}{0.4,0.4,0.4}
\definecolor{mauve}{rgb}{0.58,0,0.82}
\definecolor{darkblue}{rgb}{0.0,0.0,0.6}
\definecolor{lightblue}{rgb}{0.0,0.0,0.9}
\definecolor{cyan}{rgb}{0.0,0.6,0.6}
\definecolor{darkred}{rgb}{0.6,0.0,0.0}

\definecolor{yellow}{RGB}{255,255,153}
\definecolor{grey}{RGB}{220,220,220}
\definecolor{green}{RGB}{0,100,0}

\definecolor{KWCo
lor}{rgb}{0.37,0.08,0.25}
\definecolor{CommentColor}{rgb}{0.133,0.545,0.133}
\definecolor{StringColor}{rgb}{0,0.126,0.941}
\definecolor{commentgreen}{RGB}{2,112,10}
\definecolor{eminence}{RGB}{108,48,130}
\definecolor{weborange}{RGB}{255,165,0}
\definecolor{frenchplum}{RGB}{129,20,83}

\newcommand{\ea}{\textit{et~al.}}

\AtBeginDocument{%
  \providecommand\BibTeX{{%
    \normalfont B\kern-0.5em{\scshape i\kern-0.25em b}\kern-0.8em\TeX}}}

\setcopyright{acmcopyright}
\copyrightyear{2024}
\acmYear{2024}
\acmDOI{XXXXXXX.XXXXXXX}

\acmJournal{TOSEM}
\acmVolume{X}
\acmNumber{Y}
\acmArticle{1}
\acmMonth{12}

\begin{document}

\title{Large Language Models for Software Engineering: A Systematic Literature Review}

\author[X Hou]{Xinyi Hou}
\email{xinyihou@hust.edu.cn}
\authornotemark[1]
\affiliation{%
  \institution{Huazhong University of Science and Technology}
  \city{Wuhan}           
  \country{China}
}
\author[Y Zhao]{Yanjie Zhao}
\email{yanjie_zhao@hust.edu.cn}
\authornote{Co-first authors who contributed equally to this work.}
\affiliation{%
  \institution{Huazhong University of Science and Technology}
  \city{Wuhan}           
  \country{China}
}
\author[Y Liu]{Yue Liu}
\email{yue.liu1@monash.edu}
\affiliation{%
  \institution{Monash University}
  \city{Melbourne}
  \country{Australia}
}
\author[Z Yang]{Zhou Yang}
\email{zyang@smu.edu.sg}
\affiliation{%
  \institution{Singapore Management University}
  \country{Singapore}
}
\author[K Wang]{Kailong Wang}
\email{wangkl@hust.edu.cn}
\affiliation{%
  \institution{Huazhong University of Science and Technology}
  \city{Wuhan}
  \country{China}
}
\author[L Li]{Li Li}
\email{lilicoding@ieee.org}
\affiliation{%
  \institution{Beihang University}
  \city{Beijing}
  \country{China}
}
\author[X Luo]{Xiapu Luo}
\email{csxluo@comp.polyu.edu.hk}
\affiliation{%
  \institution{The Hong Kong Polytechnic University}
  \city{Hong Kong}
  \country{China}
}
\author[D Lo]{David Lo}
\email{davidlo@smu.edu.sg}
\affiliation{%
  \institution{Singapore Management University}
  \country{Singapore}
}
\author[J Grundy]{John Grundy}
\email{John.Grundy@monash.edu}
\affiliation{%
  \institution{Monash University}
  \city{Melbourne}
  \country{Australia}
}
\author[H Wang]{Haoyu Wang}
\authornote{Haoyu Wang is the corresponding author (haoyuwang@hust.edu.cn).}
\email{haoyuwang@hust.edu.cn}
\affiliation{%
  \institution{Huazhong University of Science and Technology}
  \city{Wuhan}     
  \country{China}
}

\begin{abstract}
Large Language Models (LLMs) have significantly impacted numerous domains, including Software Engineering (SE). Many recent publications have explored LLMs applied to various SE tasks. Nevertheless, a comprehensive understanding of the application, effects, and possible limitations of LLMs on SE is still in its early stages. To bridge this gap, we conducted a systematic literature review (SLR) on LLM4SE, with a particular focus on understanding how LLMs can be exploited to optimize processes and outcomes. We select and analyze 395 research papers from January 2017 to January 2024 to answer four key research questions (RQs). In RQ1, we categorize different LLMs that have been employed in SE tasks, characterizing their distinctive features and uses. In RQ2, we analyze the methods used in data collection, preprocessing, and application, highlighting the role of well-curated datasets for successful LLM for SE implementation. RQ3 investigates the strategies employed to optimize and evaluate the performance of LLMs in SE. Finally, RQ4 examines the specific SE tasks where LLMs have shown success to date, illustrating their practical contributions to the field. From the answers to these RQs, we discuss the current state-of-the-art and trends, identifying gaps in existing research, and flagging promising areas for future study. Our artifacts are publicly available at \url{https://github.com/xinyi-hou/LLM4SE_SLR}.
\end{abstract}



\maketitle

\section{Introduction}
\label{sec:Introduction}
In the field of language processing, traditional \textbf{Language Models (LMs)} have been foundational elements, establishing a basis for text generation and understanding~\cite{moore2010intelligent}. 
Increased computational power, advanced machine learning techniques, and access to very large-scale data have led to a significant transition into the emergence of \textbf{Large Language Models (LLMs)}~\cite{zan2023large, zhao2023survey}. 
Equipped with expansive and diverse training data, these models have demonstrated an impressive ability to simulate human linguistic capabilities, leading to a sea of changes across multiple domains.
With their capacity to learn from massive corpora and generate plausible text, LLMs are blurring the line between human and machine-produced language. They have provided researchers and engineers alike with a powerful tool to explore the complexity and richness of human communication, consequently sparking a transformational period in the field of language processing and beyond.

\textbf{Software Engineering (SE)} – a discipline focused on the development, implementation, and maintenance of software systems – is one of those areas reaping the benefits of the LLM revolution~\cite{ma2023scope}. 
The utilization of LLMs in SE primarily emerges from an innovative perspective where numerous SE challenges can be effectively reframed into data, code, or text analysis tasks~\cite{wang2022machine}.
Using LLMs to address these  SE tasks has shown a wealth of potential breakthroughs~\cite{xia2023keep,tian2023chatgpt,xia2023conversational,lajko2022towards,charalambous2023new,sobania2023analysis,cao2023study,zhang2020retrieval}. 
The applicability of LLMs is particularly pronounced in tasks such as code summarization~\cite{wan2018improving}, which involves yielding an abstract natural language depiction of a code's functionality, as well as the generation of well-structured code~\cite{yin2017syntactic} and code artifacts like annotations~\cite{liang2018automatic}. 
Codex, an LLM with 12 billion parameters, has demonstrated the ability to solve 72.31\% of complex Python programming challenges posed by humans~\cite{chen2021evaluating}.
GPT-4~\cite{openai2023gpt4}, an LLM from OpenAI, has been used with a strong performance in several SE tasks, encompassing code writing, understanding, execution, and reasoning. It not only handles real-world applications and diverse coding challenges but also shows the ability to explain results in natural language and generate code from pseudocode~\cite{bubeck2023sparks}.

Simultaneously, researchers have embarked on a series of research activities regarding LLM-related works, where a number of literature reviews or survey papers have been produced~\cite{fan2023recommender,chang2023survey,yang2023harnessing}. Table~\ref{tab:comparison} summarises some of these. 
However, these related studies have limitations. They either focus narrowly on a single SE scope, such as the application of LLMs in software testing~\cite{wang2023software} and natural-language-to-code (NL2Code) tasks~\cite{zan2023large}, or they are primarily centered on Machine Learning (ML) or Deep Learning (DL) models~\cite{wang2022machine,watson2022systematic,yang2022survey}, overlooking more advanced and recently emerged LLM applications, such as ChatGPT~\cite{openai2022chatgpt}, which are increasingly finding applications within the SE field~\cite{tian2023chatgpt,white2023chatgpt,lubowitz2023chatgpt,sridhara2023chatgpt}. 
Alternatively, they merely offer a preliminary exploration of the performance of LLMs in various SE tasks through empirical experiments~\cite{ma2023scope,sridhara2023chatgpt,xu2022systematic,dou2023towards,yuan2023evaluating}, or analyze existing partially relevant studies to reveal the challenges in this field~\cite{fan2023large} without conducting a systematic literature survey.
Furthermore, some works have investigated the application of Code LLMs in SE~\cite{zhang2023survey,zheng2023survey}, yet have not fully considered general LLMs like ChatGPT and LLaMA~\cite{touvron2023llama}, which are also widely applied to various SE tasks~\cite{huang2023codecot,shapkin2023entity,pan2023understanding,yan2023closer}.
The integration of LLMs within SE is undoubtedly a complex endeavor, requiring key considerations including the choice of the right model, comprehension of the unique features of different LLMs, devising pre-training and fine-tuning strategies, handling of data, evaluation of outcomes, and surmounting implementation challenges~\cite{zan2023large}.
Despite the burgeoning interest and ongoing explorations in the field, \textbf{a detailed and systematic review of LLMs’ application in SE has been notably absent in the current literature}. 
This gap signifies a need for understanding the relationship between LLMs and SE. In response, our research aims to bridge this gap, providing valuable insights to the community.
\begin{table}[t]
\centering
\caption{State-of-the-art surveys related to LLMs for SE.}
\label{tab:comparison}
\begin{threeparttable}
\resizebox{\linewidth}{!}{
\begin{tabular}{|l|c|c|c|c|c|c|}
\hline
\textbf{Reference} & \textbf{Year} &\textbf{Scope of models\tnote{1}} & \textbf{Scope of SE tasks} & \textbf{SLR\tnote{2}} & \textbf{Time frame} & \textbf{\# Collected Papers} \\
\hline
Zhang \ea~\cite{zhang2023survey} & 2023 & Code LLM & Automated program repair & \checkmark & 2017-2023 & 185 \\
Zheng \ea~\cite{zheng2023survey} & 2023 & Code LLM & General SE scope & \checkmark & 2021-2023 & 149 \\
Fan \ea~\cite{fan2023large} & 2023 & LLM & General SE scope & $\times$ & - & Not specified \\
Zan \ea~\cite{zan2023large} & 2023 & LLM (12M+) & NL2Code & $\times$ & 2020-2023 & Not specified \\
Wang \ea~\cite{wang2023software} & 2023 & LLM (117M+) & Software testing & \checkmark & 2019-2023 & 52 \\
Wang \ea~\cite{wang2022machine} & 2022 & ML, DL\tnote{3} & General SE scope & \checkmark & 2009-2020 & 1,209 (ML) + 358 (DL) \\
Yang \ea~\cite{yang2022survey} & 2022 & DL & General SE scope & \checkmark & 2015-2020 & 250 \\
Watson \ea~\cite{watson2022systematic} & 2022 & DL & General SE scope & \checkmark & 2009-2019 & 128 \\
\hline
Our work & 2024 & LLM & General SE scope & \checkmark & 2017-2024 & 395 \\
\hline
\end{tabular}}
\begin{tablenotes}
\footnotesize
\item[1] ``M'' means million and ``B'' means billion. The numbers in parentheses indicate the parameter sizes of LLMs. 
\item[2] SLR stands for Systematic Literature Review. This column denotes whether the paper follows an SLR process.
\item[3] ML and DL refer to Machine Learning and Deep Learning, respectively.
\end{tablenotes}
\end{threeparttable}
\end{table}

In this paper, we conduct an SLR on the utilization of LLMs in SE (LLM4SE). By mapping the current state-of-the-art, pinpointing the key strengths, weaknesses, and gaps in the existing LLM4SE literature, and proposing potential avenues for future research, our review aims to provide researchers and practitioners with a thorough guide to the convergence of LLMs and SE. We anticipate that our findings will be instrumental in guiding future inquiries and advancements in this rapidly evolving field.
This work makes the following key contributions:

\begin{itemize}
    \item We are the first to present a comprehensive SLR on 395 papers published between January 2017 and January 2024 that focus on the use of LLM-based solutions to address SE challenges. We conducted a detailed analysis of the selected papers based on publication trends, distribution of publication venues, etc.
    \item We have classified the LLMs utilized for the reported SE tasks and have provided a summary of the usage and trends of different LLM categories within the SE domain.
    \item We describe the reported data processing stages, encompassing data collection, categorization, preprocessing, and representation. 
    \item We discuss optimizers used for LLM4SE tasks, including tuning techniques, prevalent prompt engineering techniques, and commonly employed evaluation metrics.
    \item We describe the key applications of LLM4SE encompassing a diverse range of 85 specific SE tasks, grouped into six core SE activities -- requirements engineering, software design, software development, software quality assurance, software maintenance, and software management.
    \item We have summarised key challenges that using LLMs encounters within the SE field and have suggested several potential research directions for LLM4SE.
\end{itemize}

Section~\ref{sec:approach} presents our research questions (RQs) and elaborates on our SLR methodology.
The succeeding Sections~\ref{sec:rq1} to \ref{sec:rq4} are devoted to answering each of these RQs individually.
Section~\ref{Sec:limitations} discloses the potential threats to the validity of our study. 
Section~\ref{sec:Discussion} discusses the challenges yet to be overcome when employing LLMs to solve SE tasks and highlights promising opportunities and directions for future research.
Section~\ref{sec:Conclusion} concludes the whole paper.

\section{Approach}
\label{sec:approach}
This SLR follows the methodology proposed by Kitchenham~\ea~\cite{kitchenham2007guidelines,kitchenham2022segress}, used in most other SE-related SLRs~\cite{li2017static,wang2022machine,ramirez2018systematic,liu2022deep}.
Following the guidelines provided by Kitchenham~\ea, our methodology included three main steps: planning the review (i.e., Section~\ref{sec: rq_motivation}, \ref{sec:search_strategy}), conducting the review (i.e., Section~\ref{sec: study_selection}, \ref{sec:snowsearch}), and analyzing the basic review results (i.e, Section~\ref{sec:extraction}).

\subsection{Research Questions} 
\label{sec: rq_motivation}

To provide a comprehensive overview of the LLM4SE field, it is important to fully comprehend how these models are currently being applied in SE, the challenges they face, and their potential future research directions in SE. Thus, we aim to provide an SLR of the application of LLMs to software engineering. This study thus aims to answer the following research questions:

\textbf{RQ1: What LLMs have been employed to date to solve SE tasks?}
RQ1 is designed to map out the landscape of LLMs applied in the field of SE. It seeks to identify and categorize the various LLM architectures—such as decoder-only, encoder-decoder, and encoder-only models—that have been leveraged to address diverse SE challenges. This RQ aims to provide a comprehensive overview of how these models are being utilized and the implications of their usage in this field.

\textbf{RQ2: How are SE-related datasets collected, preprocessed, and used in LLMs?}
RQ2 delves into the methodologies behind the assembly, refinement, and application of datasets in the realm of LLMs for SE tasks. It aims to uncover the strategies for dataset collection, the criteria for dataset selection, and the preprocessing steps essential for making the data conducive for LLM training and application. Additionally, this question seeks to explore the types of data that are most prevalent in SE-related LLM research and how these data types influence the modeling and outcomes. 

\textbf{RQ3: What techniques are used to optimize and evaluate LLM4SE?}
RQ3 aims to explore the use of different optimization and evaluation techniques specific to LLMs in the context of SE. This includes an investigation into Parameter Efficient Fine-Tuning (PEFT) methods and various prompting techniques that are tailored to enhance LLM performance on SE tasks. Furthermore, this RQ aims to assess the range of evaluation metrics and methodologies employed to gauge the effectiveness and impact of LLMs in SE, providing insights into how these models are fine-tuned and assessed for their utility and efficiency.

\textbf{RQ4: What SE tasks have been effectively addressed to date using LLM4SE?}
This RQ aims to identify the SE tasks that have been successfully tackled using LLMs, offering a detailed view of the application spectrum of LLMs in SE. It seeks to identify the specific tasks within SE, such as code generation and program repair, where LLMs have shown significant utility, and to explore the nature and scope of these applications. 






\subsection{Search Strategy}
\label{sec:search_strategy}

\begin{figure}[t]
    \centering
    \includegraphics[width=\linewidth]{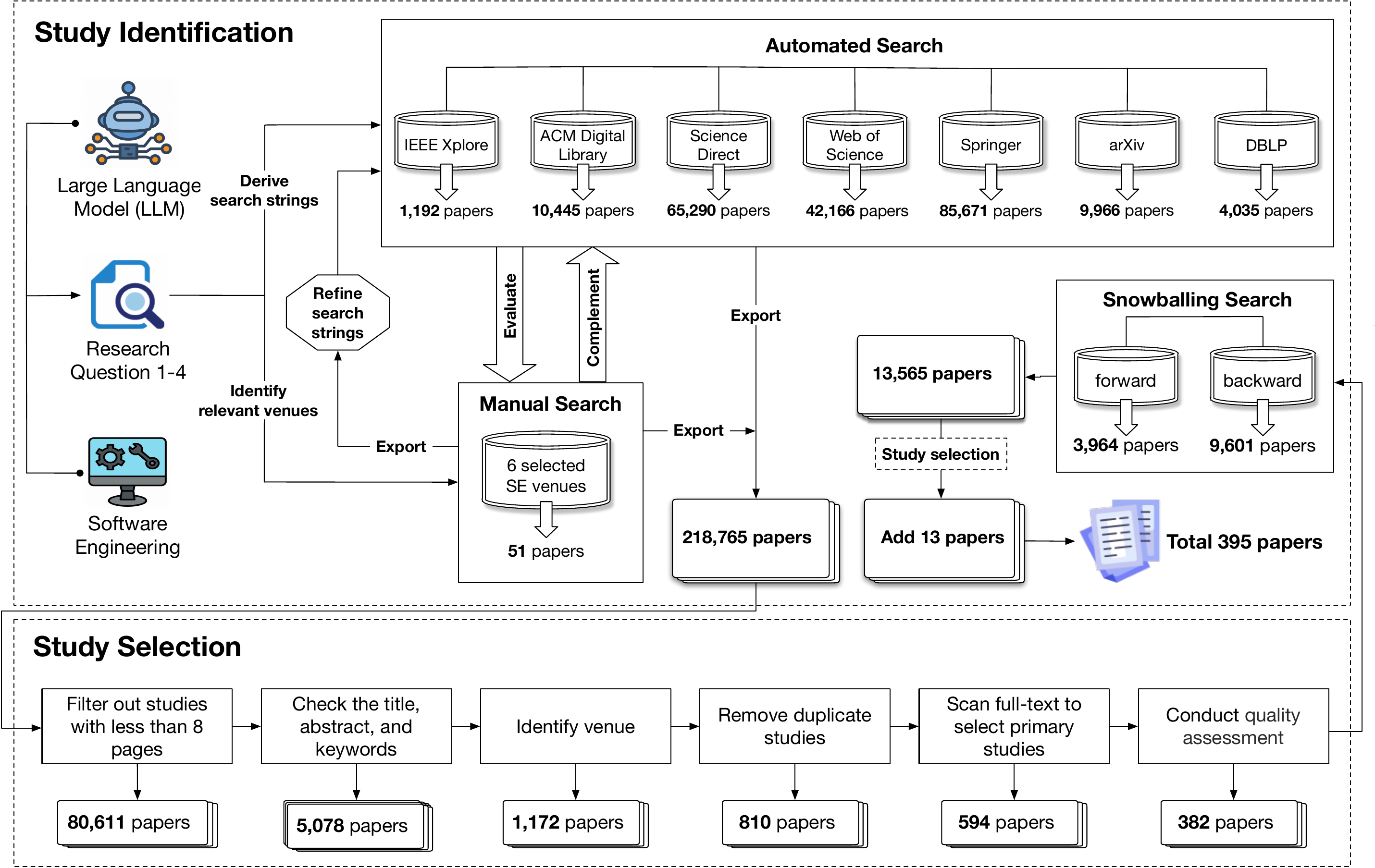}
    \caption{Study identification and selection process.}
    \label{fig:process}
\end{figure}

As shown in Fig.\ref{fig:process}, we employed the ``Quasi-Gold Standard'' (QGS)~\cite{zhang2011identifying} approach for paper search. We conducted a manual search to identify a set of relevant studies and extracted a search string from them. This search string was then used to perform an automated search, and subsequently, a snowballing search was employed to further supplement the search results. This approach ensures both search efficiency and maximum coverage, minimizing the risk of omission. Subsequently, we employed a series of relatively strict filtering steps to obtain the most relevant studies. Specifically, we followed five steps to determine the relevance of the studies:

\begin{enumerate}
    \item Select publication venues for manual search and select digital databases for automated search to ensure coverage of all the selected venues.
    \item Establish QGS: Screen all papers for manual search and filter by inclusion/exclusion criteria (defined in Table \ref{tab:criteria}).
    \item Subjectively define the search string based on domain knowledge.
    \item Conduct an automated search using the search string defined in Step (3).
    \item Conduct snowballing search after performing study selection on the results of manual search and automated search.
\end{enumerate}

\subsubsection{Search Items}
During the manual search, we selected six of the top SE conferences and journals (i.e., ICSE, ESEC/FSE, ASE, ISSTA, TOSEM, and TSE, as shown in Table~\ref{tab:search}) and searched for papers that applied LLM4SE. 
We systematically crawled a list comprising 4,618 published papers from the top venues. Following automated scanning via scripts, we manually verified and identified 51 papers that were relevant to our research objectives.
These 51 relevant papers formed the basis for constructing the Quasi-Gold Standard (QGS). 
Our search string should combine two sets of keywords: one pertaining to SE tasks, and the other related to LLMs. Only if the paper contains both types of keywords, there is a higher probability that it is the paper we need. The complete set of search keywords is as follows:

\begin{itemize}
    \item \textit{Keywords related to SE tasks: Software Engineering, Software Development, Program*\footnote{The * symbol serves as a wildcard, representing any characters or character sequence. For example, ``Program*'' can match ``Program'', ``Programming'', ``Programmer'', and so on.}, Software Testing, Software Mainten*, SE, Software Lifecycle, Software Design*, Code representation, Code generation, Code comment generation, Code search, Code localization, Code completion, Code summarization, Method name generation, Bug detection, Bug localization, Vulnerability detection, Testing techniques, Test case generation, Program analysis, Bug classification, Defect prediction, Program repair, Code clone detection, Bug report, Software quality evaluation, SATD detection, Code smell detection, Compiled-related, Code review, Software classification, Code classification, Code change, Incident detection, Requirement extraction, Requirement traceability, Requirement validation, Effort cost prediction, Mining GitHub/Github mining, Mining SO (Stack Overflow)/SO mining, Mining app/App mining, Mining tag/Tag mining, Developer-based mining}
    \item  \textit{Keywords related to LLMs: LLM, Large Language Model*, Language Model*, LM, PLM, Pre-trained, Pre-training, Natural Language Processing, NLP, Machine Learning, ML, Deep Learning, DL, Artificial Intelligence, AI, Transformer, BERT, Codex, GPT, T5, Sequence Model*, Attention Model*, Transfer Learning, Neural Network*, ChatGPT, GPT-*}
\end{itemize}

It is important to note that the list of keywords related to LLMs that we set up includes Machine Learning, Deep Learning, and other such terms that do not seem to be necessarily related to LLMs. The reason for this is that we want to avoid omitting papers related to our research as much as possible, so the process of performing automated searches expands our search scope.

\begin{table}[t]
\caption{Publication venues for manual search.}
\resizebox{0.9\linewidth}{!}{
\begin{tabular}{rl}
\hline
\textbf{Acronym}  & \textbf{Venues}    \\ \hline
ASE      & International Conference on Automated Software Engineering \\
ESEC/FSE & Joint European Software Engineering Conference and Symposium on the Foundations of Software Engineering \\
ICSE     & International Conference on Software Engineering           \\
ISSTA    & International Symposium on Software Testing and Analysis   \\
TOSEM    & Transactions on Software Engineering and Methodology       \\
TSE      & Transactions on Software Engineering                       \\ \hline
\end{tabular}}
\label{tab:search}
\end{table}

\subsubsection{Search Datasets}
After determining the search string, we conducted an automated search across seven widely used databases, which are capable of covering all published or latest papers.
Given that the first paper about the Transformer architecture~\cite{vaswani2017attention}, which forms the basis for LLMs, was published in 2017, we focused our search on papers published from that year onward\footnote{The cut-off date for the paper collection process of this version is January 31, 2024.}.
Two authors independently performed the automated search, and the search results from each database were merged and deduplicated. Specifically, we obtained 1,192 papers from IEEE Xplore, 10,445 papers from the ACM Digital Library, 62,290 papers from ScienceDirect, 42,166 papers from Web of Science, 85,671 papers from Springer, 9,966 papers from arXiv, and 4,035 papers from DBLP.

\subsection{Study Selection}
\label{sec: study_selection}
\subsubsection{Study Inclusion and Exclusion Criteria}

\begin{table}[t]
\caption{Inclusion criteria and Exclusion criteria.}
\resizebox{0.8\linewidth}{!}{
\begin{tabular}{ll}
\hline
\multicolumn{2}{l}{\textbf{Inclusion criteria}}                                        \\ \hline
1) & The paper claims that an LLM is used.                                                \\
2) & The paper claims that the study involves an SE task.                               \\
3) & The paper with accessible full text.                                               \\ \hline
\multicolumn{2}{l}{\textbf{Exclusion criteria}}                                        \\ \hline
1) & Short papers whose number of pages is less than 8.                                    \\
2) & Duplicate papers or similar studies with different versions from the same authors. \\
3) & Studies belonging to books, thesis, monographs, keynotes, panels, or venues not executing a full  \\
   &  peer-review process.                                                 \\
4) & Tool demos and editorials.                                           \\
5) & The paper is published in a workshop or a doctoral symposium.                 \\
6) & The paper is a grey publication, e.g., a technical report or thesis.          \\ 
7) & Non-English written literature.                                                    \\
8) & The paper mentions the use of LLMs without describing the employed techniques.\\      
9) & The paper leverages SE methods to enhance LLMs, rather than focusing on using LLMs for SE tasks.
\\ \hline
\end{tabular}}
\label{tab:criteria}
\end{table}
   
Based on our search strategy, we initially obtained 218,765 papers that potentially relate to our research. Next, we needed to further evaluate the relevance of these papers based on inclusion and exclusion criteria (To ensure that our inclusion and exclusion criteria were sufficiently objective and rational, we designed these criteria following several state-of-the-art SLR papers~\cite{naveed2024model,wang2022machine,watson2022systematic,yang2022survey}.), as shown in Table~\ref{tab:criteria}, so that the selected papers can directly address our research questions. The paper selection process, as illustrated in Fig.~\ref{fig:process}, consists of six phases. In the first phase, we conducted automated filtering to exclude papers with less than 8 pages~\cite{bashroush2017case,wang2022machine} (Exclusion criteria 1), reducing the number of papers to 80,611. In the second phase, we examined the titles, abstracts, and keywords of the papers to identify those that include relevant LLM-related keywords. We then expanded the search scope to avoid missing relevant papers, including ML, DL, and other related keywords that may not directly correspond to LLM. The purpose of this phase is to narrow down the scope and filter out papers directly related to LLM (Inclusion criteria 1). Papers that are filtered out in this phase are then manually reviewed in the fifth phase.
Additionally, we excluded 448 non-English written literature (Exclusion criteria 7). After the second phase, the number of papers was reduced to 5,078. 

The third phase involves identifying the venues of the papers (Exclusion criteria 3). We extracted publication information such as ``journal'', ``URL'', ``DOI'', and ``series'' to determine the publication sources. For papers from arXiv in 2023 and 2024, we chose to retain them, considering that this field is emerging and many works are in the process of submission. Although these papers did not undergo peer review, we have a quality assessment process to eliminate papers with low quality. This step resulted in 1,172 papers. 

In the fourth phase, we merged and deduplicated the remaining papers from the seven databases and the manually searched paper list (Exclusion criteria 2), resulting in 810 papers. We then reviewed the full texts of the papers and excluded 190 papers that were grey publications or were published in workshops or doctoral symposiums (Exclusion criteria 4, 5, 6). By further assessing the quality of the papers, we identified 382 papers directly relevant to our research. This phase primarily involved excluding papers that mentioned LLMs but did not directly apply them, such as papers that only discussed LLMs in future work or focused on evaluating the performance of LLM-enabled tools~\cite{wang2023software} (Exclusion criteria 8). For systematic views, survey, and review papers, we have retained them and will assess their content during the quality assessment phase to determine their relevance to our research.

\subsubsection{Study Quality Assessment}
\begin{table}[t]
\caption{Checklist of Quality Assessment Criteria (QAC) for LLM studies in SE.}
\resizebox{0.7\linewidth}{!}{
\begin{tabular}{rl}
\hline
\textbf{ID} & \textbf{Quality Assessment Criteria} \\ \hline
QAC1  & Is the study relevant to SE tasks?      \\
QAC2  & Does the study utilize LLMs? \\
QAC3  & Is the research not a secondary study, such as an SLR, review, or survey?           \\
QAC4  & Was the research published in a high-repute venue?                         \\
QAC5  & Is there a clear motivation for the research?                    \\
QAC6  & Does the study provide a clear description of the techniques used?        \\
QAC7  & Are the experimental setups, including experimental environments and           \\
      & dataset information, described in detail?                                 \\
QAC8  & Does the study clearly confirm the experimental findings?                     \\
QAC9  & Are the key contributions and limitations of the study discussed?      \\
QAC10 & Does the study make a contribution to the academic or industrial community?   \\ \hline 
\end{tabular}}
\label{tab:assessment}
\end{table}

A well-crafted quality assessment can help to prevent biases introduced by low-quality studies and can indicate to readers where caution about conclusions should be drawn~\cite{yang2021quality}.
We formulated ten Quality Assessment Criteria (QAC), as shown in Table~\ref{tab:assessment}. These aim to assess the relevance, clarity, validity, and significance of included papers.
We used a scoring system of -1, 0, 1 (irrelevant/unmet, partially relevant/met, relevant/fully met). The first three questions were designed for the remaining 382 papers in the fifth stage. 
If QAC1, QAC2, or QAC3 received a score of -1, there is no need to proceed with QAC4-QAC10, and the paper can be excluded directly. 
QAC4-QAC10 involved assessing the content of the papers using a scoring system of 0, 1, 2, 3 (poor, fair, good, excellent). 
Finally, we calculated the total score of QAC4-QAC10 for each paper. For published papers, the maximum score for QAC4-QAC10 should be 21 (3 $\times$ 7). We retained papers with a score of 16.8 (21 $\times$ 0.8) or above. For unpublished papers on arXiv, the score for QAC4 is always 0, and the maximum score for QAC5-QAC10 should be 18 (3 $\times$ 6). We retained papers with a score of 14.4 (18 $\times$ 0.8) or above.
After this quality assessment, we obtained a final set of 382 papers.

\subsection{Snowballing Search}
\label{sec:snowsearch}

To identify any additional possibly relevant primary studies, we conducted a snowballing search.
Snowballing refers to using the reference list of a paper or the citations to the paper to identify additional papers. Snowballing could benefit from not only looking at the reference lists and citations but also complementing them with a systematic way of looking at where papers are actually referenced and where papers are cited. Using the references and the citations respectively is referred to as backward and forward snowballing.

Before conducting snowballing, a set of initial papers needs to be prepared. In this study, the initial paper list consists of the remaining 382 papers after the quality assessment. We performed forward and backward snowballing, which resulted in the collection of 3,964 and 9,610 papers, respectively. After initial deduplication, we were left with 5,152 papers. We then conducted the full study selection process on these 5,152 papers, including deduplicating them with the 382 papers from performing snowballing on the initial list. As a result, we obtained an additional 13 papers.

\subsection{Data Extraction and Analysis}
\label{sec:extraction}

\begin{figure}[t]
  \centering
  \subfloat[Distribution of papers across venues.] {\includegraphics[width=2.6in]{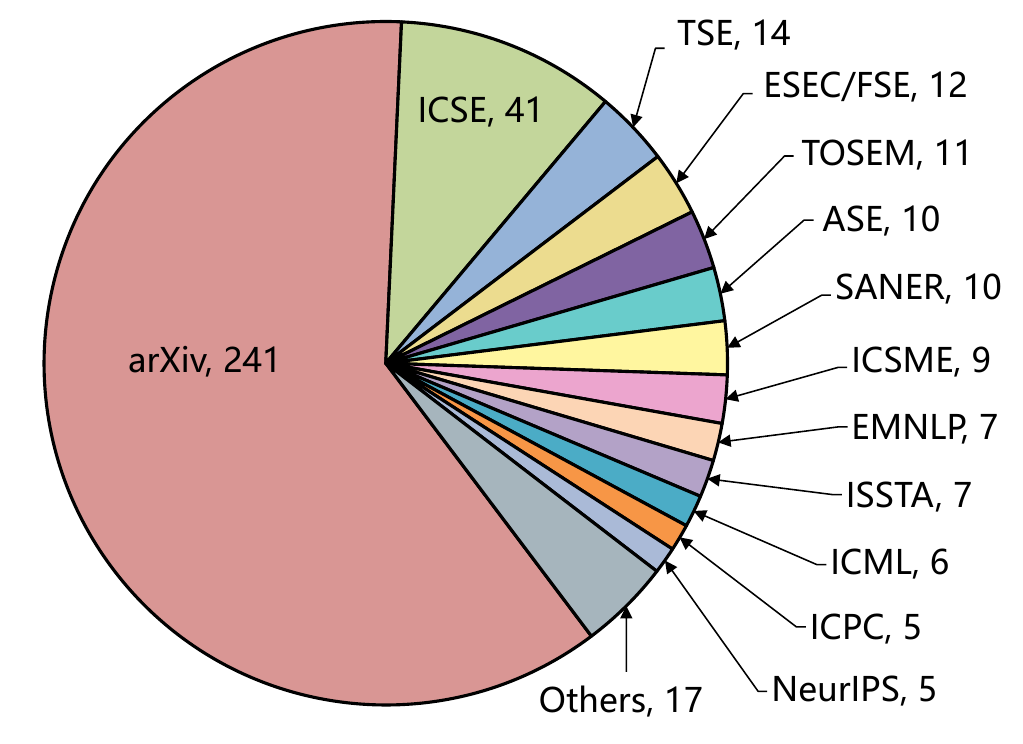}}
  \subfloat[Distribution of papers over years.]{\includegraphics[width=2.6in]{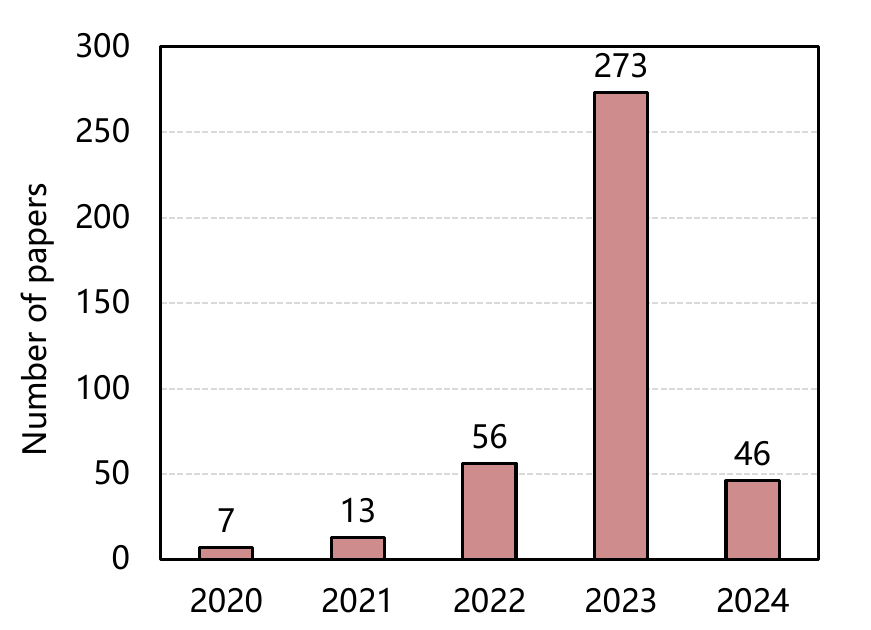}}
  \caption{Overview of the selected 395 papers' distribution.}
  \label{fig:paper_dictribution}
\end{figure}

\begin{figure}
    \centering
    \includegraphics[width=0.5\linewidth]{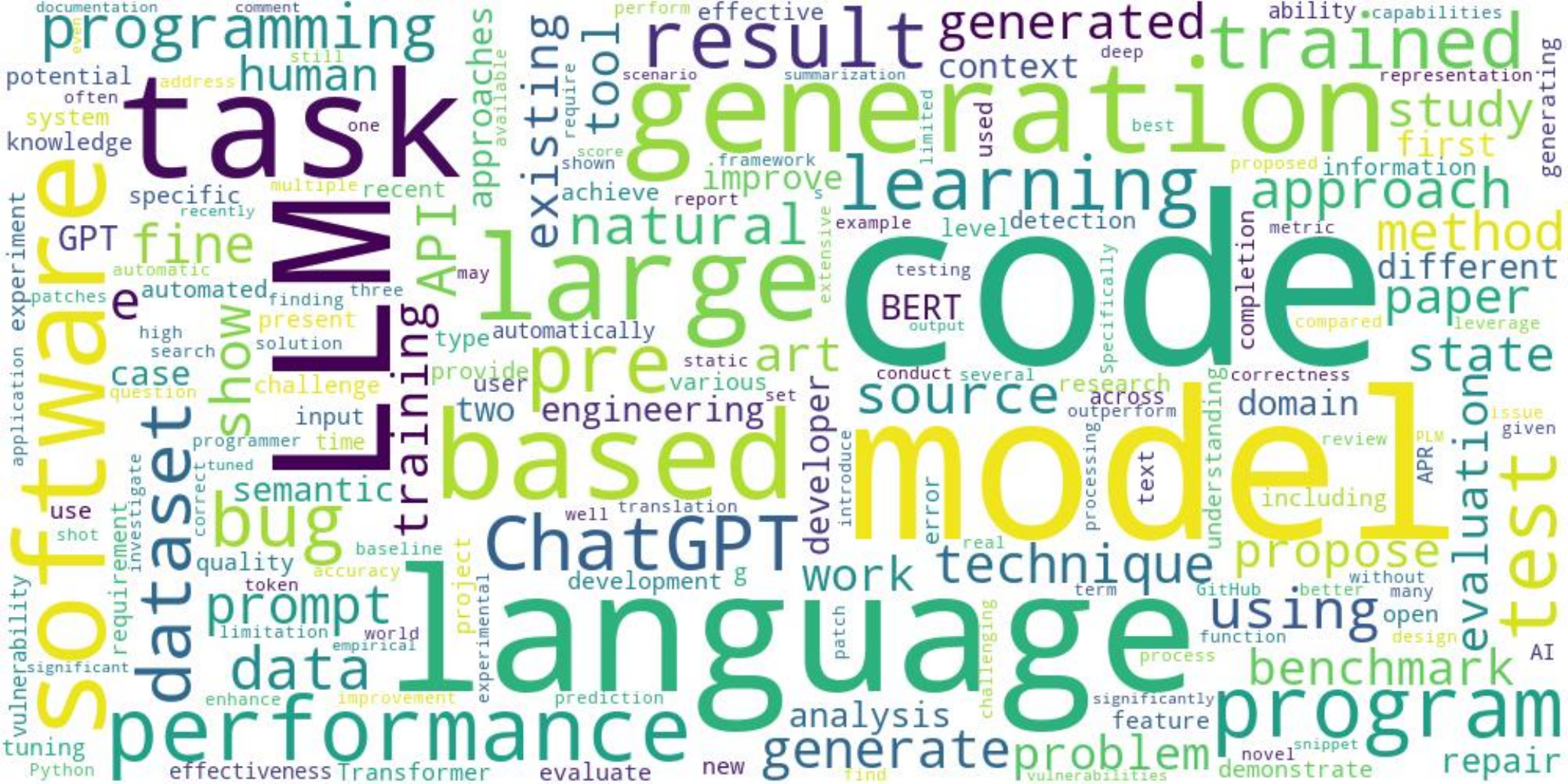}
    \caption{Topics discussed in the collected papers.}
    \label{fig:wordcloud}
\end{figure}

\begin{table}[ht!]
\caption{Extracted data items and related research questions (RQs).}
\resizebox{0.5\linewidth}{!}{
\begin{tabular}{r l}
\hline
\textbf{RQ} & \textbf{Data Item}                                                   \\ \hline
1,2,3,4 & The category of SE task                                     \\
1,2,3,4 & The category of LLM                                         \\
1,4     & Characteristics and applicability of LLMs                   \\
2       & The adopted data handling techniques                        \\
3       & The adopted weight training algorithms and optimizer        \\
3       & The selected evaluation metrics                             \\
4       & The SE activity to which the SE task belongs                \\
4       & The developed strategies and solutions                      \\
\hline
\end{tabular}}
\label{tab:questions}
\end{table}

We finally obtained 395 relevant research papers after searching and snowballing. 
Fig.~\ref{fig:paper_dictribution} presents an overview of the distribution of the included papers.
As shown in Fig.~\ref{fig:paper_dictribution} (a), 154 papers are published in peer-reviewed venues.
ICSE is the most common of these venues, with a contribution of 41 papers.
Other venues with noteworthy contributions include TSE, ESEC/FSE, and TOSEM, contributing 14, 12, and 11 papers respectively.
Meanwhile, the remaining 241 papers are published on arXiv, an open-access platform that serves as a repository for scholarly articles.
This finding is not surprising since much new LLM4SE research is rapidly emerging and thus many works are just completed and are likely in the peer review process.
Despite the non-peer-reviewed nature of these papers, we have performed a rigorous quality assessment process on all collected papers, to ensure the quality of validity of our findings.
This approach allows us to include all high-quality and relevant publications while maintaining high research standards.

Fig.~\ref{fig:paper_dictribution} (b) shows the temporal distribution of the included papers. 
The number of publications has seen a rapidly growing trend since 2020.
In 2020 and 2021, there are only 7 and 13 relevant papers, respectively.
However, by 2022, the number of papers has increased dramatically to 56.
What's surprising is that, in 2023 alone, the number of published papers has already reached 273. And within just one month in 2024, 46 relevant papers are published.
This rapid growth trend demonstrates that there is a growing research interest in the domain of LLM4SE.

In order to visualize the main content of our collection of papers, we generated a word cloud based on the abstracts of 395 papers as shown in Fig.~\ref{fig:wordcloud}. The most frequently occurring words include ``code'', ``LLM'', ``language'', ``model'', ``large'',  ``task'', ``software'',``generation'', ``performance'', and ``program'', clearly indicating the main themes explored in these papers. The terms ``code'' and ``software'' emphasize the core elements of software engineering, while ``LLM'', ``large'', ``language'' and ``model'' denote the use of large language models in a variety of tasks. The terms ``generation'', ``task'', and ``program'' emphasize the use of the LLM for automatic code generation and other SE tasks. In addition, ``performance'' reflects the evaluation and assessment of the effectiveness of LLM in SE applications. The word cloud provides further visual evidence that the literature we have collected is closely related to our research topic.

We then conducted data extraction during the full-text review.
This extraction phase collected all relevant data that would facilitate a comprehensive and insightful response to the RQs outlined in Section~\ref{sec: rq_motivation}.
As depicted in Table~\ref{tab:questions}, we extracted data including the classification of SE tasks, their corresponding activities, as well as the category, characteristics, and applicability of the LLMs.
With this collected data, we systematically analyzed the relevant aspects of LLM4SE.

\section{RQ1: What LLMs have been employed to date to solve SE tasks?}
\label{sec:rq1}

\subsection{Large Language Models (LLMs)} 
\label{sec:LLMs}
\begin{figure}[t]
    \centering
    \includegraphics[width=\linewidth]{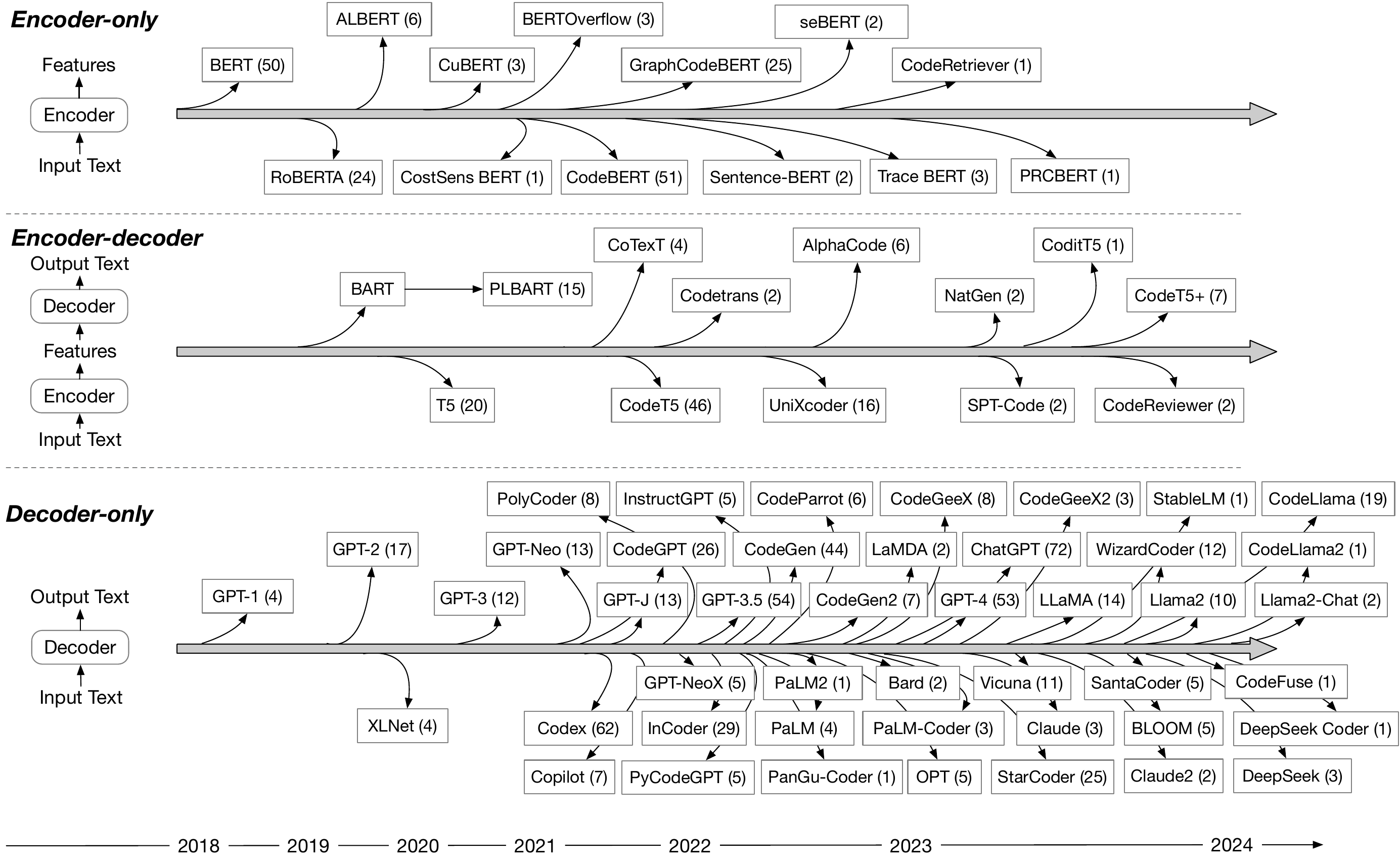}
    \caption{Distribution of the LLMs (as well as LLM-based applications) discussed in the collected papers. The numbers in parentheses indicate the count of papers in which each LLM has been utilized.}
    \label{fig:llms}
\end{figure}

Pre-trained language models (PLMs) have demonstrated impressive capabilities in solving various NLP tasks~\cite{kojima2022large,shanahan2022talking,wei2022chain,zhao2023survey}. Researchers have observed that scaling up the model sizes significantly enhances their capacity, leading to remarkable performance improvements when the parameter scale surpasses a certain threshold~\cite{shanahan2022talking,hoffmann2022training,taylor2022galactica}. 
The term ``Large Language Model'' (LLM) was introduced to distinguish language models based on their parameter size, specifically referring to large-sized PLMs~\cite{zhao2023survey}.   
However, we note that the literature lacks a formal consensus on the minimum parameter scale for LLMs, as the model's capacity is intertwined with both data size and total compute~\cite{wang2023software}. 
In this paper, we adopt the LLM scope division and taxonomy introduced by Pan \ea\cite{pan2023unifying} and categorize the mainstream LLMs investigated in this study into three groups according to their architectures: encoder-only, encoder-decoder, and decoder-only LLMs.
This taxonomy and relevant models are shown in Fig.~\ref{fig:llms}. 
We have included the LLMs used by each work and their parameter sizes (if declared in the paper) in our public repository: \url{https://github.com/xinyi-hou/LLM4SE_SLR}.
Additionally, Table~\ref{tab:llm_used_to_task} summarizes the LLMs with different architectures suitable for different types of SE tasks.

\begin{table}[ht!]
\caption{Summary of LLMs with different architectures used in SE tasks.}
\resizebox{0.6\linewidth}{!}{
\begin{tabular}{|r|c|l|}
\hline
\textbf{Model} & \textbf{Type} & \textbf{Example of SE tasks}\\ \hline
Encoder-only & Understanding & Code Understanding \\
&  & Bug localization \\
&  & Vulnerability detection \\
\hline
Encoder-Decoder & Understanding and Generation & Code summarization \\
&  & Code translation \\
&  & Program repair \\
\hline
Decoder-only & Generation & Code generation \\
& & Code completion \\
& & Test case generation \\
\hline
\end{tabular}}
\label{tab:llm_used_to_task}
\end{table}

\vspace{0.05cm}
\noindent\textbf{Encoder-only LLMs. }
Encoder-only LLMs are a type of neural network architecture that utilizes only the encoder component of the model~\cite{devlin2018bert}. The encoder's function is to process and encode the input sentence into a hidden representation, capturing the relationships between words and the overall context of the sentence. 
Notable instances of encoder-only LLMs include BERT~\cite{devlin2018bert} and its variants~\cite{feng2020codebert,guo2020graphcodebert,liu2019roberta,lan2019albert}. As an example, BERT's structure, based on the Transformer's encoder architecture, has been referenced in 50 our selected primary studies. Its distinctive bidirectional attention mechanism simultaneously considers the left and right context of each word during training.
In the SE domain, other prominent models like CodeBERT~\cite{feng2020codebert}, GraphCodeBERT~\cite{guo2020graphcodebert}, RoBERTa~\cite{liu2019roberta}, and ALBERT~\cite{lan2019albert} have been widely employed. Specialized models such as BERTOverflow~\cite{tabassum2020code} and CodeRetriever~\cite{li2022coderetriever} have been specifically developed for SE applications.
These models differ from BERT by leveraging program structure, introducing new pre-training tasks, or engaging new modalities, thereby improving the architecture's application to code-related tasks. 
For example, CodeBERT integrates a token prediction scheme to comprehend code by predicting subsequent tokens, enhancing its understanding of programming languages for tasks like code completion and bug detection~\cite{feng2020codebert}. GraphCodeBERT introduces edge-type prediction, recognizing relationships between code elements as a graph. This enables GraphCoderBERT to leverage code structure, improving its effectiveness in tasks like code summarization and program analysis~\cite{guo2020graphcodebert}.
Encoder-only LLMs have shown efficacy in tasks requiring a nuanced understanding of the entire sentence or code snippet. Examples include code review, bug report understanding, and named entity recognition pertaining to code entities~\cite{pudari2023copilot,sghaier2023multi,yang2022aspect,arakelyan2023exploring,li2023codeie,mukherjee2023stack}. 

\vspace{0.05cm}
\noindent\textbf{Encoder-decoder LLMs.}
Encoder-decoder LLMs incorporate both encoder and decoder modules~\cite{vaswani2017attention}. The encoder ingests the input sentence and encodes it into a hidden space, effectively capturing the underlying structure and semantics. This hidden representation serves as an intermediary language, bridging the gap between diverse input and output formats. Conversely, the decoder utilizes this hidden space to generate the target output text, translating the abstract representation into concrete and contextually relevant expressions.
Models such as PLBART~\cite{ahmad2021unified}, T5~\cite{raffel2020exploring}, and CodeT5~\cite{wang2021codet5} embodies this architecture. Further advancements are evident in CodeT5+~\cite{wang2023codet5+}, while AlphaCode~\cite{li2022competition} and CoTexT~\cite{phan2021cotext} showcase the architecture's adaptability to various SE tasks. The encoder-decoder design offers flexible training strategies and is proficient in handling multifaceted tasks such as summarization, translation, and question-answering.
Within the field of SE, this ability has been successfully applied to tasks like code summarization~\cite{al2023extending,gu2022assemble,mastropaolo2021studying}. The encoder module's capacity to understand and represent both the structure and semantics of code is pivotal, allowing the decoder to translate this comprehension into concise, human-readable summaries. 

\vspace{0.05cm}
\noindent\textbf{Decoder-only LLMs.}
Decoder-only LLMs exclusively utilize the decoder module to generate the target output text, following a distinct training paradigm that emphasizes sequential prediction~\cite{radford2018improving}. Unlike the encoder-decoder architecture, where the encoder processes input text, the decoder-only architecture begins with an initial state and predicts subsequent tokens, gradually building the output text. This approach relies heavily on the model's ability to understand and anticipate language structure, syntax, and context. GPT-series models, such as GPT-1~\cite{radford2018improving}, GPT-2~\cite{radford2019language}, GPT-3~\cite{brown2020language}, GPT-3.5~\cite{openai2022gpt}, GPT-4~\cite{openai2023gpt4}, as well as their notable derivative, ChatGPT~\cite{openai2022chatgpt}\footnote{ChatGPT is a conversational agent built upon the GPT architecture, with GPT-3.5 and GPT-4 being specific versions of the architecture, each representing successive advancements.}, represent their major implementations. 
More specialized versions like CodeGPT~\cite{lu2021codexglue}, InstructGPT~\cite{ouyang2022training}, Codex~\cite{chen2021evaluating}, Copilot~\cite{github2021copilot}\footnote{Copilot is an application built upon LLMs tailored for coding tasks. \textbf{For convenience, all subsequent references in this paper to LLMs and their applications, such as ChatGPT and Copilot, will collectively be referred to as LLMs.}}, and others have been fine-tuned for specific tasks in SE. Open-source models like GPT-J~\cite{wang2021gpt}, GPT-Neo~\cite{black_sid_2021_5297715}, GPT-NeoX~\cite{black2022gpt}, LLaMA~\cite{touvron2023llama}, and Vicuna~\cite{chiang2023vicuna} also follow this architecture. Decoder-only LLMs are usually more suitable for various generation tasks, such as code generation and code completion. These models can generally perform downstream tasks from a few examples or simple instructions without adding prediction heads or fine-tuning, making them valuable tools in SE research.
\textbf{2022 marked a surge in the development of decoder-only LLMs, a trend that gained further momentum in 2023, notably with the launch of commercial products by leading Internet companies.} For example, Google launched Gemini~\cite{google2023gemini}, Meta introduced LLaMA~\cite{touvron2023llama} and Llama~2~\cite{touvron2023llama2}, and Anthropic unveiled Claude~\cite{anthropic2023claude}, etc. 
Contrary to LLMs such as GPT-4 and its derivative application, ChatGPT, released by OpenAI, which were promptly integrated into SE tasks, these new additions have not yet found widespread application within the SE field. Their potential remains largely unexplored, with opportunities for further assessment and utilization in specific tasks and challenges. The continued advancement of these models emphasizes the active exploration and innovation within decoder-only architectures.

\begin{figure}[ht!]
    \centering
    \includegraphics[width=0.8\linewidth]{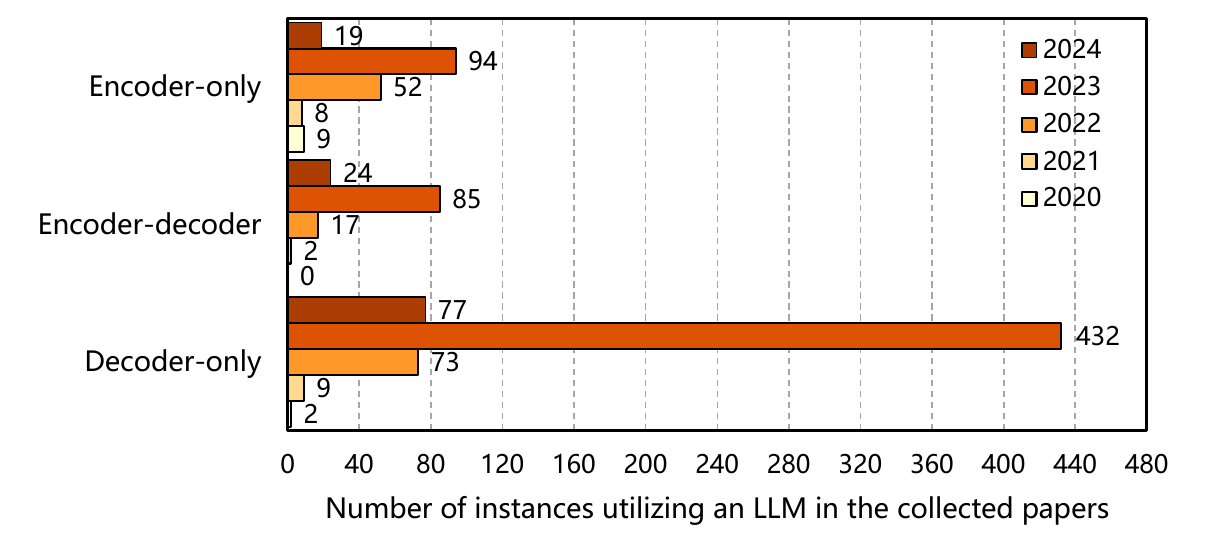}
    \caption{Trends in the application of LLMs with different architectures in SE tasks over time.}
    \label{fig:llms_time_distribution}
\end{figure}

\subsection{Trend Analysis}
\label{sec:trend analysis}

As shown in Fig.~\ref{fig:llms_time_distribution}, in the span from 2020 to 2024, the architecture of LLMs has witnessed notable shifts in preference and application within SE tasks. The specific choices between decoder-only, encoder-decoder, and encoder-only structures have shaped the direction of research and solutions in the SE domain~\cite{wong2023natural}. This analysis explores trends in the adoption of these architectures over the years, reflecting the evolving dynamics of LLM for SE tasks.

\noindent\textbf{Evolution of LLM architectures in 2021.} The year 2020 saw research papers predominantly concentrating on encoder-only LLMs for SE tasks, evidenced by a total of eight papers. Decoder-only LLMs or encoder-decoder LLMs were scarcely featured in that year's research. A marked change occurred in 2021. Out of 19 papers in 2021, nine were dedicated to decoder-only LLMs, constituting 47.37\% of the research. Additionally, two papers, or 10.53\%, focused on encoder-decoder LLMs. Encoder-only LLMs witnessed a slight decline, representing 42.1\% of the field with eight papers. 
This rapid transition can be linked to the generative capability of decoder-only LLMs. Researchers~\cite{laskar2023systematic,sadik2023analysis,sridhara2023chatgpt} found that these models, e.g., GPT series, requiring minimal fine-tuning, could produce not only syntactically correct but also functionally relevant code snippets. Their proficiency in grasping the context of code quickly made them a preferred choice.

\vspace{0.05cm}
\noindent\textbf{Diversity of LLM architectures in 2022.} 2022 experienced a significant increase in diversity, with more varied LLM architectures finding representation. Out of a total of 142 papers, 73 were centered around decoder-only LLMs, comprising 51.41\% of the studies. Encoder-decoder LLMs made their presence known in 17 papers, accounting for 11.97\%. Meanwhile, encoder-only LLMs led the field slightly with 52 papers, capturing 36.62\% of the research interest. This diverse distribution suggests an exploration phase where researchers were actively assessing and leveraging different architectures to suit varied needs and challenges. The near-equal interest across different architectures underscores the field's richness, indicating that no single approach had become the definitive choice. 

\vspace{0.05cm}
\noindent\textbf{Dominance of the decoder-only architecture in 2023.} 2023 signaled a strong shift towards decoder-only LLMs. An impressive 432 instances of utilizing decoder-only LLMs were recorded across 195 unique papers, reflecting that a single paper might employ multiple such models. These papers focusing on decoder-only LLMs constituted a significant 70.7\% of the total research this year. In comparison, encoder-decoder LLMs were the subject of 85 papers, contributing 13.91\%, while encoder-only LLMs appeared to stabilize, with 94 papers, representing 15.39\% of the 2023 research landscape.
This trend signifies a shift in focus and resources toward exploring and harnessing the decoder-only architecture as the primary approach in many current and future LLM4SE research and applications.

\vspace{0.05cm}
\noindent\textbf{Exploration of the LLM architecture in 2024.} The initial trends in January 2024 showcase the ongoing evolution of LLM architectures. Among the 120 papers examined, decoder-only LLMs continued to maintain a prominent position, with 77 papers dedicated to this architecture, constituting 64.17\% of the research. Encoder-decoder LLMs appeared in 24 papers, representing 20\% of the total, while encoder-only LLMs were featured in 19 papers, making up 15.83\%. Although there is a slight decrease in the dominance of decoder-only architectures compared to the previous year, they still hold a central role. The persistent exploration of encoder-decoder and encoder-only architectures suggests an enduring interest in diverse configurations within the SE research community.

\vspace{0.05cm}
\noindent\textbf{Criteria for LLM selection in SE tasks.} The selection of an LLM for SE tasks should involve careful consideration rather than arbitrary choice. Key factors guiding this selection encompass the model's proficiency in understanding the context of code, its ability to generate relevant content, responsiveness to fine-tuning, and demonstrated performance on SE-specific benchmarks~\cite{xie2023chatunitest,li2023enabling,li2023exploring}. Given the stringent syntactical rules and functional requirements inherent to SE tasks, models capable of seamlessly integrating these complex aspects were typically favored.

\vspace{0.05cm}
    \noindent\textbf{Task-specific fine-tuning.} A notable trend is the customization of LLMs for precise SE tasks~\cite{izadi2022codefill,li2023codeie,zhang2022coditt5}. By fine-tuning models with datasets tailored to specific functions such as bug detection or code review, researchers were able to achieve marked performance improvements~\cite{ciborowska2023too,kou2023automated}.

In conclusion, the evolution of LLMs for SE, transitioning from encoder-only to decoder-only architectures, highlights the field's vibrancy and adaptability. This shift has fundamentally altered the approach to SE tasks, reflecting the ongoing innovation within the discipline.

\begin{tcolorbox}[title=\textbf{RQ1} - Summary, left=2pt, right=2pt,top=2pt,bottom=2pt]
(1) There are more than 70 different LLMs used for SE tasks in our selected primary studies. Based on the underlying architecture or principles of different LLMs, we classified the summarized LLMs into three categories, i.e., decoder-only, encoder-decoder, and encoder-only LLMs.

(2) We observed that each LLM architecture serves a specific purpose in SE tasks, with encoder-only LLMs focusing on comprehensive understanding, encoder-decoder LLMs used for tasks requiring understanding of input information followed by content generation, and decoder-only LLMs being more suitable for generation tasks.

(3) We analyzed the trend of LLM usage for SE tasks. The most widely used LLMs are with decoder-only architectures. There are over 45 LLMs in the decoder-only category and 195 papers have researched the application of decoder-only LLMs to SE tasks.

\end{tcolorbox}

\section{RQ2: How are SE-related datasets collected, preprocessed, and used in LLMs?}
\label{sec:rq2}
Data plays a crucial role in the model training phase~\cite{sun2022importance}.
First, data is collected to obtain diversity and richness to ensure that the model can cope with different scenarios and situations. Second, data is classified to clarify the training objectives of the model and avoid confusion and misinformation. The preprocessing of data is indispensable to clean and transform the data to improve its quality. Finally, data is formatted into a structure suitable for model processing, allowing the LLM to learn the data’s features and patterns effectively.
We analyze the reported processes of data collection, data classification, data preprocessing, and data representation in our selected primary studies on LLM4SE.

\subsection{How are the datasets for training LLMs sourced?}
Data is an indispensable and critical factor in training LLMs, which determines the generalization ability, effectiveness, and performance of the models~\cite{sun2022importance}. Adequate, high-quality, and diverse data is critical to allow models to fully learn features and patterns, optimize parameters, and ensure reliability in validation and testing. 
We first investigate the methods used to obtain the dataset. By analyzing the methods of data collection, we divided the data sources into four categories: open-source datasets, collected datasets, constructed datasets, and industrial datasets. 
\textit{Open-source datasets}~\cite{chen2023improving,khakhar2023pac,wang2023evaluating,zeng2022extensive} refer to publicly accessible collections of data that are often disseminated through open-source platforms or repositories. For example, datasets like HumanEval~\cite{chen2021evaluating}, which consists of 164 manually crafted Python problems, each accompanied by its respective unit tests. The open-source nature of these datasets ensures their credibility and allows for community-driven updates, making them a reliable resource for academic research.
\textit{Collected datasets}~\cite{huang2018api,tian2023chatgpt,sghaier2023multi,mastropaolo2022using} are those that researchers compile directly from a multitude of sources, including but not limited to, major websites, forums, blogs, and social media platforms. For instance, researchers~\cite{chan2023transformer,salza2022effectiveness,weyssow2023usage,yang2022aspect} often scrape data from Stack Overflow~\cite{stackoverflowStackoverflow} threads or GitHub~\cite{githubGithub} issues comments to create a dataset tailored to their specific research questions.
\textit{Constructed datasets}~\cite{ezzini2022automated,koide2023detecting,kang2022large,zhang2022beqain} are specialized datasets that researchers create by modifying or augmenting collected datasets to better align with their specific research objectives. These modifications can be carried out through manual or semi-automatic methods and may include the generation of domain-specific test sets, annotated datasets, or synthetic data. For example, researchers often take a collected dataset of code snippets and manually annotate them with bug types to create a constructed dataset for studying automated program repair techniques~\cite{fan2023automated,jin2023inferfix,wu2023effective}.
\textit{Industrial datasets}~\cite{alhamed2022evaluation,moharil2022identification,wang2020deep} are those obtained from commercial or industrial entities and often contain proprietary business data, user behavior logs, and other sensitive information. These datasets are particularly valuable for research that aims to address real-world business scenarios. However, the acquisition of such datasets is often complicated by issues related to business confidentiality and data privacy. 
For example, in a collaborative effort with China Merchants Bank (CMB), Wang~\ea~\cite{wang2020deep} were able to access 21 projects from CMB's repositories.
Access to such data would likely require non-disclosure agreements and other legal safeguards to protect business interests.
Each of these dataset types offers unique advantages and challenges, and the choice between them should be guided by the specific requirements and constraints of the research project at hand.

\begin{figure}[ht!]
    \centering
\includegraphics[width=0.55\linewidth]{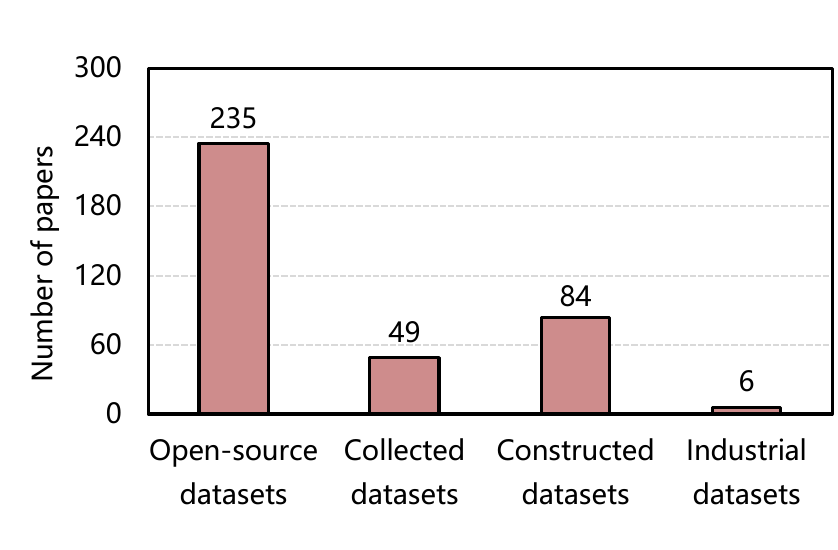}
    \caption{The collection strategies of LLM-related datasets.}
    \label{fig:data_collection}
\end{figure}

Fig.~\ref{fig:data_collection} shows the collection strategies of LLM-related datasets. As can be seen from the data in the figure, \textbf{235 studies used open-source datasets for training LLMs}. 
One of the main reasons for using open-source datasets in LLM training is their authenticity and credibility. Open-source datasets usually contain real-world data collected from various sources (such as relevant studies that have been conducted), which makes them highly reliable and representative of real-world scenarios. This helps LLMs learn from real examples to better understand real-world applications and improve their performance. 
Second, since LLMs are a topic that has just recently emerged, a lack of suitable training sets does exist. 
Therefore, researchers often collect data from sites such as Stack Overflow and GitHub and build datasets to make the data more composite for SE tasks. 
\textbf{Among the 395 papers we studied, we discovered that merely six studies utilized industrial datasets.} 
This suggests a potential misalignment between the properties of datasets used in academic research and those encountered in real-world industrial contexts. This divergence underscores the need for future research to investigate industrial datasets, thereby ensuring that LLMs are applicable and robust across both academic and industrial scenarios.

Note that some papers use multiple datasets that span different categories, e.g., Xu \ea~\cite{xu2022systematic} evaluated the performance of Codex, GPT-J, GPT-Neo, and other LLMs on SE tasks, and Mastropaolo \ea~\cite{mastropaolo2021studying} investigated the use of T5 in several code-related tasks such as fixing bugs and generating code comments. For different LLMs or different SE tasks, researchers may use different training datasets. On the other hand, some papers focus on exploring how existing LLMs (e.g., ChatGPT) are used in SE tasks~\cite{white2023chatgpt} and do not specify the dataset used for model training, as these LLMs like ChatGPT often do not require users to prepare training data themselves for general usage scenarios.

\subsection{What types of SE datasets have been used in existing LLM4SE studies?}

\begin{table}[t]
\caption{Data types of datasets involved in prior studies.}
\resizebox{\linewidth}{!}{
\begin{tabular}{|r|ll|c|}
\hline
\textbf{Category} & \multicolumn{2}{c|}{\textbf{Data type}}  & \textbf{Total} \\ \hline
Text-based
& Programming tasks/problems (42) & Prompts (33) & 151 \\
datasets
& SO (i.e. Stack Overflow) posts (12) & Bug reports (11) & \\
& Requirements documentation (9) & APIs/API documentation (8) & \\
& Q\&A pairs (6) & Vulnerability descriptions (4) & \\
& Reviews (4) & Logs (3) & \\
& Methods (3) & Project issues (3) & \\
& Code comments (2) & Theorems (2) & \\
& Buggy text (1) & Dockerfiles (1) & \\
& Outage descriptions (1) & Semantic merge conflicts (1) & \\
& Site text (1) & Software development tasks (1) & \\
& User intents (1) & Software specifications (1) & \\
& User reviews (1) & & \\
\hline
Code-based
& Source code (60) & Bugs/Buggy code (16) & 103 \\
datasets
& Vulnerable source code (8) & Patches (4) & \\
& Code changes (3) & Test suites/cases (3) & \\
& Bug-fix pairs (2) & Error code (2) & \\
& Error-fix pairs (1) & Flaky test cases (1) & \\
& Identifiers (1) & Labeled clone pairs (1) & \\
& Packages (1) & & \\
\hline
Graph-based
& GUI Images (1) & & 1 \\
datasets 
& & & \\
\hline
Software 
& Code repository (9) & Android apps (3) & 20 \\
repository
& Issues and commits (3) & Pull-requests (2) & \\
-based datasets
& Industrial projects (1) & Open-source projects (1) & \\
& Web applications (1) & & \\
\hline
Combined
& Programming tasks and test suites/cases (17) & Source code and comments (12) & 55 \\
datasets
& Programming tasks and solutions (8) & Source code and description (3) & \\
& Code-text pairs (2) & Souce code and API usage sequences (2) & \\
& Source code and test suites/cases (2) & Bug report and test suites/cases (1) & \\
& Buggy code and comments (1) & Buggy code and solutions (1) & \\
& Code files and summaries (1) & Binary code and related annotations (1) & \\
& Failing test code and error messages (1) & Source code and Q\&A pairs (1) & \\
& Source code, methods, and logs (1) & Vulnerable code and description (1) & \\
\hline
\end{tabular}}
\begin{tablenotes}
\footnotesize
\item *See \autoref{app:datasets} for the full table including references. 
\end{tablenotes}
\label{tab:datatypes}
\end{table}

Data types play a pivotal role in shaping the architecture and selection of LLMs, as they directly influence the extraction of implicit features and subsequent model decisions\cite{chan2023transformer,ghadhab2021augmenting,yang2023syntax,shi2022cross}.
The choice of data types can significantly impact the overall performance and generalization ability of the LLMs. We examine and classify the types of SE datasets employed in LLM4SE studies. By investigating the relationship between data types, model architectures, and performance, we seek to shed light on the critical role of data types in the success of LLM4SE applications.

\noindent\textbf{Data type categorization.} We classified the data types of all datasets into five categories: code-based, text-based, graph-based, software repository-based, and combined data types. Table~\ref{tab:datatypes} describes the specific data included in the data types corresponding to the datasets we summarized from the 395 studies. We can find that \textbf{most of the studies used text-based datasets, accounting for a total of 151}. The dominance of text-based datasets in training LLMs for SE tasks highlights the models' exceptional natural language processing capabilities. These LLMs excel in understanding and processing textual data, making them an ideal choice for tasks that involve code comprehension, bug fixing, code generation, and other text-oriented SE challenges. Their ability to process and learn from vast amounts of text data enables them to provide powerful insights and solutions for various SE applications.

\textbf{The most prevalent type of data utilized in training LLMs for SE tasks is programming tasks/problems with 42 instances observed among the surveyed papers.} This dominance can be attributed to the diverse and challenging nature of programming problems, which provide LLMs with opportunities to generalize knowledge and skills across various SE challenges, fostering a robust understanding of software concepts and enhancing performance across a wide range of tasks, including code generation, code completion, and code summarization, etc.
Prompts follow closely behind programming tasks, with 33 instances observed in the surveyed papers, providing task-specific guidance to LLMs, serving as cues or instructions for the models, and helping them understand the context and requirements of SE tasks. 
This combination helps the models develop a robust understanding of software concepts and perform well in a wide range of tasks. There are also SO (i.e., Stack Overflow) posts~(12), bug reports~(11), etc., which are among the more numerous data types in text-based datasets. 

The predominance of source code~(60) as the most abundant data type in code-based datasets can be attributed to its fundamental role in SE. Source code serves as the foundation of any software project, containing the logic and instructions that define the program's behavior. Therefore, having a large volume of source code data is crucial for training LLMs to understand the intricacies of software development, enabling them to effectively generate, analyze, and comprehend code in various SE tasks. There are also common data types, such as bugs/buggy code (16) and patches (4), for program repair tasks. Additionally, vulnerable source code~(8) is used for vulnerability detection tasks. Graph-based datasets are used in some research studies for SE tasks, e.g., Kolthoff \ea~\cite{kolthoff2023data} used a dataset composed of screenshots from Google Play Android applications to construct a graphical user interface (GUI) repository in their study on LLM for the rapid prototyping task. These datasets represent code using graph structures, capturing relationships and dependencies between code components. 

Software repository-based datasets are compilations of data extracted from version control systems, such as Git repositories, containing code, documentation, and related artifacts. This data includes Code repository~(3), issues and commits~(3), and so on. The data in software repositories can provide a wealth of information covering all aspects of the software development process, including code evolution history, records of issue fixes and feature improvements, code quality assessments, and so on. These data are valuable for studying behaviors and trends in the software development process, improving software quality and development efficiency, and evaluating the performance of software engineering techniques. Therefore, many studies have used software repository-based datasets for empirical analysis and model training.

Some studies employed combined datasets containing multiple datatypes. Among them, the most common type is ``programming tasks and test suites/cases''.  Other combinations of data types include ``source code and comments'', ``programming tasks and solutions'', ``source code and description '', ``code-text pairs'', etc.

\subsection{How do data types influence the selection of data-preprocessing techniques?}

For the training and application of LLMs, the raw dataset needs to be subjected to data processing to obtain a clean and suitable dataset for model training. The data processing steps~\cite{manh2023vault,lee2022light} involve operations such as data cleaning, noise removal, normalization, etc.
To ensure consistency and quality of the data, different data types may require different processing methods to improve the performance and effectiveness of LLMs in SE tasks. In this section, we aim to detail the data preprocessing procedures for the two most used types of datasets, i.e., text-based datasets and code-based datasets.

\begin{figure}[ht!]
    \centering
    \includegraphics[width=\linewidth]{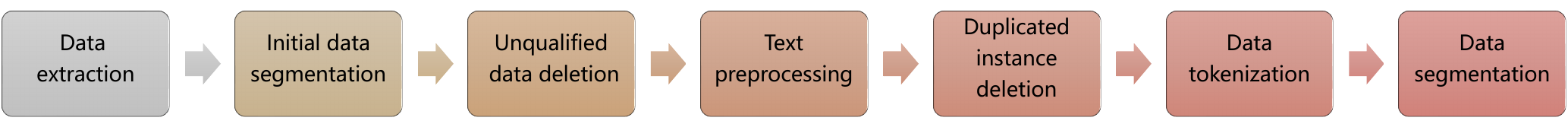}
    \caption{The data preprocessing procedure for text-based datasets.}
    \label{fig:text_based}
\end{figure}

\begin{figure}[ht!]
    \centering
    \includegraphics[width=\linewidth]{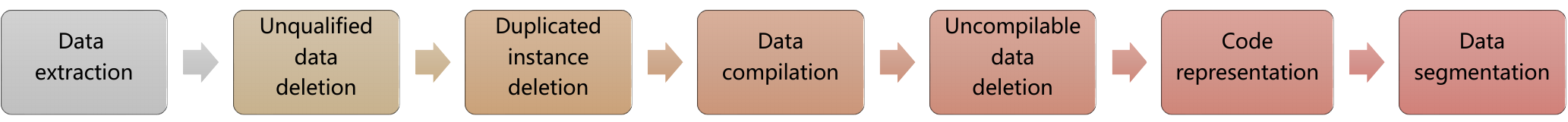}
    \caption{The data preprocessing procedure for code-based datasets.}
    \label{fig:code_based}
\end{figure}


\noindent\textbf{The data preprocessing procedure for text-based datasets.} As displayed in  Fig. ~\ref{fig:text_based}, the steps of text-based dataset preprocessing consist of seven steps in total, yet there are some differences from the code-based dataset preprocessing steps. The process begins with data extraction~\cite{yang2023syntax,ciborowska2022fast,ezzini2022automated,ciborowska2023too}, where relevant text is carefully extracted from SE documentation from a variety of sources, including bug reports~\cite{ciborowska2023too}, requirements documents~\cite{kolthoff2023data}, code comments~\cite{prenner2021making}, and API documentation~\cite{khan2021automatic}. This step ensures that the dataset captures diverse, task-specific textual information. After data extraction, the text is initially segmented and categorized according to the specific requirements of the research task. For example, the text can be segmented into sentences or further broken down into individual words as needed for analysis~\cite{he2023representation,kou2023automated}. To ensure the quality and relevance of the dataset, substandard data deletion is performed to eliminate any invalid or irrelevant text. For example, the dataset used by Lee \ea~\cite{lee2022light} was constructed from bug reports, and in the ``unqualified data deletion'' process the researchers filtered out bug reports with fewer than 15 words because the text was too short to contain contextual information. Next, preprocessing operations are performed on the text to standardize and clean it. Common preprocessing steps include removing certain symbols, stop words, and special characters~\cite{rahmani2023improving,wang2020deep}. This standardized form of text facilitates the efficient processing of LLMs. 
To avoid introducing bias and redundancy in the dataset, researchers eliminated duplicate instances by removing any duplicate text samples~\cite{he2023representation,kou2023automated,xu2022systematic}. 
This step enhances the diversity of the dataset and helps the model to generalize better to new inputs. ``Data tokenization'' is a key step in preparing the text for LLMs~\cite{luo2022prcbert}. 
Text is labeled into smaller units, such as words or subwords, so that LLMs are easier to manage and process efficiently. Finally, the preprocessed dataset is partitioned into different subsets, usually including a training set, a validation set, and a test set.

\noindent\textbf{The data preprocessing procedure for code-based datasets.} 
We now summarize the process of preprocessing a code-based dataset, which consists of seven steps.
Fig. ~\ref{fig:code_based} describes the individual data processing steps in detail and gives examples. The first step is data extraction, which involves retrieving relevant code segments from different sources such as software repositories or version control systems~\cite{kang2023explainable,yang2023syntax}. Depending on the requirements of the research task~\cite{mastropaolo2021studying,yuan2023no}, code segments can be extracted at different levels of granularity, ranging from individual methods and functions to entire source code files or even complete software projects. The next step is to remove any code segments that do not meet predefined criteria or quality standards~\cite{li2021toward,shi2022cross,prenner2021making}. This filtering process ensures that the extracted code is relevant to the specific SE task under study, thus eliminating incomplete or irrelevant code snippets.
To avoid introducing bias and redundancy during model training, the third step involves removing duplicate instances~\cite{zhao2021impact,ciniselli2021empirical,xu2022systematic}. Any duplicate code instances are identified and removed from the dataset, thus increasing the diversity and uniqueness of the data. 
After the data extraction and filtering steps, the fourth step, data compilation, comes into play. The extracted and filtered code segments are merged and compiled into a unified code dataset. This compilation process simplifies data storage and access and facilitates subsequent analysis and model training~\cite{chan2023transformer,mastropaolo2022transfer}. In the fifth step, the problem of invalid or non-executable code is solved by removing data that cannot be compiled. Any code segments that cannot be compiled or executed are removed from the dataset to ensure that the remaining code instances are valid and usable during model training and evaluation. 
The sixth step is code representation, which consists of converting the code segments into a suitable representation that can be processed by the LLMs. 
This conversion can take different forms: token-based representation involves tokenizing the source or binary code into distinct tokens; tree-based representation parses the code into Abstract Syntax Trees (AST); and graph-based representation generates a Program Dependence Graph (PDG), encompassing Control Flow Graphs (CFG) and Call Graphs (CG). 
Finally, in the ``data segmentation'' step, the preprocessed dataset is partitioned into different subsets for training, validation, and testing~\cite{ciniselli2021empirical,weyssow2023usage}. The training set is used to train the LLM, the validation set helps to tune the hyperparameters and optimize the model performance, and the testing set evaluates the model's ability on unseen data. By strictly adhering to these seven preprocessing steps, researchers can create structured and standardized code-based datasets, thus facilitating the effective application of LLMs for a variety of SE tasks such as code completion, error detection, and code summarization. 

It is worth emphasizing that the order of these steps is not fixed and can be adjusted based on the specific research task and its associated requirements. Researchers need to carefully consider the objectives, characteristics of the dataset, and the desired outcomes when determining the optimal sequence for these preprocessing techniques.

\subsection{What input formats are the datasets for LLM training converted to?} 

Once suitable datasets have been carefully chosen and clean data has been achieved through the preprocessing steps, the next critical aspect is the transformation of the data into appropriate formats that can effectively serve as inputs for LLMs.
Table~\ref{tab:input-forms} shows four distinct data input types that emerged during the research: Token-based input, Tree/Graph-based input, Pixel-based input, and Hybrid-based input. We now detail each as follows:

\begin{table}[ht!]
\caption{The various input forms of LLMs proposed in prior studies. See \autoref{app:inputs} for the full table including references.}
\resizebox{0.8\linewidth}{!}{
\begin{tabular}{|r|ll|c|}
\hline
\textbf{Category} & \multicolumn{2}{c|}{\textbf{Input forms}}  & \textbf{Total} \\ \hline
Token-based input
& Text in tokens (150) & Code in tokens (118) & 347 \\
& Code and text in tokens (78) & & \\
\hline
Tree/Graph-based input
& Code in tree structure (2) & Code in graph structure (3) & 5 \\
\hline
Pixel-based input & Pixel (1) & & 1 \\ \hline
Hybrid-based input & Hybrid input forms (2) & & 2 \\ \hline
\end{tabular}}
\label{tab:input-forms}
\end{table}

\noindent\textbf{Token-based input.} 
Token-based input~\cite{ahmed2024automatic,al2023extending,arakelyan2023exploring} involves representing code and text as sequences of tokens, which are smaller units like words or subwords. Text in tokens refers to the tokenization of textual data, such as documentation, bug reports, or requirements, enabling the LLMs to process and analyze natural language descriptions effectively. Code and text in tokens combine both code and its associated textual context, allowing the model to capture the relationships between code elements and their descriptions. Code in tokens refers to the representation of code snippets broken down into meaningful tokens, allowing the LLMs to understand programming language syntax and semantics at a fine-grained level.

\noindent\textbf{Tree/Graph-based input.} Tree-based input~\cite{ma2023scope,ochs2023evaluating,zhang2023neural} represents code as hierarchical tree structures, capturing the syntactic relationships between code elements. Each node in the tree represents a code element, and the edges represent the hierarchical nesting of control flow statements and other code structures. This form of input allows the LLMs to understand the code's hierarchical structure and perform tasks like code completion and bug fixing. Graph-based input represents code as a graph structure, where nodes represent code elements and edges represent the relationships between them. Unlike trees, graphs allow more flexible and complex relationships between code elements, enabling the model to capture non-linear dependencies in the code. This form of input is used in tasks like code summarization and vulnerability detection by considering the code's intricate relationships.

\vspace{0.1em}
\noindent\textbf{Pixel-based input.} Pixel-based input~\cite{nasir2023llmatic} visualizes code as images, where each pixel represents a code element or token. This visual representation allows the LLMs to process and understand code through image-based learning. In this input form, LLMs learn from the visual patterns and structures in the code to perform tasks like code translation or generating code visualizations.

\vspace{0.1em}
\noindent\textbf{Hybrid-based input.} Hybrid-based input~\cite{niu2022spt} combines multiple modalities to provide LLMs with diverse perspectives for better code comprehension. For example, a hybrid input may combine code in tokens with visual representations of code, allowing the model to learn both from the fine-grained details in the tokenized code and from the overall visual structure of the code. This approach enhances the model's ability to understand complex code patterns and improve performance in tasks such as code comprehension and code generation.

During our investigation of LLM-based models for SE tasks, we observed distinct trends in the usage of different input forms during the training process. \textbf{Token-based input forms, namely code in tokens and text in tokens were the most prevalent, collectively constituting approximately 97.75\% of the studies\footnote{This refers to studies that explicitly state input forms of LLMs, i.e., a total of 355 papers as shown in Table~\ref{tab:input-forms}.}.} Specifically, code in tokens was widely adopted in 118 studies, accounting for approximately 33.24\% of the total studies, demonstrating its popularity as a primary choice for representing code snippets. This approach allowed LLMs to grasp programming language syntax and semantics effectively, making it suitable for a wide range of code-related tasks.
Similarly, text in tokens was utilized in 150 studies, comprising around 42.25\% of the total studies. This input form allowed LLMs to process natural language descriptions, bug reports, and documentation with greater efficiency and accuracy. The popularity of token-based input forms underscores their significance in leveraging the power of LLMs for software engineering applications.

In contrast, \textbf{tree/graph-based input forms, such as code in tree-structure, were used in only seven studies, making up approximately 1.4\% of the total}. Although less prevalent, this input type emerged as a promising choice to represent the hierarchical structure and syntactic relationships within code. Its adoption indicated an ongoing exploration of tree-based representations in specialized tasks, such as code completion and bug fixing.

\textbf{Pixel-based input and hybrid-based input were relatively less common, each found in one study, contributing approximately 0.28\% of the total studies each}. While their adoption rates were lower, these input forms presented intriguing possibilities for specific applications. Pixel-based input offered a unique visual representation of code, potentially advantageous for code translation tasks. Meanwhile, hybrid-based input, combining multiple modalities (e.g., code in tree structure and text
in tokens in Niu \ea's work~\cite{niu2022spt}), showcased the potential for enhancing code comprehension tasks by offering diverse perspectives for the models to learn from.

In summary, the trends in input form usage reveal a strong preference for token-based input, demonstrating its versatility and effectiveness in various SE tasks. However, ongoing exploration of other input forms, such as tree/graph-based, pixel-based, and hybrid-based, suggests a dynamic and evolving landscape in the application of LLMs for SE, with potential for further innovation and improvement in specialized domains.
Each of these input forms caters to specific characteristics of the SE tasks being addressed, enabling LLMs to perform effectively across a wide range of code-related applications with a more comprehensive understanding of the input data.

\begin{tcolorbox}[title=\textbf{RQ2} - Summary, left=2pt, right=2pt,top=2pt,bottom=2pt]
(1) We divided the datasets into four categories based on the source of data: open-source, collected, constructed, and industrial datasets. \textbf{The use of open-source datasets is the most prevalent}, constituting approximately 62.83\% of the 374 papers that explicitly state the dataset.

(2) We categorized the data types within all datasets into five groups: code-based, text-based, graph-based, software repository-based, and combined. \textbf{Text-based and code-based types are the most frequently used in applying LLMs to SE tasks.} This pattern indicates that LLMs are particularly adept at handling text and code-based data in SE tasks, leveraging their natural language processing capabilities.

(3) We summarized the data preprocessing procedures for different data types and found several common preprocessing procedures, i.e., \textit{data extraction}, \textit{unqualified data deletion}, \textit{duplicated instance deletion}, and \textit{data segmentation}.
\end{tcolorbox}

\section{RQ3: What techniques are used to optimize and evaluate LLM4SE?}
\label{sec:rq3}

\subsection{What tuning techniques are used to enhance the performance of LLMs in SE tasks?}

Through surveying research related to LLM4SE, we found that while many general-purpose LLMs (e.g., ChatGPT) can be directly applied to software engineering tasks such as code generation~\cite{dong2023self,liu2023improving,yeticstiren2023evaluating}, code summarization~\cite{shi2023sotana,sun2023prompt,yang2023enhancing}, and program repair~\cite{charalambous2023new,gao2023constructing,xia2023keep} without fine-tuning, the hidden potential of LLMs often needs to be realized through tuning to be fully exploited. Specifically, this requires training LLMs with task-specific data to learn knowledge relevant to the task context to perform better. 
We observed that out of 83 studies, LLMs were fine-tuned using \textbf{full fine-tuning} techniques to adapt to downstream SE tasks, with the majority being BERT series models~\cite{ciniselli2021empirical,ezzini2022automated,fatima2022flakify,jesse2022learning,kou2023automated,lee2022light,lin2021traceability,luo2022prcbert,salza2022effectiveness,von2022validity,wang2022your,wei2022clear,zhang2022beqain}.
The cost of training these LLMs is expensive, requiring a large amount of computational resources and massive amounts of data. It is also costly to train and deploy the fine-tuned models separately for each downstream task, as the traditional fine-tuning approach would need to copy a model and perform full-parameter fine-tuning for each downstream task~\cite{cassano2023multipl,deng2023large,ezzini2022automated,izadi2022codefill,jesse2022learning,lee2022light}.

To reduce this computational burden, some researchers have previously used \textbf{In-Context Learning (ICL)}~\cite{gao2023constructing,geng2024large,hu2024leveraging,huang2023chain,jiang2023self}, which feeds the model with manually designed ``prompts'' that are overly reliant on human design and do not require updating model parameters at all. However, ICL only operates at the time of inference and does not involve learning task-specific parameters, which experimentally proved to give the model limited improvement in downstream tasks~\cite{liu2022few}.
To address this problem, researchers have begun to apply \textbf{Parameter Efficient Fine-Tuning (PEFT)}~\cite{houlsby2019parameter} techniques to LLMs. 
PEFT aims to improve the performance of pre-trained models on new tasks by optimizing the subset of parameters fine-tuned, thereby reducing the overall computational complexity. This approach maintains the majority of the pre-trained model's parameters in a fixed state, focusing fine-tuning efforts on a minimal yet impactful set of parameters~\cite{weyssow2023exploring}.
Prior code intelligence research has demonstrated the capabilities of PEFT techniques, frequently revealing their superiority over full fine-tuning on a variety of tasks~\cite{weyssow2023exploring}. Four common techniques of PEFT include Low-Rank Adaptation (LoRA)~\cite{hu2021lora}, prompt tuning~\cite{lester2021power}, prefix tuning~\cite{li2021prefix}, and adapter tuning~\cite{houlsby2019parameter}. We now elaborate on each as follows:

\vspace{0.1em}
\noindent\textbf{Low-Rank Adaptation (LoRA).} LoRA injects low-rank trainable matrices into the attention layers of the Transformer architecture to significantly reduce the number of trainable parameters. We observed that eight studies~\cite{arakelyan2023exploring,lu2023llama,pan2023stelocoder,shestov2024finetuning,shi2023sotana,silva2023repairllama,wang2023codet5+,zhang2023toolcoder} utilized LoRA to enhance the performance of LLMs in SE tasks. For instance, Pan \ea~\cite{pan2023stelocoder} trained SteloCoder, specifically designed for translating multiple programming languages into Python code, which is based on the StarCoder LLM. LoRA technology was employed during the modification of the StarCoder model architecture to adjust the parameter count. Additionally, Silva \ea~\cite{silva2023repairllama} applied LoRA to LLaMA, resulting in a highly effective ``program repair adapter'' for fixing bugs through fine-tuning.

\vspace{0.1em}
\noindent\textbf{Prompt tuning.} 
Prompt tuning involves appending learnable tokens to the model's input, guiding it towards better task performance. This method keeps the model's architecture unchanged, leveraging adaptable prompts to influence outputs without altering internal parameters.
In the surveyed papers, three research works~\cite{lu2023llama,wang2023codet5+,zhu2023automating} utilized prompt tuning. For instance, Zhu \ea~\cite{zhu2023automating} proposed a method named AUMENA, which automates method naming tasks through context-aware prompt tuning.

\vspace{0.1em}
\noindent\textbf{Prefix tuning.} 
Prefix tuning adapts pre-trained language models by adding trainable tokens not just to the input but also across internal layers, affecting the model's intermediate representations. This approach modifies the model's processing with minimal changes to its original parameters, allowing for task-specific customization.
This technique was utilized in the following two studies: Lu \ea~\cite{lu2023llama} fine-tuned LLaMA-Reviewer for automating code review, while Wang \ea~\cite{wang2023codet5+} fine-tuned CodeT5+ for multiple downstream tasks such as code completion, code generation, and code search.

\vspace{0.1em}
\noindent\textbf{Adapter tuning.} Adapter tuning adds small neural network modules to the original model, then fine-tuning them on specific tasks without altering the original model's parameters. Agarwal \ea~\cite{agarwal2024structured} fine-tuned LLMs using adapter tuning techniques to make them suitable for code representation tasks. Wang \ea~\cite{wang2023one} indicated that LLMs refined through adapter tuning perform exceptionally well in code search and code summarization tasks.

In addition to the above-mentioned tuning methods, other techniques have been used for tuning LLMs in the LLM4SE domain, such as \textbf{Reinforcement Learning (RL)}~\cite{islam2024code,islam2024llm,jain2023coarse,steenhoek2023reinforcement,yang2023syntax}, \textbf{Supervised Fine Tuning (SFT)}~\cite{dong2023abilities,islam2024code,mastropaolo2022using,steenhoek2023reinforcement,yang2023syntax}, an unsupervised data augmentation method called \textbf{syntax fine-tuning}~\cite{qi2023sut}, \textbf{knowledge preservation fine-tuning}~\cite{siddiq2023lightweight}, and \textbf{task-oriented fine-tuning}~\cite{sun2023prompt}, etc.

\subsection{What prompt engineering techniques are applied to improve the performance of LLMs in SE tasks?}

\begin{figure}[ht!]
    \centering
    \includegraphics[width=0.9\linewidth]{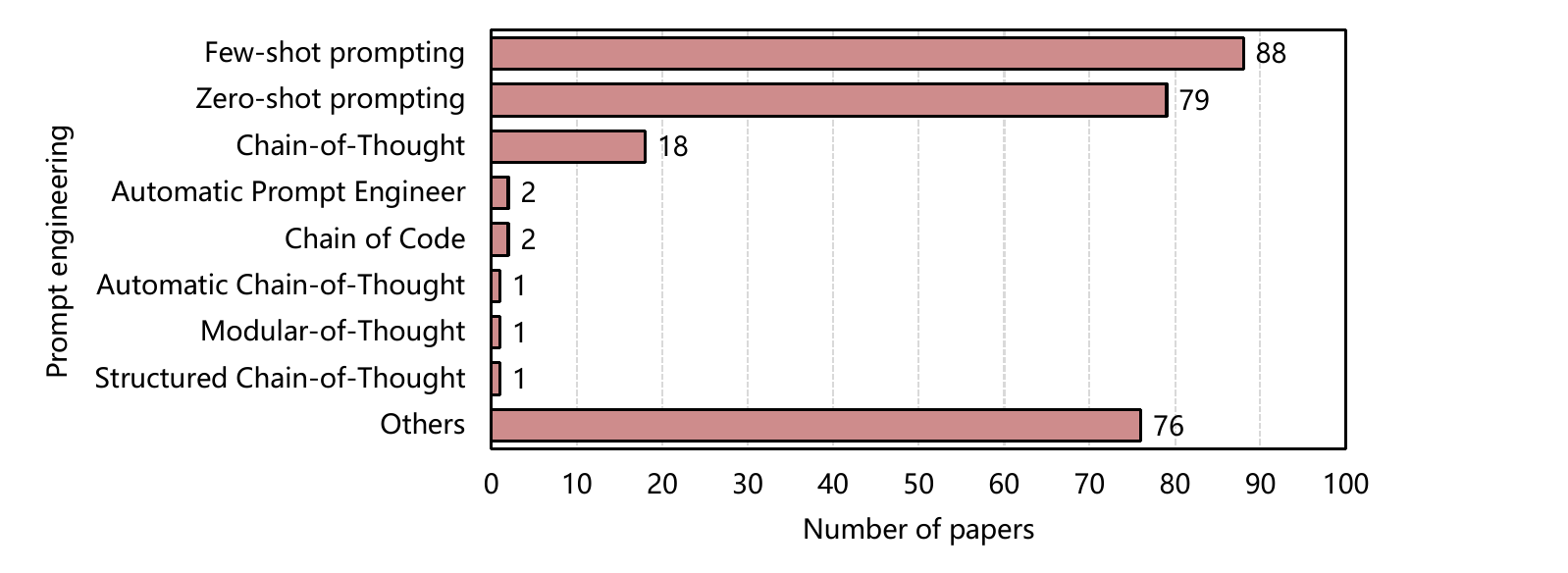}
    \caption{The prompting engineering techniques used in LLMs for SE tasks. See \autoref{app:prompting} for the full table including references.}
    \label{fig:prompt}
\end{figure}

Prompt engineering is a method of enhancing model performance by using task-specific instructions, known as prompts, without modifying the core model parameters. This approach enables LLMs to seamlessly integrate into downstream tasks solely based on the given prompts, guiding model behavior without the need to update model parameters~\cite{sahoo2024systematic}. Fig.~\ref{fig:prompt} presents eight prompt engineering techniques currently applied in the LLM4SE domain. 

\vspace{0.1em}
\noindent\textbf{Few-shot prompting.} Few-shot prompting involves providing a limited number of examples or instructions to the model to perform a specific task. The model learns from these examples and generalizes to similar tasks with minimal training data. In the surveyed LLM4SE research, 88 studies utilized few-shot prompting~\cite{ahmed2024automatic,feng2023prompting,first2023baldur,geng2024large,kang2022large,wei2022clear,xu2024unilog,zhang2023revisiting}. For instance, Geng \ea~\cite{geng2024large} adopted an in-context learning paradigm and providing a specific number of prompts simultaneously significantly outperformed state-of-the-art supervised learning methods in generating comments with multiple intents.

\vspace{0.1em}
\noindent\textbf{Zero-shot prompting.} In zero-shot prompting~\cite{radford2019language}, the model is expected to perform a task without any explicit training on that task. Instead, it relies on the prompt provided during inference to generate the desired output. Following few-shot prompting in terms of usage frequency, 79 studies adopted zero-shot prompting~\cite{kou2023automated,li2023codeeditor,luo2022prcbert,pei2023can,schafer2023empirical,weyssow2023usage,wu2023effective,yan2023closer}. 
For example, Li \ea~\cite{li2023codeeditor} introduced CodeEditor, a pre-trained model specifically designed for code editing, and demonstrated its effectiveness in automatic code editing under zero-shot settings.

\vspace{0.1em}
\noindent\textbf{Chain-of-Thought (CoT) prompting.} Wei \ea~\cite{wei2022chain} introduced a prompting technique called Chain-of-Thought (CoT), which involves each prompt building upon the preceding one, resulting in a coherent chain of reasoning that enhances the model's ability to generate well-structured and thoughtful responses. Huang \ea~\cite{huang2023ai} proposed a novel method leveraging the fault-tolerance and comprehension capabilities of pre-trained LLMs to generate Control Flow Graphs. This method involves a Chain-of-Thought (CoT) with four steps: structural hierarchy extraction, nested code block extraction, CFG generation for nested code blocks, and merging of CFGs for all nested code blocks. Tian \ea~\cite{tian2023test} also introduced the first test case-driven code generation technique, named TCoT, to further enhance LLMs' capabilities in code generation. Including the two studies mentioned earlier, a total of 18 studies applied CoT to improve LLMs' performance in SE tasks~\cite{deng2023language,feng2023prompting,huang2023ai,huang2023chain,li2023nuances,li2023structured,liu2024interpretable,mu2023clarifygpt}.

\vspace{0.1em}
\noindent\textbf{Automatic Prompt Engineer (APE).} Inspired by classical program synthesis and human prompt engineering methods, Zhou \ea~\cite{zhou2023large} introduced an Automatic Prompt Engineer (APE) for automatic instruction generation and selection. APE is a system designed to automatically generate effective prompts for LLMs based on the desired task. It aims to simplify the process of prompt engineering by automating the generation of task-specific instructions. Sharing a similar concept of automated prompts, Sun \ea~\cite{sun2023prompt} proposed a new prompt learning framework called PromptCS. PromptCS trains a prompt agent that can generate continuous prompts to fully explore LLMs' potential in code summarization tasks. Continuous prompts, generated under the guidance of LLMs, are easier for LLMs to comprehend compared to manually written discrete prompts.

\vspace{0.1em}
\noindent\textbf{Chain of Code (CoC) prompting.} CoC prompting~\cite{li2023chain} is similar to CoT prompting but is specifically tailored for programming tasks. It involves providing a sequence of prompts or code snippets to guide the model's code generation process. Huang \ea~\cite{huang2023codecot} proposed CodeCoT, and Le \ea~\cite{le2023codechain} proposed CodeChain, both of which are reasoning frameworks that better guide LLMs in code generation.

\vspace{0.1em}
\noindent\textbf{Automatic Chain-of-Thought (Auto-CoT) prompting.} Auto-CoT~\cite{zhang2022automatic} is an automated version of CoT prompting where the sequence of prompts is generated automatically based on the input and desired task. Paranjape \ea~\cite{paranjape2023art} introduced a framework, Automatic. ART, for generating intermediate reasoning steps automatically. ART can select multi-step reasoning and tools from a task library based on given tasks at any time and has been experimentally proven effective in code tasks.

\vspace{0.1em}
\noindent\textbf{Modular-of-Thought (MoT) prompting.} In code generation tasks, LLMs often generate solutions in the form of a single block of code, limiting their effectiveness in handling complex problems. To overcome this limitation, Li \ea~\cite{li2023motcoder} proposed the Modular-of-Thought Coder (MoTCoder). They introduced a new MoT prompting optimization framework to facilitate task decomposition into logical subtasks and submodules. Experimental results demonstrate that MoTCoder significantly improves the modularity and correctness of solutions generated by LLMs in programming tasks.

\vspace{0.1em}
\noindent\textbf{Structured Chain-of-Thought (SCoT) prompting.} Considering that source code contains rich structural information, Li \ea~\cite{li2023structured} proposed SCoT prompting specifically for code generation tasks. Researchers enable LLMs to use program structure to construct CoTs (i.e., intermediate natural language reasoning steps) to obtain SCoTs. Then, LLMs generate the final code based on SCoTs. Compared to CoT prompts, SCoT prompts explicitly constrain LLMs to consider how to address requirements from the source code perspective. Evaluations across multiple benchmarks show that SCoT significantly enhances LLMs' performance in code generation.

In addition to the eight prompting techniques mentioned above, we identified 76 studies where researchers, although not explicitly mentioning the application of any of the aforementioned prompting techniques, carefully designed prompts or proposed new strategies based on prompts to apply LLMs to SE tasks better. For instance, Ren \ea~\cite{ren2023misuse} proposed a code generation method based on knowledge-driven prompt chains. Li \ea~\cite{li2023nuances} applied differential prompting to ChaGPT to better identify test cases that cause failures in buggy programs. Ahmed \ea~\cite{ahmed2024automatic} enhanced the performance of LLMs in code summarization tasks using automatic semantic augmentation prompts.\looseness=-1

\subsection{How are evaluation metrics utilized to assess the performance of LLM4SE tasks?}

Evaluating the performance of LLM4SE is a crucial aspect of their development and deployment~\cite{kang2022large}. Benchmarking against existing datasets and using baselines are common practices to evaluate the effectiveness of LLMs~\cite{cassano2023multipl}. However, given the diversity of SE tasks, a single evaluation metric may not suffice to capture the model's performance comprehensively. Thus, researchers often employ a range of evaluation metrics tailored to specific problem types~\cite{mastropaolo2021studying,niu2022spt,salza2022effectiveness}. 
We categorize the SE tasks summarized from 395 papers into four categories according to their addressed problem types, i.e., regression, classification, recommendation, and generation tasks, as displayed in Fig.~\ref{fig:llm4se} (b).  
The selection of evaluation metrics depends on the target problem types. 
For example, MAE (Mean Absolute Error) has been used for regression tasks~\cite{fu2022gpt2sp}.
We summarize the most frequently used evaluation metrics for each task type.

\begin{table}[ht!]
\caption{Evaluation metrics for different types of tasks.}
\resizebox{\linewidth}{!}{
\begin{tabular}{|r|ll|c|}
\hline
\textbf{Problem Type} & \multicolumn{2}{c|}{\textbf{Metric}}  & \textbf{Total} \\ \hline
Regression
& MAE (Mean Absolute Error) (1) & & 1 \\
\hline
Classification
& Precision (35) & Recall (34) & 147 \\
& F1-score (33) & Accuracy (23) & \\
& AUC (Area Under the ROC Curve) (9) & ROC (Receiver Operating Characteristic) (4) & \\
& FPR (False Positive Rate) (4) & FNR (Falser Negative Rate) (3) & \\
& MCC (Matthews Correlation Coefficient) (2) & & \\
\hline
Recommendation
& MRR (Mean Reciprocal Rank) (15) & Precision/Precision@k (6) & 39 \\
& MAP/MAP@k (6) & F-score/F-score@k (5) & \\
& Recall/Recall@k (4) & Accuracy (3) & \\
\hline
Generation
& BLEU/BLEU-4/BLEU-DC (62) & Pass@k (54) & 338 \\
& Accuracy/Accuracy@k (39) & EM (Exact Match) (36) & \\
& CodeBLEU (29) & ROUGE/ROUGE-L (22) & \\
& Precision (18) & METEOR (16) & \\
& Recall (15) & F1-score (15) & \\
& MRR (Mean Reciprocal Rank) (6) & ES (Edit Similarity) (6) & \\
& ED (Edit Distance) (5) & MAR (Mean Average Ranking) (4) & \\
& ChrF (3) & CrystalBLEU (3) & \\
& CodeBERTScore (2) & MFR (Mean First Ranking) (1) & \\
& PP (Perplexity) (1) & & \\
\hline
\end{tabular}}
\begin{tablenotes}
\footnotesize
\item *See \autoref{app:evaluation} for the full table including references. 
\end{tablenotes}
\label{tab: evaluation}
\end{table}

For classification tasks, the most commonly used metrics are Precision~\cite{biswas2020achieving,chen2023diversevul,ezzini2022automated,fatima2022flakify,he2022ptm4tag}, Recall~\cite{biswas2020achieving,chen2023diversevul,ezzini2022automated,fatima2022flakify,he2022ptm4tag,hey2020norbert} and F1-score~\cite{alhamed2022evaluation,biswas2020achieving,chen2023diversevul,ezzini2022automated,fatima2022flakify,he2022ptm4tag}, with 35, 34, and 33 tudies, respectively, employing these metrics. For example, in the study conducted by Khan~\ea~\cite{khan2021automatic}, F1-score is utilized to evaluate the performance of an automatic bug-fixing model. Similarly, Sharma~\ea~\cite{sharma2022exploratory} use Precision and Recall to assess the effectiveness of a transformer-based model for code summarization. These metrics are essential for evaluating the model's ability to correctly classify code snippets~\cite{fatima2022flakify} or identify specific SE properties~\cite{chen2023diversevul}.

For recommendation tasks, MRR (Mean Reciprocal Rank) is the most frequent metric, used in 15 studies~\cite{ciborowska2022fast,izadi2022codefill,li2021toward,lin2021traceability,rahmani2023improving,salza2022effectiveness,shi2022cross,wei2022clear}. MRR is employed to measure the effectiveness of recommendation systems for code completion, as demonstrated in the study by Ciborowska \ea~\cite{ciborowska2022fast}. Precision@k~\cite{he2023representation,ciborowska2022fast,lin2021traceability,zhu2023automating} and F1-score@k~\cite{he2023representation,lin2021traceability,zhu2022enhancing,zhu2023automating} are also utilized in recommendation tasks, with 6 studies each. These metrics are used to evaluate the precision and F1-score of the recommended code snippets or code completions.

In generation tasks, metrics like BLEU, along with its variants BLEU-4 and BLEU-DC~\cite{ahmed2024automatic,al2023extending,arakelyan2023exploring,chen2022transferability,ciniselli2021empirical}, and Pass@k~\cite{bui2023codetf,cassano2023multipl,chen2023improving,chen2021evaluating,dibia2022aligning,doderlein2022piloting} are the most commonly used, appearing in 62 and 54 studies, respectively. For instance, Wang \ea~\cite{wang2023codet5+} employed BLEU to evaluate a code-to-code translation model. Pass@k is used in the research by Jiang \ea~\cite{jiang2023self} to assess code generation models, measuring the proportion of generated code snippets that match the reference solutions. Additionally, ROUGE/ROUGE-L~\cite{ahmed2024automatic,al2023extending,gao2023constructing,geng2024large,li2022auger,mastropaolo2021studying,mastropaolo2022using,niu2022spt,zan2023private,li2023exploring}, METEOR~\cite{ahmed2024automatic,al2023extending,chen2022transferability,gao2023constructing,niu2022spt,geng2024large}, EM~(Exact Match)~\cite{al2023extending,gao2023constructing,gupta2023grace,murali2023codecompose,wang2023codet5+,weyssow2023usage,ye2023generating,zhang2023neural}, and ES~(Edit Similarity)~\cite{liu2023repobench} are used in specific studies to evaluate the quality and accuracy of generated code or natural language code descriptions.

\begin{tcolorbox}[title=\textbf{RQ3} - Summary, left=2pt, right=2pt,top=2pt,bottom=2pt]


(1) We discovered a range of tuning techniques gradually becoming widely adopted in the LLM4SE domain. Among these, Parameter Efficient Fine-Tuning (PEFT) techniques, including Low-Rank Adaptation (LoRA), prompt tuning, prefix tuning, and adapter tuning, are gaining prominence for optimizing LLMs while minimizing computational complexity.

(2) We identified a diverse set of eight prompting techniques, including few-shot prompting, zero-shot prompting, Chain-of-Thought (CoT), Automatic Prompt Engineer (APE), Chain of Code (CoC), Automatic Chain-of-Thought (Auto-CoT), Modular-of-Thought (MoT), and Structured Chain-of-Thought (SCoT), applied in the LLM4SE domain to enhance model performance. These techniques leverage task-specific instructions, known as prompts, to guide LLMs without modifying core model parameters, providing a promising avenue for improving LLM capabilities in software engineering tasks.

(3) We summarized the most widely used evaluation metrics according to four problem types, i.e., regression, classification, recommendation, and generation. Nineteen different evaluation metrics appeared in the generation task, while nine metrics were used for the classification task.

\end{tcolorbox}

\section{RQ4: What SE tasks have been effectively addressed to date using LLM4SE?}
\label{sec:rq4}

\subsection{What are the distributions SE activities and problem types addressed to date with LLM4SE?}
\label{sec:problem_types}

\begin{figure}[ht!]
  \centering
  \subfloat[Distribution of LLM usages in SE activities.] {\includegraphics[width=2.6in]{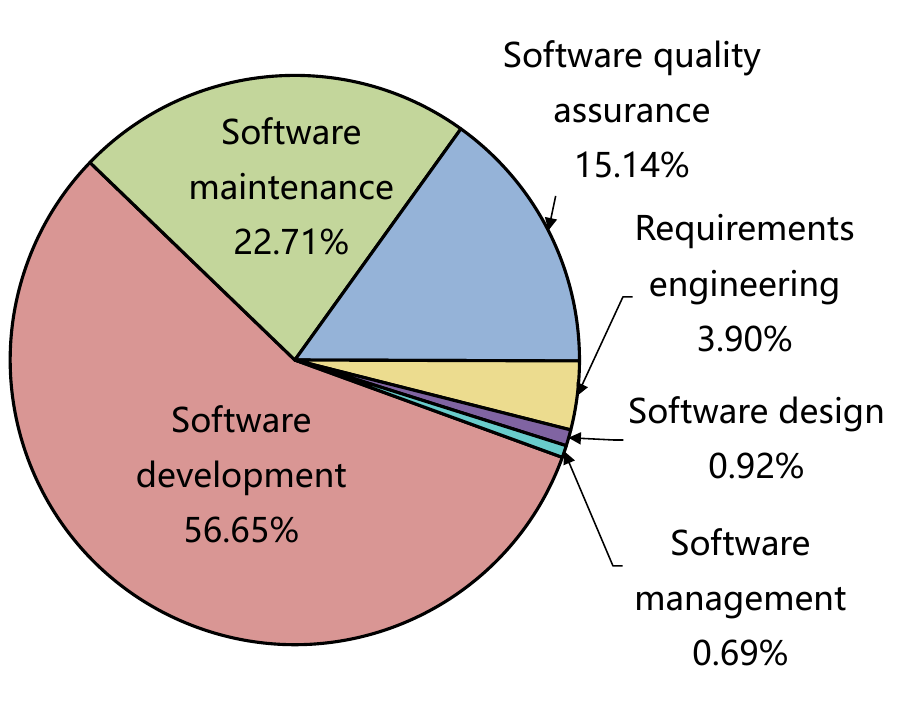}}
  \subfloat[Problem classification based on collected studies.]{\includegraphics[width=2.6in]{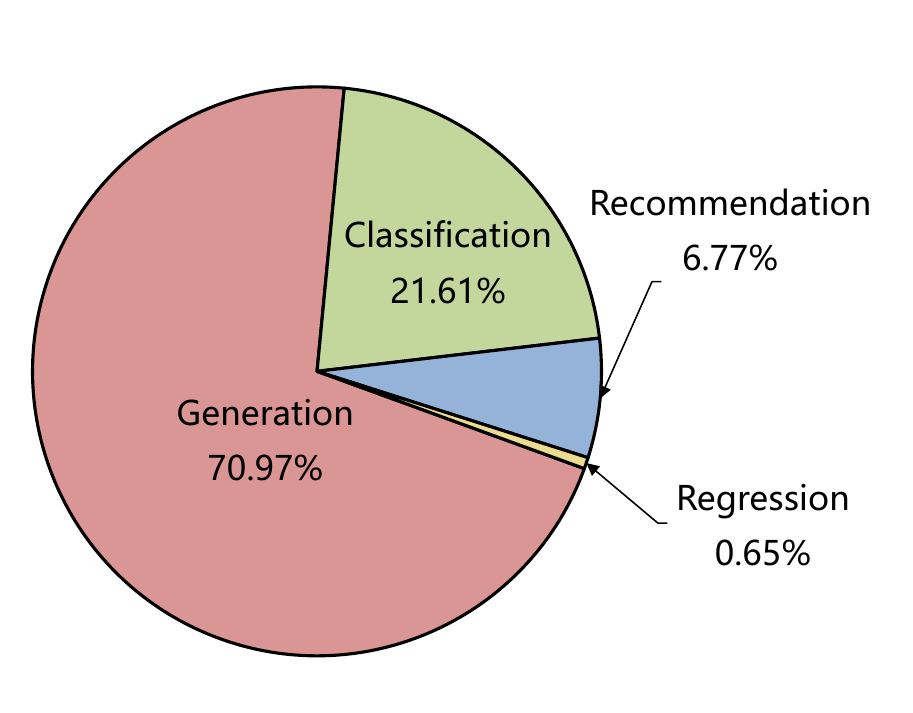}}
  \caption{Distribution of LLM utilization across different SE activities and problem types.}
  \label{fig:llm4se}
\end{figure}

In this section, we provide a detailed analysis of the use of LLMs in different SE tasks.
We summarise reported SE tasks~\cite{yang2022survey} addressed with LLMs, following the six phases of the \textbf{Software Development Life Cycle (SDLC)} (i.e., requirements engineering, software design, software development, software quality assurance, software maintenance, and software management). Fig.\ref{fig:llm4se} (a) describes the distribution of LLMs in these six activities. Table~\ref{tab:setasks} shows a detailed count of studies reporting specific SE tasks addressed with LLMs. 

\textbf{The highest number of studies is observed in the software development domain, constituting approximately 56.65\% of the total research volume}. This underscores the primary focus to date on utilizing LLMs to enhance coding and development processes. Software maintenance tasks account for about 22.71\% of the research share, highlighting the significance of LLMs in aiding software updates and improvements. The software quality assurance domain holds approximately 15.14\% of the research proportion, indicating a growing interest in automating testing procedures. In contrast,  requirements engineering and software design activities represent approximately 3.9\% and 0.92\% of the research share, respectively, suggesting relatively limited exploration so far in these areas. The software management domain has the least research representation, accounting for a tiny 0.69\% proportion. This distribution underscores the vital focus on development and maintenance tasks while also indicating potential avenues for further research in testing, design, and management domains.

In our collection of LLM studies for SE tasks, we've classified them based on the type of problems they address (shown in Fig.\ref{fig:llm4se} (b)). The distribution reveals that \textbf{the majority of studies, about 70.97\%, center around generation tasks}, showcasing the significance of LLMs in producing code or text. Following this, around 21.61\% of studies fall under classification tasks, indicating the relevance of LLMs in categorizing software elements. Additionally, roughly 6.77\% of studies are related to recommendation tasks, demonstrating the utility of LLMs in suggesting solutions. Lastly, a smaller portion, around 0.65\%, is allocated to regression tasks, reflecting the limited exploration of LLMs for predictive modeling. \textbf{This distribution underscores the broad applicability of LLMs across different SE challenges, with a notable emphasis on code generation and classification tasks}.

\begin{table}[ht!]
\caption{Distribution of SE tasks over six SE activities.}
\resizebox{\linewidth}{!}{
\begin{tabular}{|r|ll|c|}
\hline
\textbf{SE Activity} & \multicolumn{2}{c|}{\textbf{SE Task}}  & \textbf{Total} \\ \hline
Requirements
& Anaphoric ambiguity treatment (4) & Requirements classification (4) & 17 \\
engineering
& Requirement analysis and evaluation (2) & Specification generation (2) & \\
& Coreference detection (1) & Requirements elicitation (1) & \\
& Specification formalization (1) & Traceability automation (1) & \\
& Use cases generation (1) & & \\
\hline
Software design 
& GUI retrieval (1) & Rapid prototyping (1) & 4 \\
& Software specification synthesis (1) & System design (1) & \\
\hline
Software development 
& Code generation (118) & Code completion (22) & 247 \\
& Code summarization (21) & Code search (12) & \\
& Code translation (12) & Code understanding (8) & \\
& API inference (5) & Program synthesis (6) & \\
& API recommendation (5) & Code editing (5) & \\
& Code representation (3) & Code comment generation (2) & \\
& Method name generation (2) & Code recommendation (2) & \\
& Agile story point estimation (1) & API documentation augment (1) & \\
& API documentation smells (1) & API entity and relation extraction (1) & \\
& Data analysis (1) & Fuzz driver generation (1) & \\
& Control flow graph generation (1) & Identifier normalization (1) & \\
& Instruction generation (1) & Type inference (1) & \\
& Others (14) & & \\
\hline
Software quality 
& Vulnerability detection (18) & Test generation (17) & 66 \\
assurance
& Bug localization (5) & Verification (5) & \\
& Testing automation (4) & Fault localization (3) & \\
& Defect detection (2) & GUI testing (2) & \\
& Static analysis (2) & Binary taint analysis (1) & \\
& Compiler fuzzing (1) & Decompilation (1) & \\
& Invariant prediction (1) & Malicious code localization (1) & \\
& Mobile app crash detection (1) & Resource leak detection (1) & \\
& Test prediction (1) & & \\
\hline
\multirow{9}{*}{Software maintenance}  
& Program repair (35) & Code clone detection (8) & 99 \\
& Code review (7) & Debugging (4) & \\
& Bug reproduction (3) & Review/commit/code classification (3) & \\
& Duplicate bug report detection (3) & Logging (3) & \\
& Log parsing (3) & Code revision (2) & \\
& Sentiment analysis (3) & Vulnerability repair (2) & \\
& API misuses repair (1) & Bug prediction (1) & \\
& Bug triage (1) & Code coverage prediction (1) & \\
& Code review explained (1) & Code-Review defects repair (1) & \\
& Crash bug repair (1) & Dockerfile Repair (1) & \\
& Incivility detection (1) & Patch correctness prediction (1) & \\
& Patch detection (1) & Program merge conflicts repair (1) & \\
& Rename Refactoring (1) & Tag recommendation (1) & \\
& Technical debt payback (1) & Traceability recovery (1) & \\
& Web test repair (1) & Type error repair (1) & \\
& Others (5) & & \\
\hline
Software management 
& Effort estimation (2) & Software tool configuration (1) & 3 \\
\hline
\end{tabular}}
\begin{tablenotes}
\footnotesize
\item *See Appendix~\ref{app:tasks} for the full table including references. 
\end{tablenotes}
\label{tab:setasks}
\end{table}

\subsection{How are LLMs used in requirements engineering?}
This section explores the utilization of LLMs in the domain of requirements engineering. It encompasses tasks such as anaphoric ambiguity treatment, requirements classification, coreference detection, requirements elicitation, and software traceability.

\noindent\textbf{Anaphoric ambiguity treatment.} Ambiguity in software requirements arises when a single reader can interpret a natural language (NL) requirement in multiple ways, or when different readers have varying understandings of the same requirement. Unclear and ambiguous NL software requirements can lead to suboptimal software artifacts during later development stages. 
Moharil~\ea~\cite{moharil2023tabasco} and Ezzini~\ea~\cite{ezzini2022automated} have empirically demonstrated the significant role of LLMs such as BERT and SpanBERT in effectively addressing anaphoric ambiguity.
Sridhara \ea~\cite{sridhara2023chatgpt} revealed that ChatGPT excels in addressing anaphoric ambiguity in software requirements. Through researchers’ analysis of ten English requirement specifications~\cite{ezzini2022automated} containing anaphora-related challenges, ChatGPT consistently demonstrated its remarkable capability to accurately identify antecedents. This empirical evidence emphasizes the valuable role ChatGPT can play in enhancing the clarity and precision of software requirements, ultimately contributing to more effective software development processes by reducing interpretational uncertainties.

\noindent\textbf{Requirements classification.} Originating in NL documents, requirements demand effective classification, especially for early-stage project discernment, like security-related ones~\cite{knauss2011supporting,li2014non}. Automated processing hinges on identifying these requisites. 
Categorizing into functional (FR) or non-functional (NFR) requirements, with quality constraints, benefits automated approaches~\cite{li2014non}.
Hey \ea\cite{hey2020norbert} employ BERT for requirement classification, where it excels in categorizing both FR and NFR requirements using a fine-tuning transfer learning technique, outstripping traditional methods. Luo \ea\cite{luo2022prcbert} introduce a BERT-based software requirement classification method, demonstrating remarkable transferability and generalization, especially in zero-shot scenarios.

\noindent\textbf{Requirements term identification.} 
Moharil \ea~\cite{moharil2022identification} propose a technique for identifying terms used in different contexts within the same domain or in interdisciplinary projects. Using BERT, which reads entire word sequences for deeper language understanding, and K-means clustering, they create and group vectors for each term in the corpora. The method has been validated on large Computer Science and multi-domain corpora comprising eight different fields.

\noindent\textbf{Coreference detection.} Requirements, authored by diverse stakeholders, continually evolve, leading to terminology differences and inconsistencies across domains. Entity coreference in Requirement Engineering (RE), where various expressions refer to the same real-world entity, can cause confusion and affect comprehensibility. Wang \ea~\cite{wang2020deep} offer a novel application of the BERT model for coreference detection.

\noindent\textbf{Traceability automation.} 
Software and system traceability refers to the ability to establish and maintain relationships between software artifacts, such as requirements, design definitions, code, and test cases, for product querying and development support~\cite{rierson2017developing}. Lin \ea~\cite{lin2021traceability} found that T-BERT can effectively migrate knowledge from code search to NLA-PLA (i.e., Natural Language Artifacts to Programming Language Artifacts) traceability, even with limited training instances. It outperforms existing techniques in accuracy and can be adapted to different domains without intermediate training for each project, offering a promising step toward practical, trustworthy traceability.

\noindent\textbf{Others.} 
In addition to the four requirement engineering tasks detailed above, LLMs can also be applied to requirement analysis and evaluation~\cite{poudel2023leveraging,ronanki2023chatgpt}, specification generation~\cite{ma2024specgen,xie2023impact}, requirements elicitation~\cite{white2023chatgpt}, specification formalization~\cite{endres2023formalizing}, and use case generation~\cite{zhang2024experimenting}.
    
\subsection{How are LLMs used in software design?}

\noindent\textbf{GUI (Graphical User Interface) retrieval.} Kolthoff \ea~\cite{kolthoff2023data} present the application of BERT in the task of GUI retrieval in SE. The authors fine-tune a BERT-based learning-to-rank (LTR) model for this task. GUIs, which are not standard well-structured text documents, present unique challenges for text-based ranking tasks. The BERT model is prepared by concatenating the natural language query and the GUI document text, and then this input is used to train different BERT-LTR models. The models are evaluated based on their performance in NL-based GUI ranking.

\noindent\textbf{Rapid prototyping.} Rapid prototyping enables developers to quickly visualize and iterate on software designs, thereby accelerating the development process and ensuring alignment with user needs. White \ea~\cite{white2023chatgpt} investigate the role of LLMs in augmenting this process. The study introduces prompt design techniques, organized into patterns, providing a structured methodology to tackle prevalent challenges in LLM4SE. This research indicates that the realm of rapid prototyping stands to benefit from deeper integration with advanced machine learning techniques, thereby creating opportunities for additional research and refinement aimed at producing more intuitive and user-centric software designs.

\noindent\textbf{Software specification synthesis.} 
Software configuration is vital for system behavior, but managing configurations and specifications becomes complex with larger systems. Mandal \ea~\cite{mandal2023large} introduce SpecSyn, a framework using an LLM for automatic software specification synthesis from natural language sources. This end-to-end approach treats the task as a sequence-to-sequence learning problem, surpassing the previous state-of-the-art tool by 21\% in F1 score, and can find specifications from both single and multiple sentences.

\subsection{How are LLMs used in software development?}

Our analysis identifies wide-ranging applications of LLMs for software development, encompassing tasks such as code generation, code completion, and code summarization.

\noindent\textbf{Code generation.} Code generation has long been a task of interest: there is extensive work on program synthesis using symbolic and neural-semiotic approaches~\cite{alur2013syntax,wu2023ai}. Recently, LLMs trained for text generation have demonstrated the ability to complete programs~\cite{brown2020language,black2022gpt}. Since 2020, several code generation models have been trained or fine-tuned on programming language text~\cite{nijkamp2022conversational,chen2021evaluating,fried2022incoder,xu2022systematic,feng2020codebert,clement2020pymt5}. Unlike traditional program synthesis techniques, neurolinguistic models can be conditioned on natural language (e.g., code annotations) as well as generate programming language text. Researchers have experimentally demonstrated that LLMs like GPT-4~\cite{bareiss2022code,liu2023your,jiang2023selfevolve,gilbert2023semantic}, GPT-2/GPT-3/GPT-3.5~\cite{azaria2023chatgpt,yeticstiren2023evaluating,ke2023discriminating,liu2023your,nascimento2023comparing,li2023enabling,wang2023evaluating,liu2023improving,dong2023self}, BERT series~\cite{zeng2022extensive,lai2023ds}, Codex~\cite{dibia2022aligning,bareiss2022code,yu2023codereval,chen2021evaluating,gupta2023grace,madaan2022language,kuznia2022less}, CodeGen~\cite{dibia2022aligning,jones2022capturing,zan2022language}, InCoder~\cite{murali2023codecompose,kou2023model,liu2023your,wang2022recode}, Copilot~\cite{wu2023ai} and CodeGeeX~\cite{zheng2023codegeex}, play a key role in code generation. By pre-training on large-scale text data, these models learn rich linguistic knowledge and semantic representations that enable them to understand the meaning and structure of natural language. LLMs can automate code generation by converting natural language descriptions into code~\cite{jiang2023self}. These models generate program code from natural language descriptions, enhancing code-writing efficiency and accuracy. They show excellent performance in code completion, automatic code generation, and conversion of natural language annotations to code, providing software developers with powerful auxiliary tools and promoting further automation and intelligence in the code writing and development process.

Within the domain of LLMs applied to software development tasks, studies centered on code generation distinctly dominate the academic landscape. As reflected in Table~\ref{tab:code_generation}, \textbf{the GPT series, particularly GPT-4, emerged as a key focus, with many more studies using them in the realm of code generation}~\cite{dong2023self, du2023classeval, li2023enabling, liu2023your}. 
Analyzing these studies, several noteworthy findings surface:
\begin{itemize}
    \item\textbf{Programming thinking in LLMs.} Techniques that evoke ``programming thinking'' within LLMs, such as the TIP (i.e., Thinking in Programming)~\cite{li2023enabling} methodology, have shown promising strides. By guiding LLMs to first craft a high-level code sketch before delving into detailed implementations, the synthesized code exhibits higher accuracy and robustness.
    \item\textbf{Class-level vs. Method-level generation.} LLMs, while adept at method-level code generation, present varied performance metrics when tasked with class-level generation~\cite{du2023classeval}. This divergence underscores the evolving nature of challenges as the granularity of code synthesis shifts.
    \item \textbf{Expanding LLM capabilities.} The next frontier in this discipline seems to lie in harmoniously integrating LLMs with established SE tools and practices. The emergence of frameworks like EvalPlus~\cite{dong2023self} indicates a trend towards enhancing the evaluation and accuracy of LLM-generated code, possibly ushering in an era where human developers and LLMs collaboratively craft software solutions.
\end{itemize}

\begin{table}[t]
\caption{The state-of-the-art applications of LLMs in code generation task.}
\resizebox{\linewidth}{!}{
\begin{tabular}{|p{0.1\linewidth}|p{0.45\linewidth}|p{0.18\linewidth}|l|l|l|}
\hline
\textbf{Model} & \textbf{Baseline} & \textbf{Benchmark} & \textbf{Metric} & \textbf{Date} & \textbf{Reference} \\ 
\hline
GPT-3.5 & Codex, CodeGen, CodeGeeX, LLaMA, InCoder, PyCodeGPT, CodeParrot, GPT-2 & HumanEval, MBPP, MBCPP & Pass@k & May 11, 2023 & \cite{li2023enabling} \\ 
\hline
GPT-4 & PaLM Coder, Codex, CodeGen-Mono, Incoder, CodeGeeX, AlphaCode & HumanEval, HumanEval-ET, MBPP, MBPP-ET & Pass@k & May 24, 2023& \cite{dong2023self} \\ 
\hline
GPT-4 & GPT-3.5, StarCoder, CodeGen, CodeGen2, Vicuna, SantaCoder, Incoder, GPT-J, GPT-Neo, PolyCoder, StableLM & HumanEval, HumanEval+, HumanEval-mini & Pass@k & Jun 12, 2023 & \cite{liu2023your} \\ 
\hline
GPT-4 & GPT-3.5, WizardCoder, Instruct-StarCoder, SantaCoder, Instruct-CodeGen, CodeGeeX, InCoder, Vicuna, ChatGLM, PolyCoder & ClassEval,     HumanEval & Pass@k & Aug 3, 2023 & \cite{du2023classeval}\\ 
\hline
\end{tabular}}
\label{tab:code_generation}
\end{table}
    
\noindent\textbf{Code completion.} Code completion is an assistive feature provided by many integrated development environments (IDEs) and code editors. Its purpose is to automatically display possible code suggestions or options as developers write code~\cite{amann2016study}. This innovation has been advanced by Language Models (LMs), evolving from n-gram and RNN models to transformer-based models like Copilot~\cite{github2021copilot} and CodeGPT~\cite{Judini2023codegpt}, pre-trained on extensive code datasets. Recent LLMs equipped with billions of parameters, excel in generating code snippets. 
These models are trained on vast amounts of natural language text, equipping them with powerful semantic understanding capabilities. In the context of code completion, LLMs such as Codex~\cite{li2022cctest,pearce2021examining,doderlein2022piloting,chen2021evaluating}, BERT series~\cite{khan2022automatic}, GitHub Copilot~\cite{li2022cctest,pudari2023copilot,doderlein2022piloting}, CodeParrot~\cite{li2022cctest,xu2022systematic}, GPT series~\cite{xu2022systematic,ochs2023evaluating}, T5~\cite{ciniselli2021empirical}, InCoder~\cite{fried2022incoder}, PolyCoder~\cite{xu2022systematic}, CodeGen~\cite{ding2023static,dinh2023large,li2022cctest,nijkamp2022codegen}, and other LLMs~\cite{izadi2022codefill,ochs2023evaluating}, can generate accurate and intelligent code suggestions based on code context and syntax structures. They comprehend the developer's intent, predict the next possible code snippet, and provide appropriate recommendations based on the context.

With the support of LLMs, code completion achieves significant improvements in efficiency and accuracy. Developers can save time by avoiding manual input of lengthy code and reducing the risk of code errors. LLMs also learn from extensive code repositories, acquiring knowledge and best practices to offer more intelligent and precise suggestions, aiding developers in better understanding and utilizing code~\cite{ciniselli2021empirical}. Additionally, these models can provide personalized code recommendations based on developers' coding styles and preferences, further enhancing the effectiveness and user experience of code completion~\cite{liu2023repobench}. 

\noindent\textbf{Code summarization.} 
Code summarization is a task that attempts to understand the code and automatically generate descriptions directly from the source code. It can also be viewed as an extended form of documentation. Successful code summarization not only facilitates the maintenance of source code~\cite{iyer2016summarizing,nguyen2017automatic} but can also be used to improve the performance of code search using natural language queries~\cite{nie2016query,yang2016query} and code classification~\cite{nguyen2017automatic}. LLMs play a significant role in code summarization by analyzing code structures and contexts to generate informative natural language summaries.
Specifically, LLMs such as Codex~\cite{gao2023constructing,arakelyan2023exploring,ahmed2023improving}, CodeBERT~\cite{chen2022transferability,gu2022assemble,gao2023constructing}, and T5~\cite{mastropaolo2021studying,mastropaolo2022using} comprehend the functionality and logic of the code, producing easily understandable human language descriptions. 
For example, Arakelyan \ea~\cite{arakelyan2023exploring} rigorously evaluate the efficacy of CodeT5 and Codex across code generation and summarization tasks, shedding light on their performance under distribution shifts. It unveils practical adaptation techniques, underscoring Codex's commendable performance. Additionally, the study demonstrates that while adapted models exhibit proficiency in code generation, their generality can present trade-offs in the context of code summarization.
As a result, code summarization with the support of LLMs enhances code readability, improves software documentation quality, and accelerates code comprehension and collaboration among developers. This advanced approach to code summarization demonstrates great potential for automating and streamlining various aspects of software development in modern SE practices with the employment of LLMs.

\noindent\textbf{Code search.} Code search, or code retrieval, is the task of retrieving source code from a large code base, usually based on a user's natural language query.
Despite the success of neural models in code search, such models are relatively shallow and are not capable of learning large amounts of data~\cite{salza2022effectiveness}. In recent years, some bimodal pre-training models based on the BERT neural architecture have been proposed to capture semantic links between natural and programming languages~\cite{feng2020codebert,guo2020graphcodebert,roziere2021dobf,wang2021syncobert}, such as CodeBERT~\cite{feng2020codebert} and GraphCodeBERT~\cite{guo2020graphcodebert}. Bimodal pre-training models learn generic representations from large amounts of data in an unsupervised manner by designing pre-training goals. Salza \ea~\cite{salza2022effectiveness} explored the effectiveness of LLMs such as BERT~\cite{salza2022effectiveness} and RoBERTa~\cite{chen2022transferability} in understanding natural language and code semantics and enhancing code search and retrieval. These studies show that pre-training tasks alone may not be sufficient for code search, which emphasizes the need for a multimodal understanding of data~\cite{shi2022cross}, including both natural language and code. In addition, research has shown that the use of code generation models such as Codex~\cite{li2022generation} can enhance code retrieval by generating code snippets from natural language documents, thereby improving semantic similarity and obtaining state-of-the-art results on benchmark datasets.

\noindent\textbf{Code understanding.} In contrast to code summarization, which focuses on automatically generating human-readable descriptions from source code, code understanding involves a deep analysis of source code to comprehend its logic, structure, functionality, and dependencies, as well as understanding the programming languages, frameworks, and libraries used~\cite{shen2022benchmarking}. LLMs can assist in code understanding by leveraging their powerful natural language processing capabilities to interpret code-related text, such as comments and documentation~\cite{wang2023codet5+,kanade2020learning}. They aid developers in grasping code functionality, identifying dependencies, and generating relevant code documentation~\cite{shen2022benchmarking,ma2023scope}. Through their ability to comprehend both code and natural language, LLMs enhance the efficiency and accuracy of code understanding, empowering developers to maintain, optimize, and integrate code effectively~\cite{kanade2020learning}.

\noindent\textbf{Program synthesis.} 
Program synthesis is the automated process of generating code that satisfies a given specification or set of constraints, emphasizing the derivation of functional properties of the code~\cite{chen2017towards,chen2021latent,manna1980deductive,srivastava2010program,parisotto2016neuro}. It differs from code generation, which primarily translates higher-level representations into target code without necessarily deriving its functionality from scratch~\cite{siddiq2023lightweight,zhang2023self,zheng2023codegeex}.
Several studies have demonstrated that LLMs can be used for program synthesis tasks. LLMs have a significant impact on program synthesis due to their advanced language understanding and generation capabilities. LLMs can effectively interpret natural language descriptions, code comments, and requirements, and then generate corresponding code snippets that fulfill the given specifications. This helps developers rapidly prototype code and automate repetitive coding tasks~\cite{kuznia2022less,gandhi2023natural}. When applied to program synthesis, LLMs enhance productivity and reduce the burden on developers by automating the code-writing process based on high-level input~\cite{jain2022jigsaw}. Their ability to understand the nuances of both natural language and programming languages makes them valuable tools in advancing the field of SE and streamlining the development lifecycle.

\noindent\textbf{API recommendation.}
Several methods have been proposed to automate API (Application Programming Interface) recommendations~\cite{gu2016deep,huang2018api,liu2018effective,nguyen2016api}, falling into two orthogonal approaches: information retrieval-based (IR-based) and neural-based. In this context, our focus is on the latter.
Wei \ea~\cite{wei2022clear} introduced CLEAR, an API recommendation method that employs the BERT sentence embedding model to represent queries, capturing continuous semantic information. Through contrast training, CLEAR enables BERT to learn precise semantic representations of queries, independent of their lexical content.
Recently, Zhang \ea~\cite{zhang2023toolcoder} developed ToolCoder, which combines API search tools with existing models to aid in code generation and API selection. This approach involves an automated data annotation method using ChatGPT, adding tool usage information to the source code data, followed by fine-tuning the code generation model. During inference, an API search tool is integrated into the generation process, allowing the model to utilize the tool for suggestions when selecting APIs automatically.

\noindent\textbf{API inference.} 
The automated generation of application programming interface calls, known as API synthesis, plays a crucial role in bridging human intent with machine execution. In recent studies, Wang \ea~\cite{wang2023measuring} and Patil \ea~\cite{patil2023gorilla} have both explored the potential of LLMs in this realm. Utilizing models like GPT-4 and LLaMA-based architectures, these researchers showcase the prowess of LLMs in generating accurate API calls and adapting to real-time documentation changes, effectively addressing challenges like hallucination and inaccurate input arguments. The integration of LLMs in API synthesis signifies a paradigm shift, promising enhanced accuracy, adaptability, and reliability in code generation. As illuminated by these studies, the future of API synthesis may be deeply anchored in advanced machine learning, heralding new research avenues and refinements for more seamless human-machine interactions.

\noindent\textbf{Code representation.} 
Code representation learning (also known as code embedding) aims to encode the code semantics into distributed vector representations and plays a key role in recent deep-learning-based models for code intelligence. Code representation can be used to support a variety of downstream tasks, such as code completion~\cite{raychev2014code}, code search~\cite{gu2018deep,wan2019multi}, and code summarization~\cite{wan2018improving,zhang2020retrieval}. Niu \ea~\cite{niu2022spt} propose a novel sequence-to-sequence pre-training model that utilizes structural information from source code to enhance its representation learning. The model is trained on a large corpus of source code, which enables it to capture the complex patterns and dependencies inherent in programming languages. 
Wan \ea~\cite{wan2022they} show through their research that attention is highly consistent with the syntactic structure of the code, that pre-trained code language models can preserve the syntactic structure of the code in the intermediate representations of each converter layer, and that pre-trained code models have the ability to induce a syntactic tree of the code. These revelations suggest that incorporating the syntactic structure of the code into the pre-training process results in better code representations.

\noindent\textbf{Code comment generation.} 
Code comment generation, the automatic creation of comments for source code, serves to elucidate code functionality, implementation logic, and input-output details, thereby enhancing readability and maintainability~\cite{geng2024large}. As code complexity grows, manually crafting these comprehensive and accurate comments can become burdensome and prone to errors. Automation in this domain can markedly enhance the efficiency and quality of code documentation.
LLMs such as Codex~\cite{geng2024large} and T5~\cite{mastropaolo2021empirical} have been effectively applied to code comment generation. These models are pre-trained on vast amounts of data and possess powerful natural language processing and semantic understanding capabilities. During comment generation, LLMs analyze the structure, semantics, and context of the source code to automatically generate high-quality comments that correspond to the code's functionality and logic. Addressing the often observed disconnect between code evolution and its accompanying documentation, Mastropaolo \ea~\cite{mastropaolo2021empirical} explore the potential of LLMs, particularly the T5 architecture, in assisting developers with code comment completion. Their empirical study juxtaposes the performance of the T5 model against an n-gram model, revealing T5's superior capabilities, though the n-gram model remains a competitive alternative. The research underscores the significance of open-source datasets for training and highlights the scant use of industrial datasets in current studies.

\noindent\textbf{Method name generation.} 
Method names significantly affect program comprehensibility, serving as a brief summary of the source code and indicating the developer's intent~\cite{ko2006exploratory}. The importance of method names in program comprehension is further evidenced by recent studies showing that some programmers even write down important method names to help them figure out the procedures of an application~\cite{roehm2012professional}.
Zhu \ea~\cite{zhu2023automating} present AUMENA, a novel approach using the CodeT5 model for context-aware method naming in SE. AUMENA first learns the contextualized representation of programming and natural language, then leverages LLMs with prompt tuning to detect inconsistent method names and suggest accurate alternatives. This method avoids previous generate-then-compare consistency checking limitations, modeling the task as a two-class classification problem. 

\noindent\textbf{Agile story point estimation.} 
Agile story point estimation, representing the total work needed to implement a product backlog item, is a complex task in agility. Story points are typically estimated by team consensus, using methods like plan poker and expert judgment, and considering factors like workload and complexity. However, subjective estimates may introduce uncertainty. Fu \ea~\cite{fu2022gpt2sp} present GPT2SP, a Transformer-based approach that overcomes limitations of a previous method called Deep-SE. Unlike Deep-SE, which restricts language models to known words within a trained project, GPT2SP employs a broader context, making it transferable across projects. GPT2SP's performance is comparable to Deep-SE in within-repository evaluations and surpasses it in 62.5\% of cases, with improvements ranging from 3\% to 46\% across various projects.

\noindent\textbf{API documentation smell detection.} 
APIs, vital for modern software development, are often accompanied by official documentation. Good documentation is key to proper API use, while poor quality can hinder adoption and negatively impact developers' productivity~\cite{aghajani2020software,robillard2009makes,robillard2011field}. Khan \ea~\cite{khan2021automatic} identified five API documentation smells and presented a benchmark of 1,000 API documentation units containing the five smells found in the official API documentation. The authors developed classifiers to detect these odors, with BERT showing the best performance, demonstrating the potential of LLMs in automatically monitoring and warning about API documentation quality.

\noindent\textbf{API entity and relation extraction.}
Extracting APIs and their semantic relationships from unstructured text (e.g., data from Stack Overflow) is a fundamental task in SE, but existing methods require labor-intensive manual rule creation or data labeling. Huang \ea~\cite{huang2023api} present an innovative approach, AERJE, that leverages LLMs for this task. AERJE consists of a BERT-based dynamic hint generator and a T5-based joint entity-relationship extractor, which together enable efficient extraction of API entities and relationships without manual effort. The approach achieved an F1 score of 96.51\% for API entity extraction and 81.2\% for API relationship extraction, offering a significant advancement over traditional methods.

\noindent\textbf{Code recommendation.} 
Zhou \ea~\cite{zhou2019lancer} pointed out that software developers tend to write similar code examples several times due to the need to implement similar features in different projects. Therefore, during the software development process, recommender systems can provide programmers with the most pertinent and high-quality examples written by other programmers, thus helping them to complete their tasks quickly and efficiently~\cite{di2021development}. Open-source projects and informal documentation are the two main sources of information that developers rely on to perform programming tasks. For example, open-source projects on GitHub provide code examples and code resources for various tasks. 
Rahmani \ea~\cite{rahmani2023improving} introduce a methodology to improve code example recommendations for Java programming language on Stack Overflow using BERT and Query-Aware Locality-Sensitive Hashing (LSH). They employ BERT to convert code into numerical vectors and then apply two LSH variants, Random Hyperplane-based, and Query-Aware, to identify Approximate Nearest Neighbors (ANN). 

\noindent\textbf{Control flow graph generation.} Control Flow Graphs (CFGs) are a cornerstone of SE that illustrate program behavior by showing sequences of statements and their execution order conditions~\cite{allen1970control}. As a graphical representation of program behavior, CFGs are critical in many SE tasks, including code search~\cite{guo2020graphcodebert,chen2019capturing}, code clone detection~\cite{wang2020detecting,hu2018deep,wei2017supervised} and code classification~\cite{wang2020modular,zhang2019novel}.
Huang \ea~\cite{huang2023ai} presented a novel approach for generating behaviorally correct CFGs of statically typed partial code by leveraging the error-tolerant and understanding ability of LLMs. The approach involves a Chain of Thoughts (CoT) with four steps: structure hierarchy extraction, nested code block extraction, CFG generation of nested code blocks, and fusion of all nested code blocks' CFGs~\cite{le2022autopruner}. The CoT is broken down into an AI chain according to the single responsibility principle, along with effective prompt instructions. This results in superior node and edge coverage compared to traditional program analysis-based methods and the original CoT method. 

\noindent\textbf{Identifier normalization.} Identifiers usually consist of multiple words, and a certain number of identifiers contain abbreviations~\cite{jiang2020automated}. Consequently, the lexical meaning of identifiers and the overall functionality of source code written by one developer may be challenging for other developers to comprehend. In addition, the source code cannot match the vocabulary in other software artifacts described in natural language, thus invalidating some automated algorithms. Therefore, there is a strong need to normalize identifiers with the aim of aligning the vocabulary in identifiers with the natural language vocabulary in other software artifacts. 
Zhang \ea~\cite{zhang2022beqain} addressed this by introducing BEQAIN, an approach for identifier normalization. BEQAIN combines BERT with a Question and Answering (Q\&A) system and Conditional Random Fields (CRF), treating identifier splitting as sequence labeling and abbreviation expansion as a Q\&A task. It uses programming context to refine expansion results when multiple expansions are possible, aligning identifier vocabulary with natural language and enhancing software development comprehension and automation.

\noindent\textbf{Type inference.} 
Type inference, the automated process of determining data types in programming, plays a crucial role in enhancing readability, maintainability, and reducing runtime errors~\cite{hellendoorn2018deep,pierce2000local}. TypeScript, with its unique blend of optional typing, presents a nuanced challenge, especially when navigating the vast landscape of user-defined types. Addressing this complexity, Jesse \ea~\cite{jesse2022learning} introduced an approach that leverages the capabilities of a BERT-style pre-trained model. Their solution, DIVERSETYPER, adeptly infers types for user-defined classes and interfaces by uniquely correlating class and interface declarations with their respective usage contexts. Beyond merely filling the gaps of previous methodologies, DIVERSETYPER sets a new benchmark in type inference, especially for user-defined types.

\noindent\textbf{Others.} 
In addition to the 18 software development tasks detailed above, LLMs can also be applied to code translation~\cite{jana2023attention,pan2023stelocoder,pan2023understanding,qi2023sut,yan2023codetransocean,yang2023assessing}, code editing~\cite{bairi2023codeplan,gupta2023grace,li2023codeeditor,moon2023coffee,shypula2023learning}, API documentation augment~\cite{yang2023apidocbooster}, data analysis~\cite{cheng2023gpt}, fuzz driver generation~\cite{zhang2023understanding}, instruction generation~\cite{zhou2023large}.

\subsection{How are LLMs used in software quality assurance?}

Within the domain of software quality assurance, LLMs have emerged as valuable tools with diverse applications for various tasks, including vulnerability detection, test generation, bug localization, verification, test automation, etc. 

\noindent\textbf{Vulnerability detection.} 
The number of software vulnerabilities is rapidly increasing, as shown by the vulnerability reports from Common Vulnerabilities and Exposures (CVEs)~\cite{anon2022national} in recent years. As the number of vulnerabilities increases, there will be more possibilities for cybersecurity attacks, which can cause serious economic and social harm. Therefore, vulnerability detection is crucial to ensure the security of software systems and protect social and economic stability. Traditional static detection methods are based on static analysis and predefined matching rules, which rely on developers' expertise and make it difficult to detect unknown vulnerabilities. 
With the assistance of LLMs~\cite{thapa2022transformer,chan2023transformer,chen2023diversevul}, Tang \ea~\cite{tang2023csgvd} introduced novel approaches using LLMs to enhance vulnerability detection. One of their proposed models, CSGVD, combines sequence and graph embedding for function-level vulnerability detection, outperforming other deep learning-based models on a real-world benchmark dataset. Their study also explores the application of CodeT5 for vulnerability detection, highlighting the importance of code-specific pre-training tasks.

\noindent\textbf{Test generation.} Test generation involves automating the process of creating test cases to evaluate the correctness and functionality of software applications. It encompasses various aspects, including test case generation~\cite{zhang2023algo}, unit test generation~\cite{tang2023chatgpt,yuan2023no,schafer2023adaptive,xie2023chatunitest,siddiq2023exploring}, etc. 
LLM application in test generation offers several advantages, including the ability to automatically generate diverse test cases, improving test coverage~\cite{schafer2023adaptive,siddiq2023exploring} and identifying potential defects~\cite{xie2023chatunitest}. 
LLMs can also assist in generating test cases based on natural language descriptions, fostering better collaboration between developers and testers. Additionally, they help identify areas lacking test coverage and suggest relevant test cases, ensuring comprehensive testing and reducing the risk of undiscovered issues~\cite{zhang2023algo}. By enhancing test efficiency and effectiveness, LLMs contribute to producing more reliable and high-quality software products.

\noindent\textbf{Bug localization.} Bug localization refers to the process of identifying the specific source code files, functions, or lines of code that are responsible for a reported bug or software defect.
Bug localization typically involves analyzing bug reports or issue descriptions provided by users or testers and correlating them with the relevant portions of the source code. This process can be challenging, especially in large and complex software projects, where codebases can contain thousands or even millions of lines of code. Traditional bug localization methods often rely on heuristics, code metrics, or stack trace analysis, which may not always provide precise results. 
Ciborowska \ea~\cite{ciborowska2023too} investigated data augmentation techniques to enhance bug localization models. They introduce a pipeline applying token-level operations such as dictionary replacement, insertion, random swapping, and deletion, along with paragraph-level back-translation to bug reports. By employing augmented data to train BERT-based models for bug localization, they demonstrate that these techniques can substantially expand the training data and boost the models' performance.

\noindent\textbf{Verification.} 
Verification techniques, including prominent methods such as formal verification, hold a pivotal role in the domain of software quality assurance~\cite{charalambous2023new,tihanyi2023formai}. 
These techniques validate the correctness of software systems, improving their reliability and security against potential threats. Utilizing mathematical and logical principles in the verification process facilitates thorough error detection and correction before deployment, ensuring stable and secure performance in different operational contexts.
Charalambous \ea~\cite{charalambous2023new} leverage LLMs, particularly the GPT-3.5, in the realm of formal verification. Their approach combines LLMs with bounded model checking (BMC) to automatically repair software based on formal methods, showcasing the model's capability to understand intricate software structures and generate accurate repairs. 

\noindent\textbf{Test automation.} Automated testing methodologies offer a comprehensive array of tools and strategies designed for the evaluation of software applications' accuracy, reliability, and performance. These methodologies encompass various techniques, such as mutation testing~\cite{khanfir2023efficient} and fuzzing~\cite{deng2023large,deng2023language}. 
LLMs have been used for mutation testing, introducing faults to the codebase to assess the effectiveness of test suites in identifying and detecting errors~\cite{khanfir2023efficient}. Furthermore, LLMs can aid in fuzzing, generating valid and diverse input programs that help identify vulnerabilities and bugs, particularly in challenging domains like deep learning libraries~\cite{deng2023large}. By incorporating LLMs into test techniques, software engineers benefit from improved test coverage, reduced manual effort, and enhanced bug detection~\cite{deng2023language}, leading to more robust and reliable software systems.

\noindent\textbf{Fault localization.} Test suites typically include two types of test cases: pass-through test cases and fault-inducing test cases~\cite{li2023finding}. In practice, there are far more pass test cases for faults than fault-inducing test cases, which hinders the effectiveness of program debugging. However, in practice, it is difficult to find fault-inducing test cases. This is because developers first need to find test inputs that trigger program faults, and the search space for such test inputs is huge~\cite{fraser2015does}. Moreover, developers need to build a test oracle to automatically detect program faults, and building a test oracle is often an undecidable problem~\cite{ibrahimzada2022perfect}. 
Li \ea~\cite{li2023finding} investigated the application of ChatGPT to the task of finding fault-inducing test cases in SE. While recognizing ChatGPT's potential, they initially observed suboptimal performance in pinpointing these cases, particularly when two versions of a program had similar syntax. The authors identified this as a weakness in ChatGPT's ability to discern subtle code differences. To enhance its performance, they devised a novel approach blending ChatGPT with difference testing. Leveraging ChatGPT's strength in inferring expected behavior from erroneous programs, they synthesized programs that amplified subtle code differences. The experimental results reveal that this approach greatly increases the probability of finding the correct fault-inducing test case.

\noindent\textbf{Others.}
In addition to the six software quality assurance tasks detailed above, LLMs can also be applied to defect detection~\cite{sun2023dexbert,wong2023natural}, GUI testing~\cite{yoon2023autonomous,liu2023fill}, static analysis~\cite{hao2023v,mohajer2023skipanalyzer}, binary taint analysis~\cite{liu2023harnessing}, compiler fuzzing~\cite{quan2023xgv}, decompilation~\cite{xu2023lmpa}, invariant prediction~\cite{pei2023can}, malicious code localization~\cite{sun2023dexbert}, mobile app crash detection~\cite{liu2023testing}, and resource leak detection~\cite{wang2023boosting}.

\subsection{How are LLMs used in software maintenance?}

Within the context of software maintenance, LLMs have been leveraged for bug prediction, program repair, code review, debugging, and an array of other activities. 

\noindent\textbf{Program repair.} The goal of automated program repair (APR) is to automatically identify and fix bugs or defects in software~\cite{zhang2023neural}. It involves leveraging automated techniques to analyze buggy code and generate correct patches to address the identified issues. LLMs, such as BERT~\cite{zhang2023boosting,tian2023best}, CodeBERT~\cite{le2023invalidator}, CodeT5~\cite{paul2023automated}, Codex~\cite{fan2022automated,jin2023inferfix,wu2023effective}, PLBART~\cite{paul2023automated,wu2023effective}, T5~\cite{yuan2022circle,mastropaolo2022using} and GPT series~\cite{xia2023keep,tian2023chatgpt,xia2023conversational,lajko2022towards,charalambous2023new,sobania2023analysis,cao2023study}, have shown effectiveness in generating syntactically correct and contextually relevant code. Leveraging LLMs for program repair can achieve competitive performance in generating patches for various types of bugs and defects~\cite{xia2023keep}. These models can effectively capture the underlying semantics and dependencies in the code~\cite{charalambous2023new}, leading to the production of accurate and effective patches~\cite{zhang2023boosting,xia2023conversational}. Moreover, LLMs can be fine-tuned on specific code repair datasets~\cite{mastropaolo2022using}, further improving their ability to generate high-quality patches for real-world software projects. The application of LLMs in program repair not only accelerates the bug-fixing process but also enables software developers to focus on more complex tasks, leading to enhanced software reliability and maintainability.

\begin{table}[t]
\caption{The state-of-the-art applications of LLMs in program repair task.}
\resizebox{\linewidth}{!}{
\begin{tabular}{|p{0.1\linewidth}|p{0.38\linewidth}|p{0.23\linewidth}|p{0.15\linewidth}|l|l|}
\hline
\textbf{Model} & \textbf{Baseline} & \textbf{Benchmark} & \textbf{Metric} & \textbf{Date} & \textbf{Reference} \\ 
\hline
Codex & GPT-Neo, GPT-J, GPT-NeoX, CodeT5, InCoder & QuixBugs-Python and Java, Defects4J 1.2 and 2.0, ManyBugs & Correct / plausible patches & May 20, 2023 & \cite{xia2023automated} \\ 
\hline
Codex & CodeT5, CodeGen, PLBART, InCoder & Vul4J, VJBench, & Correct / plausible patches & May 29, 2023 & \cite{wu2023effective} \\ 
\hline
ChatGPT & Codex, CodeGen-16B, CodeGen-6B, CodeGen-2B, CodeGen-350M & QuixBugs-Python and Java & Correct / plausible patches &  Jan 30, 2023 & \cite{xia2023conversational} \\ 
\hline
ChatGPT & Codex, CodeBERT, SelfAPR, RewardRepair, Recoder, TBar, CURE, CoCoNuT & QuixBugs-Python and Java, Defects4J 1.2 and 2.0 &  Correct fixes & Apr 1, 2023 & \cite{xia2023keep} \\ 
\hline
\end{tabular}}
\label{tab:program_repair}
\end{table}

In recent research, program repair has emerged as a prevalent application. Among the LLMs, as shown in Table~\ref{tab:program_repair}, Codex~\cite{wu2023effective,xia2023automated} and ChatGPT~\cite{xia2023conversational} have particularly distinguished themselves in the program repair domain. \textbf{ChatGPT edges ahead due to its inherent interactive design, enabling a continuous feedback loop that yields refined and contextually apt patches}~\cite{xia2023conversational,xia2023keep}. Such conversational dynamics, coupled with rigorous comparisons across diverse baselines, underscore its superior adaptability and efficiency. 

Summarising several key findings from research on LLMs for program repair:
\begin{itemize}
    \item\textbf{Interactive feedback.} Incorporating an interactive feedback loop, as observed with ChatGPT, significantly augments the accuracy of program repair~\cite{xia2023conversational}. This dynamic interplay between patch generation and validation fosters a deeper understanding of the software's semantics, leading to more effective repairs.

    \item\textbf{Domain-specific integration.} Merging the capabilities of LLMs with domain-specific knowledge and techniques further enhances their performance. Customized prompts, project-specific fine-tuning, and leveraging SE techniques~\cite{xia2023automated,wang2023software} can dramatically elevate the efficacy of LLM-driven program repairs.

    \item\textbf{Comparative analysis.} Rigorous evaluation against diverse baselines reveals the versatility and adaptability of LLMs, especially ChatGPT. This wide-ranging comparison not only establishes their superiority but also underscores areas for potential improvement~\cite{xia2023keep}.
\end{itemize}

\noindent\textbf{Code clone detection.} Code clones are code samples that are identical to each other~\cite{baxter1998clone,karampatsis2020scelmo}. These code samples can have structural or semantic equivalence~\cite{svajlenko2014towards}. Sharma \ea~\cite{sharma2022exploratory} investigate BERT's application in code clone detection through an exploratory study. Analyzing BERT's attention to code markers, they found that identifiers received higher attention, advocating their use in clone detection. This insight enhanced clone detection across all layers, and the implications extended beyond BERT. The researchers suggest that these findings could lead to the development of smaller models with performance akin to larger ones, thus mitigating computational accessibility issues.

\noindent\textbf{Code review.} Code review is a critical quality assurance practice used to inspect, assess, and validate the quality and consistency of software code~\cite{sghaier2023multi}.
Code review aims to identify potential errors, vulnerabilities, and code quality issues, while also improving code maintainability, readability, and scalability.
LLMs like BERT~\cite{sghaier2023multi}, ChatGPT~\cite{sridhara2023chatgpt}, and T5~\cite{tufano2022using,li2022auger}, trained on massive code repositories, possess the ability to understand and learn the semantics, structures, and contextual information of code~\cite{zhang2022coditt5}. In the code review process, LLMs assist reviewers in comprehensively understanding code intent and implementation details, enabling more accurate detection of potential issues and errors. Moreover, these models can generate suggestions for code improvements and optimizations, providing valuable insights and guidance to reviewers. By combining the intelligence of LLMs with the expertise of human reviewers, code review becomes more efficient and precise, further enhancing software quality and reliability.

\noindent\textbf{Debugging.} Debugging targets identifying, locating, and resolving software defects or errors, commonly known as bugs. The debugging process involves scrutinizing the code, tracing the execution flow, and isolating the root cause of the problem to correct the error effectively. LLMs, such as BERT and other converter-based architectures, excel at utilizing contextual information and natural language understanding. In terms of debugging, LLMs can be used to simulate the scientific debugging process, such as AutoSD proposed by Kang \ea~\cite{kang2023explainable}. This model generates hypotheses about code problems and extracts relevant values to identify potential problems. In addition, the SELF-DEBUGGING method proposed by Chen \ea~\cite{chen2023teaching} enables LLM to debug its own generated code by learning a small number of presentations and explanations, which effectively improves the accuracy and sampling efficiency of code generation. Using LLMs in debugging not only improves fixing performance by generating competitive fixes but also provides insights into and explanations of the model's decision-making process, making it an important tool for improving software quality and developer productivity.

\noindent\textbf{Bug reproduction.} Bug reports are crucial for software maintenance, allowing users to inform developers of problems encountered while using the software. Therefore, researchers have invested significant resources in automating error playback to speed up the software maintenance process. The success of current automated approaches depends heavily on the characteristics and quality of error reports, as they are limited by manually created schemas and predefined vocabularies. Inspired by the success of the LLMs in natural language understanding, Feng \ea~\cite{feng2023prompting} propose AdbGPT, which utilizes natural language understanding and logical reasoning capabilities of the LLM to extract Steps to Reproduce (S2R) entities from bug reports and guide the bug replay process based on the current graphical user interface (GUI) state. The researchers describe how cue engineering, a small amount of learning, and thought chain reasoning can be utilized to leverage the knowledge of the LLM for automated error replay. This approach is significantly lightweight compared to traditional approaches, which utilize a single LLM to address both phases of S2R entity extraction and guided replay through novel hint engineering.

\noindent\textbf{Duplicate bug report detection.} 
In large software projects, multiple users may encounter and report the same or similar bugs independently, resulting in a proliferation of duplicate bug reports~\cite{isotani2021duplicate}. Duplicate bug report detection involves analyzing the textual content of bug reports and comparing them to find similarities and redundancies. 
LLM models, such as BERT~\cite{isotani2021duplicate}, ChatGPT~\cite{sridhara2023chatgpt}, and other transformer-based architectures, are well-suited for natural language understanding and contextual representation. When applied to this task, LLMs can effectively capture the semantic similarities between bug reports, even in cases with slight variations in language or phrasing. 
The utilization of LLMs in this context not only enhances efficiency in managing bug reports but also contributes to improving the overall software development and maintenance workflow, reducing redundancy, and ensuring prompt bug resolution~\cite{zhang2023duplicate}.

\noindent\textbf{Logging.} Logging involves the systematic recording of events, messages, or information during the operation of a software application. It provides valuable information for understanding the behavior, performance, and potential problems of an application. Developers strategically insert logging statements throughout the code base to capture relevant data such as variable values, function calls, and error messages. These logs are an important tool for testing~\cite{chen2018automated,chen2019experience}, debugging~\cite{satyanarayanan1992transparent}, monitoring~\cite{harty2021logging,hasselbring2020kieker}, and analyzing the behavior of software operations, helping developers identify and diagnose bugs, performance bottlenecks, and other critical issues. Mastropaolo \ea~\cite{mastropaolo2022using} introduce LANCE, a system for automatically generating and injecting full log statements into Java code using the T5 model.  
Sridhara \ea~\cite{sridhara2023chatgpt} present that ChatGPT performs well in the log summarization task, generating aggregated results that are better than the current state of the art.

\noindent\textbf{Sentiment analysis.} Sentiment analysis involves determining emotions in text data related to software products, such as user feedback or comments~\cite{guzman2014sentiment,jongeling2015choosing,islam2017leveraging}. The goal of sentiment analysis is to automatically classify the sentiment of the text as positive, negative, or neutral, providing valuable insights into how users perceive and react to software applications. Zhang \ea~\cite{zhang2020sentiment} conducted a study comparing pre-trained Transformer models like BERT, RoBERTa, XLNet, and ALBERT with existing SA4SE tools across six datasets. The results show that the Transformer models outperformed previous tools by 6.5\% to 35.6\% in macro/micro-averaged F1-scores, albeit with a trade-off in runtime efficiency. However, this accuracy boost comes with some runtime costs, indicating that while Transformer models are less efficient than existing SA4SE approaches, their runtime cost is not prohibitively high.

\noindent\textbf{Vulnerability repair.} Vulnerability repair is the process of identifying and fixing security holes or weaknesses in software applications.
Pearce \ea~\cite{pearce2021examining} investigate how to use LLMs for software zero-point vulnerability remediation. The authors explore the challenges faced in designing hints to induce LLMs to generate fixed versions of insecure code. It shows that while the approach is promising, with LLMs capable of fixing 100{\%} of synthetic and hand-created scenarios, a qualitative assessment of the model's performance on a corpus of historical real-life examples reveals challenges in generating functionally correct code. It is concluded that despite the potential for future targeted LLM applications in this area, challenges remain. For a complete end-to-end system, the full system needs to be evaluated in conjunction with error localization and an improved testbed.

\noindent\textbf{Bug prediction.} 
Gomes \ea~\cite{gomes2023bert} conduct a BERT and TF-IDF (Term Frequency-Inverted Document Frequency) application for long-lived bug prediction in Free/Libre Open-Source Software (FLOSS) study to compare their accuracy in predicting long-lived errors. The results show that BERT-based feature extraction consistently outperforms TF-IDF, demonstrating BERT's ability to capture the semantic context in error reports. In addition, smaller BERT architectures also show competitive results, highlighting the effectiveness of LLMs in bug prediction. This approach promises to enable more accurate error detection in FLOSS projects and improve software quality and maintenance.

\noindent\textbf{Bug triage.} 
Bug triage is pivotal for effective issue management in large projects. It entails prioritizing bugs and assigning appropriate developers for resolution. While bug triage is straightforward for smaller projects, scalability brings complexity. Finding the right developers with the needed skills becomes intricate as bugs vary in expertise requirements. Some even demand combined skills, amplifying the intricacy. Lee \ea~\cite{lee2022light} introduce the Light Bug Triage framework (LBT-P). This innovative approach employs BERT to extract semantic information from bug reports. To surmount challenges with LLMs in bug triage, the researchers employ techniques like model compression, knowledge preservation fine-tuning, and a new loss function.


\noindent\textbf{Program merge conflicts repair.} 
Program merge conflicts repair addresses the challenges faced when integrating individual code changes, which can lead to textual or semantic inconsistencies. Zhang \ea~\cite{zhang2022using} explored the potential of using k-shot learning with LLMs like GPT-3 to automate this repair process. While these models showed promise in resolving semantic conflicts for Microsoft Edge, they didn't fully replace the benefits of domain-specific languages for certain synthesis patterns.

\noindent\textbf{Tag recommendation.} 
Improper tagging in software Q\&A sites can lead to redundancy and other issues such as tag explosion. He \ea~\cite{he2022ptm4tag} introduced PTM4Tag, a framework utilizing PLMs with a triplet architecture to recommend tags for posts. By separately modeling the title, description, and code snippets of posts, PTM4Tag was compared using five popular PLMs, including BERT, CodeBERT, etc. The SE-specialized CodeBERT showed the best performance, notably surpassing CNN-based methods. An ablation study revealed that while the title was crucial in tag prediction, using all post components achieved the optimal result.

\noindent\textbf{Traceability recovery.}
Traceability recovery focuses on re-establishing lost or unclear connections between related software artifacts, thereby facilitating coherent software evolution and maintenance~\cite{gethers2011integrating}. While traditional methods have offered some solutions, the integration of LLMs has recently emerged as a promising avenue for enhancing the accuracy and efficiency of this task. Zhu \ea~\cite{zhu2022enhancing} present TRACEFUN, a traceability link recovery framework enhanced with unlabeled data, serves as a testament to this potential, leveraging LLMs to bridge the gap between labeled and unlabeled data, thereby refining traceability link predictions. 

\noindent\textbf{Others.} 
In addition to the 14 software maintenance tasks detailed above, LLMs can also be applied to review/commit/code classification~\cite{ghadhab2021augmenting,kou2023automated,yang2022aspect}, log parsing~\cite{liu2024interpretable,ma2024knowlog,yu2023log}, code revision~\cite{kabir2023empirical,wadhwa2023frustrated}, API misuses repair~\cite{zhang2023evaluating}, Code coverage prediction~\cite{tufano2023predicting}, code review explained~\cite{widyasari2023explaining}, Code-Review defects repair~\cite{zhao2023right}, crash bug repair~\cite{du2023resolving}, dockerfile Repair~\cite{henkel2021shipwright}, incivility detection~\cite{ferreira2024incivility}, patch correctness prediction~\cite{zhang2024appt}, patch detection~\cite{tang2023just}, rename Refactoring~\cite{liu2023refbert}, technical debt payback~\cite{mastropaolo2023towards}, web test repair~\cite{xu2023guiding}, type error repair~\cite{chow2024pyty}, etc.

\subsection{How are LLMs used in software management?}

Research papers describing the utilization of LLMs in software management are still limited. 

\noindent\textbf{Effort estimation.} Effort estimation refers to the process of predicting the amount of time, resources, and manpower required to complete a software development project. Alhamed \ea~\cite{alhamed2022evaluation} conduct an evaluation of the application of BERT in the task of effort estimation for software maintenance. Their study underscores BERT's potential to offer valuable insights and aid in the decision-making process while also highlighting the associated challenges and need for further investigation.

\begin{tcolorbox}[title=\textbf{RQ4} - Summary, left=2pt, right=2pt,top=2pt,bottom=2pt]
(1) We categorized SE tasks into six activities: requirements engineering, software design, software development, software quality assurance, software maintenance, and software management. Subsequently, we summarized the specific applications of LLMs in these SE activities.

(2) We identified a total of 85 SE tasks and found that LLMs are most widely used in software development, with 229 papers mentioning over 24 SE tasks. The least applied area, software management, was mentioned in only three studies.

(3) \textbf{Code generation and program repair are the most prevalent tasks for employing LLMs in software development and maintenance activities.} We analyze the top-performing LLMs repeatedly validated in these tasks and summarize novel findings.

\end{tcolorbox}

\section{Threats to Validity}
\label{Sec:limitations}

\textbf{Paper search omission.} One key limitation is the possibility of omitting relevant papers during the search process. When gathering papers related to LLM4SE tasks from various publishers, it is possible to miss some papers due to incomplete summarization of keywords for software engineering tasks or LLMs. To address this concern, we adopted a comprehensive approach, combining manual search, automated search, and snowballing techniques, to minimize the risk of missing relevant papers. For manual search, we systematically searched for LLM papers related to SE tasks in six top-tier SE venues and extracted authoritative and comprehensive SE tasks and LLM keywords from these sources. Using these constructed search strings, we conducted automated searches on seven widely used publisher platforms. Additionally, to further augment our search results, we employed both forward and backward snowballing.

\noindent\textbf{Study selection bias.} Another limitation is the potential study selection bias. We established inclusion and exclusion criteria to perform the initial selection of papers, followed by manual verification based on quality assessment criteria (QAC). This process involves a combination of automated and manual procedures. The automated selection process may result in mislabeling of papers due to incomplete or ambiguous information in their corresponding BibTeX records. To mitigate this issue, any papers that cannot be confidently excluded are temporarily retained for manual verification. However, the manual verification stage could be influenced by the subjective judgment and biases of the researchers, affecting the accuracy of the quality assessment of papers. To address these concerns, we invited two experienced reviewers in the fields of SE and LLM research to conduct a secondary review of the study selection results. This step aims to enhance the accuracy of our paper selection and minimize the likelihood of omission or misclassification. By using these measures, we strive to ensure that the selected papers are accurate and comprehensive, minimizing the impact of study selection bias and enhancing the reliability of our systematic literature review.
We additionally provide a replication package\footnote{\url{https://github.com/xinyi-hou/LLM4SE_SLR}} for others to view.

\noindent\textbf{Empirical knowledge bias.} 
This SLR, along with 395 relevant studies in the LLM4SE field, answers four RQs. This implies the need for manual analysis and understanding of each study. In this process, there may be biases introduced by subjective judgments and experiential knowledge. To minimize potential errors in this regard, we have made the following efforts. Firstly, in determining the RQs, as the first comprehensive overview of the LLM4SE field, we aim to provide a comprehensive interpretation of the current state and trends in this domain. Considering the commonality in AI4SE research, we referred to Yang et al.'s survey on DL4SE~\cite{yang2022survey} during our RQ formulation. We finally decided to focus on LLM types, datasets, tuning, evaluation, and targeted SE tasks. Secondly, for the understanding and analysis of each study, to ensure accurate comprehension of paper details, before addressing each RQ, we extensively reviewed relevant literature to predefine the approximate categories and details for each RQ. For example, in RQ3, based on prior work~\cite{sahoo2024systematic,weyssow2023exploring,zhao2023survey}, we identified differences between tuning techniques for LLMs and those commonly used in traditional machine learning, such as prompt engineering and PEFT.

\section{Challenges and Opportunities}
\label{sec:Discussion}

\subsection{Challenges}
\label{sec:challenges}

\subsubsection{Challenges in LLM Applicability.} \hfill

\noindent\textbf{Model size and deployment.} 
The size of LLMs has seen a marked increase over time, moving from GPT-1's 117M parameters to GPT-2's 1.5B, and further to GPT-3's 175B parameters~\cite{yang2023harnessing}.
The billions and even trillions~\cite{Google2021Brain} of parameters pose significant storage, memory, and computational challenges, which can hinder LLMs in resource-limited and real-time scenarios, especially when developers lack access to powerful GPUs or TPUs.
CodeBERT~\cite{feng2020codebert}, a pre-trained model proposed in 2019, has 
a total of 125M parameters, resulting in a large model size of 476 MB. Recently proposed models like Codex~\cite{chen2021evaluating} and CodeGen~\cite{nijkamp2022codegen}, have over 100 billion parameters and over 100 GB in size.
The large sizes also require more computational resources. As pointed out by Hugging Face team~\cite{bloomtraining}, training a 176B model (i.e., BLOOM~\cite{scao2022bloom}) on 1.5 TB datasets consumes an estimated 1,082,880 GPU hours.
Similarly, the training of the GPT-NeoX-20B model~\cite{black2022gpt} on the Pile dataset~\cite{gao2020pile}, encompassing over 825 GiB of raw text data, requires the deployment of eight NVIDIA A100-SXM4-40GB GPUs. Each of these GPUs comes with a price tag of over 6,000 dollars~\cite{nvidiagpuprice}, and the training extends to 1,830 hours or approximately 76 days.
Moreover, even training a relatively smaller model like the PolyCoder (2.7B)~\cite{xu2022systematic}, employing eight NVIDIA RTX 8000 GPUs on a single machine, demands a commitment of around 6 weeks. These examples illustrate the significant computational costs associated with training LLMs. These also have significant energy costs with predictions of massively increased energy usage by LLM-based platforms \cite{rillig2023risks}.
Fortunately, there are preliminary studies on reducing code models' size and improving their efficiency. Shi \ea~\cite{compressor} use a genetic algorithm to compress CodeBERT into only 3 MB and reduce its response latency by more than 70\%.
Overall, the challenge of increasing model sizes and efficient deployment requires further attention from the communities.

\vspace{0.05cm}
\noindent\textbf{Data dependency.} In Section \ref{sec:rq2}, we provide a detailed analysis of the datasets used in 395 studies and the data preprocessing process, finding that LLMs rely heavily on a large number of different datasets for training and fine-tuning, posing the data dependency challenge. The quality, diversity, and quantity of data directly affect the performance and generalizability of the models. Given their size, LLMs often require large amounts of data to capture nuances, but obtaining such data can be challenging. Relying on limited or biased datasets may cause the model to inherit these biases, resulting in biased or inaccurate predictions. In addition, the domain-specific data required for fine-tuning can be a bottleneck. Due to the relatively short period of time since the emergence of LLM, such large-scale datasets are still relatively rare, especially in the SE domain. 
Another issue is the risk of benchmark data contamination, where training and test data overlaps could lead to inflated performance metrics~\cite{zhao2021impact}. For instance, Brown \ea~\cite{brown2020language} discovered a code bug that prevented them from fully removing all overlapping data. They were unable to afford retraining and resorted to using ``cleaned'' variants of the benchmarks to mitigate the issue. 
Moreover, there are grave concerns around the inclusion of Personally Identifiable Information (PII) in pre-training corpora. Instances of PII, such as phone numbers and email addresses, have led to privacy leaks during the prompting process~\cite{Amit2021Copilot,elmhamdi2023impossible}. 

\vspace{0.05cm}
\noindent\textbf{Ambiguity in code generation.} Ambiguity in code generation poses a significant challenge for LLMs in SE tasks. When code intent is unclear (e.g., multiple valid solutions exist), LLMs may struggle to produce accurate and contextually appropriate code. This can lead to syntactically correct but functionally incorrect code, impacting the reliability and effectiveness of LLM-based code generation. Addressing this issue requires exploring techniques to incorporate additional context, domain-specific knowledge, or multi-model ensembles to improve LLMs' ability to handle ambiguity and generate precise code, ensuring their successful integration into real-world software development processes.

\subsubsection{Challenges in LLM Generalizability}
The generalizability of LLMs refers to the ability of these models to consistently and accurately perform tasks in different tasks, datasets, or domains outside their training environment. While LLMs are trained on massive amounts of data, ensuring extensive knowledge capture, their performance is sometimes problematic when confronted with specific or idiosyncratic tasks outside the scope of their training. This challenge is particularly evident in the SE domain, where we present the application of LLMs to 85 SE tasks in Section~\ref{sec:rq4}. We observed that the context and semantics of code or documents vary greatly across projects, languages, or domains. Ensuring that the LLM generalizes well requires careful fine-tuning, validation on different datasets, and continuous feedback loops. Without these measures, models run the risk of over-adapting their training data, thus limiting their usefulness in a variety of real-world applications.
Recent studies have shown that the LLMs cannot generalize their good performance to inputs after semantic-preserving transformations.
For example, Yang \ea~\cite{alert} show that the performance of CodeBERT on different tasks decreases significantly after substituting the variables' names in the input.

\subsubsection{Challenges in LLM Evaluation}
We summarized key evaluation metrics used in different types of SE tasks according to four task types: regression, classification, recommendation, and generation (Section~\ref{sec:rq4}). We found that when applying LLMs in the software engineering domain, the methodology for evaluating the performance of the models is usually based on a set of predefined metrics. Unfortunately, these metrics (e.g., Accuracy, Recall, or F1-score), while useful in some cases, may not fully capture all the effects and impacts of a model in a given SE task. For example, a model may perform well in terms of accuracy but may fail in processing specific types of inputs or in some specific situations. In addition, these metrics may not capture certain qualitative aspects of the model, such as its interpretability, robustness, or sensitivity to specific types of errors. Some of the most recent studies on LLM4SE tasks~\cite{agrawal2023monitor,hu2023augmenting,singla2023evaluating,xu2023lmpa,yuan2023evaluating,zhang2023self}, in which researchers customized some evaluation metrics to assess the performance of models, also further illustrate the limitations of some of the widely used evaluation metrics in the field of LLM.

\subsubsection{Challenges in LLM Interpretability, Trustworthiness, and Ethical Usage}
Interpretability and trustworthiness are crucial aspects in the adoption of LLMs for SE tasks. The challenge lies in understanding the decision-making process of these models, as their black-box nature often makes it difficult to explain why or how a particular code snippet or recommendation is generated. 
Recent studies~\cite{advdoor,you-see,CodePoisoner} also show that LLM of code trained on low-quality datasets can have vulnerabilities (e.g., generating insecure code).
The lack of interpretability and trustworthiness can lead to uncertainty and hesitation among developers, who may be hesitant to rely on LLM-generated code without a clear understanding of how it was derived. 
Establishing trust in LLMs requires efforts to develop techniques and tools that provide insights into the model's internal workings and enable developers to comprehend the reasoning behind the generated outputs. Enhancing interpretability and trustworthiness can ultimately promote the widespread adoption of LLMs in SE, leading to more efficient and effective development practices.
Many LLMs are not open and it is unclear what data they have been trained on, both quality and representativeness but also ownership of the source training data. This brings into question ownership of the derivative data, e.g., generated designs, code, or test cases. There is also potential for various adversarial attacks e.g. deliberately seeding LLMs with code vulnerabilities so that automatically generated code snippets have subtle but vulnerable aspects.

\subsection{Opportunities}
\label{sec:Opportunities}

\subsubsection{Optimization of LLM4SE}\hfill

\noindent\textbf{The advent of code-specialized LLMs in SE.}
The recent emergence of code-specialized LLMs, such as GitHub Copilot~\cite{github2021copilot}, Amazon's CodeWhisperer~\cite{AmazonCodeWhisperer}, OpenAI Code Interpreter~\cite{codeinterpreter} integrated into ChatGPT, and Code Llama~\cite{Meta2023Codellama} from Meta's Llama family, signals a transformative phase in LLM4SE. These specialized LLMs, fine-tuned on code-specific datasets, are not merely incremental improvements but paradigm shifts in code understanding, generation, and efficiency. They offer new avenues for automated coding, personalized developer assistance, enhanced code review, and quality assurance, among other tasks, setting the stage for groundbreaking advancements in the SE domain.

\noindent\textbf{Influence and applications of ChatGPT.} ChatGPT's popularity in recent academic research, as evidenced by its large presence in our 395 analyzed papers, emphasizes its escalating influence and acceptance within academia. Researchers' preference for ChatGPT over other LLMs and LLM-based applications since its release can be attributed to its computational efficiency, adaptability to various tasks, and potential cost-effectiveness~\cite{laskar2023systematic,li2023enabling,xia2023conversational}. Its applications extend beyond mere code efficiency and debugging, fostering a collaborative era in development. This paradigm shift signifies a broader move towards integrating advanced natural language understanding into conventional coding practices~\cite{laskar2023systematic,ma2023scope,sadik2023analysis}. By thoughtfully analyzing these dynamics and trends, we can foresee the potential pathways for LLMs and LLM applications like ChatGPT in shaping more robust, efficient, and collaborative software development procedures. Such insights stand as a promising indication of the future revolutionary impact of LLMs on SE.

\noindent\textbf{Performance enhancement from task-specific model training.} 
The choice between leveraging commercially available pre-trained models like GPT-4 and building upon open-source frameworks such as Llama~2~\cite{touvron2023llama2}, Gemma~\cite{gemma}, and Mistral~\cite{mistral} provides a nuanced set of options for individual or organizational customization in specialized tasks. The distinction between these two approaches lies in the degree of control and customization. Pre-trained models like GPT-4 are generally not designed for large-scale retraining due to their proprietary nature, but they allow quick task-specific adaptations with limited data, thereby minimizing computational overhead. On the other hand, frameworks like LLaMA offer an open-source foundation for more extensive customization. While they come pre-trained, organizations often modify the source code and retrain these models on their own large-scale datasets to meet specialized requirements~\cite{chinesellama,llama-efficient-tuning}. This process is computationally intensive, leading to greater resource allocation and cost, but affords the advantage of creating highly domain-specific models.
Hence, the primary trade-off is between the ease of use and quick deployment offered by models like GPT-4, and the deep customization capabilities but higher computational demands associated with open-source frameworks like LLaMA.

\noindent\textbf{Collaborative LLMs.} From our review it is evident that LLMs have made significant strides in addressing various SE challenges. However, as the complexity of SE tasks continues to grow, there's an emerging need for more sophisticated and tailored solutions. One promising direction is the concept of Collaborative LLMs. This approach involves integrating multiple LLMs~\cite{dong2023self,zhao2023automatic} or combining LLMs with specialized machine-learning models~\cite{ezzini2022automated,zhang2022beqain} to enhance their efficacy for SE tasks. By harnessing the collective strengths of different models, we believe that the SE community can achieve more precise and efficient outcomes, from code completion to bug detection.

\subsubsection{Expanding LLM's NLP Capabilities in More SE Phases.}\hfill 

\noindent\textbf{Integration of new input forms.} In our analysis, we observed that the predominant input forms were code-based datasets and text-based datasets. However, there was a noticeable scarcity of graph-based datasets~\cite{kolthoff2023data} (Section~\ref{sec:rq2}). Leveraging new input forms of natural language, such as spoken language, diagrams, and multimodal inputs, presents an opportunity to enhance the LLMs' ability to understand and process diverse user requirements. Integrating spoken language could improve interactions between developers and models, enabling more natural and context-rich communication. Diagrams can facilitate visual representations of code and requirements, offering a complementary perspective for code generation. Furthermore, multimodal inputs that combine text, audio, and visual cues could offer a more comprehensive context understanding, leading to more accurate and contextually appropriate code generation. Additionally, exploring graph-based datasets could be crucial for addressing complex code scenarios, as graphs capture the structural relationships and dependencies in code, allowing LLMs to better comprehend code interactions and dependencies. 

\noindent\textbf{Widening LLM applications across SE phases.} We observed a pronounced emphasis on the application of LLMs in software development and maintenance. These areas have undoubtedly benefited from the capabilities of LLMs, leading to enhanced code completion~\cite{izadi2022codefill,li2022cctest,liu2023repobench}, bug detection~\cite{ciborowska2023too,feng2023prompting,kang2023explainable}, and other related tasks. 
The current application of LLMs in requirements engineering, software design, and software management remains relatively sparse. This presents a significant opportunity: by expanding the use of LLMs to these under-explored areas, we can potentially improve how requirements are elicited, how software designs are conceptualized, and how projects are managed.

\subsubsection{Enhancing LLM's Performance in Existing SE Tasks}\hfill

\noindent\textbf{Tackling domain-specific challenges.} Many SE domains, including safety-critical systems and specific industries, suffer from a scarcity of open-source datasets, hindering the application of LLMs in these specialized areas. Future research can focus on creating domain-specific datasets and fine-tuning LLMs to cater to the unique challenges and intricacies of these fields~\cite{biswas2020achieving,sun2023gpt}. Collaboration with domain experts and practitioners is vital to curate relevant data, and fine-tuning LLMs on this data can enhance their effectiveness and ensure better alignment with the specific requirements of each domain, paving the way for LLMs to address real-world challenges~\cite{bubeck2023sparks} in diverse software engineering domains~\cite{li2023finding}.

\noindent\textbf{Establishing a comprehensive evaluation framework for LLM4SE.} The necessity for a universal, yet adaptable, evaluation framework for LLM4SE is pressing for both academic and industrial sectors. In academia, such a framework enables streamlined assessments of LLM performance, efficacy, and limitations, serving as a benchmark to verify the models' practical readiness. On the industrial side, collaborations with real-world development teams using this framework yield empirical insights into LLMs' utility, including their impacts on productivity, code quality, and team collaboration, while also revealing challenges like model biases, misinterpretation of code semantics, and context-specific limitations. Establishing this framework is critical for standardizing assessments and facilitating responsible LLM adoption in both academic research and practical applications~\cite{biswas2020achieving,gong2023intended}.

\subsection{Roadmap}
\label{sec:roadmap}
We provide a roadmap for future development in leveraging Large Language Models for Software Engineering (LLM4SE), with an additional high-level perspective that acknowledges the reciprocal relationship and emerging exploration of Software Engineering for Large Language Models (SE4LLM).

\noindent\textbf{Automated coding, development and personalized developer assistance.} 
The pursuit of automation in coding encompasses the auto-generation of code snippets, bug fixes, system optimization, and the creation of intelligent, personalized assistance for developers that is context-aware and adaptable to individual needs. LLM's generative capabilities can be leveraged to help developers better understand requirements and generate syntactically and semantically correct code, thereby accelerating development cycles and improving software quality. Leveraging LLM's natural language processing to develop context-aware tools allows for interaction with developers in a more intuitive and responsive manner. Additionally, fine-tuning LLMs for specific coding tasks and developer assistance can further enhance their accuracy and efficiency, customizing the automation process to suit the unique demands of different projects and individuals.

\noindent\textbf{Advancing testing and analysis.} 
The inclusion of LLMs in software testing methods opens up avenues for enhanced test case generation, bug classification, and defect prediction, thereby improving the precision and efficiency of the software testing process. For instance, LLMs show potential to be fine-tuned to a project's specific requirements to generate customized test cases, which elevates the likelihood of early detection of subtle bugs or security vulnerabilities. Furthermore, the integration of LLMs with traditional SE techniques, including both static and dynamic program analysis presents a compelling direction for more rigorous code analysis. The potential for utilizing LLMs in formal analysis methodologies, including formal verification, is another area that merits investigation~\cite{charalambous2023new}. These advancements not only facilitate the early discovery of complex errors but also lead to reduced development costs and quicker time-to-market, ultimately contributing to the robustness and reliability of the software products.

\noindent\textbf{Integrating programming knowledge into LLMs.}
One critical future direction lies in the integration of specialized code representation methods and programming domain knowledge into LLM4SE~\cite{wan2022they,ma2023code}. This integration aims to enhance the capability of LLMs to generate code that is not only functionally accurate but also secure and compliant with programming standards. Leveraging advanced techniques in code embedding, syntax tree parsing, and semantic analysis could significantly refine the generation capabilities of LLMs. Moreover, embedding domain-specific rules and best practices into these models would enable them to auto-generate code that adheres to industry or language-specific guidelines for security and style.

\noindent\textbf{Enhanced code review and quality assurance.} The transformation of the code review process can be supported by employing  LLMs to analyze code context, perform intelligent comparisons, and offer insights that go beyond traditional automated review systems. The application of fine-tuned LLMs for code review can allow for more precise error detection and tailored feedback, offering a more nuanced understanding of code quality and potential improvements.

\noindent\textbf{Extracting insights from data mining.} LLMs can play a critical role in mining insights from platforms like GitHub, StackOverflow, and app stores. Through the application in tasks such as requirement extraction, traceability, validation, and various types of mining (tag, app, developer-based), LLMs can provide valuable insights that inform development strategies and decision-making. By automating and enhancing these mining tasks, LLMs contribute to a deeper understanding of user needs, emerging trends, and the efficiency of development practices.

\noindent\textbf{Empowering predictive analytics and decision support.} Leveraging LLMs for effort cost prediction, software classification, code classification, incident detection, and software quality evaluation may support better data-driven insights and predictive analytics. This empowers organizations to make informed decisions throughout the development lifecycle. LLMs' ability to model and analyze vast amounts of data enables more accurate forecasts of project timelines, resource needs, and potential risks.

\noindent\textbf{LLMs in software security.} The growing impact of LLM4SE offers both unparalleled opportunities and challenges in the domain of software security. 
On the one hand, LLMs offer promising solutions for automated security audits, compliance verifications, and vulnerability detection. These models can potentially be leveraged for automated code reviews to ensure compliance with industry standards and legal regulations, while also identifying potential security vulnerabilities~\cite{ferrag2023securefalcon,ahmad2023fixing,feng2023prompting,pearce2023examining,deng2023pentestgpt,happe2023getting}. For instance, Ferrag \ea~\cite{ferrag2023securefalcon} showcased the efficacy of LLMs in cyber reasoning tasks related to software security.
On the other hand, the usage of LLMs introduces novel security concerns. Their complexity makes them susceptible to attacks, demanding novel strategies to fortify the models themselves~\cite{wu2023defending, rao2023tricking, elizondo2023langkit, ramly2023preventing,deng2023jailbreaker,liu2023prompt}. As an example, Wu \ea~\cite{wu2023defending} delve into methods to secure LLMs against jailbreak attacks.
An intriguing direction for future research lies in enabling LLMs to automatically identify and rectify their own vulnerabilities. Specifically, the focus could be on equipping LLMs to generate self-applied patches to their underlying code, thereby enhancing their inherent security, as opposed to merely implementing application-layer restrictions.
Given this landscape, future research should adopt a balanced approach, aiming to exploit LLMs for automating and enhancing existing software security protocols while concurrently developing techniques to secure the LLMs themselves. This dual focus is crucial for fully realizing the potential of LLMs in enhancing the security and compliance assurance of software systems.

\noindent\textbf{Software Engineering for Large Language Models (SE4LLM).} As the capabilities and complexities of LLMs continue to expand, there arises a reciprocal need for specialized SE practices tailored for the development, optimization, and maintenance of these models. SE4LLM encompasses a range of challenges and opportunities, including the design of scalable and maintainable architectures, the creation of efficient training algorithms, the development of rigorous testing frameworks for model robustness and fairness, and the implementation of ethical guidelines and compliance mechanisms. The convergence of SE with LLMs not only facilitates the growth of more sophisticated and adaptable models but also opens up new avenues for interdisciplinary research and innovation, bringing together the expertise of both the AI and SE communities. This aligns with a broader vision where SE practices become an integral part of the lifecycle of LLMs, ensuring their robustness, efficiency, and ethical alignment with societal values.

\section{Conclusion}
\label{sec:Conclusion}

LLMs are bringing significant changes to the field of SE. The potential of these models to handle complex tasks can fundamentally reshape many SE practices and tools.
In this SLR, we analyzed the emerging utilization of LLMs for software engineering, encompassing papers published since the inception of the first LLM (BERT). 
We examined the diverse LLMs that have been employed in SE tasks and explored their distinct features and applications (RQ1). We then investigated the processes involved in data collection, preprocessing, and usage, emphasizing the significant role well-curated datasets play in the successful application of LLMs to solve SE tasks (RQ2). Following this, we investigated the various strategies utilized to optimize and assess the performance of LLMs for SE tasks (RQ3). Lastly, we reviewed the wide range of SE tasks where LLMs have been applied to date, shedding light on the practical contributions LLMs have made (RQ4). 
We summarised some key existing challenges of LLM4SE and provided a research roadmap, outlining promising future research directions.


\bibliographystyle{ACM-Reference-Format}
\bibliography{main}

\newpage
\appendix
\section{Data Types}
\label{app:datasets}

 We classified the data types of all datasets into five categories: code-based, text-based, graph-based, software repository-based, and combined data types, as shown in Table~\ref{tab:all_datatypes}. 

\begin{table}[h]
\caption{Data types of datasets involved in prior studies.}
\resizebox{\linewidth}{!}{
\begin{tabular}{r|p{0.3\linewidth}|c|p{0.55\linewidth}}
\hline
\textbf{Category}& \textbf{Data type} & \textbf{{\#} Studies} & \textbf{References}\\ \hline
Text-based datasets
& Programming tasks/problems & 42 & \cite{buscemi2023comparative} \cite{chen2023improving} \cite{chen2021evaluating} \cite{chen2023teaching} \cite{dakhel2023effective} \cite{hendrycks2021measuring} \cite{hong2023metagpt} \cite{hu2024leveraging} \cite{jain2023coarse} \cite{ji2023benchmarking} \cite{jones2022capturing} \cite{kuznia2022less} \cite{lahiri2022interactive} \cite{lai2023ds} \cite{le2023codechain} \cite{li2023enabling} \cite{li2023motcoder} \cite{li2023structured} \cite{li2023think} \cite{liu2023better} \cite{liu2023refining} \cite{nijkamp2023codegen2} \cite{patel2023evaluating} \cite{ren2023misuse} \cite{sadik2023analysis} \cite{sakib2023extending} \cite{shi2023towards} \cite{shin2023prompt} \cite{sobania2023analysis} \cite{tian2023test} \cite{wang2023chatcoder} \cite{wang2023leti} \cite{wang2024oop} \cite{wei2023magicoder} \cite{xiong2023program} \cite{yen2023coladder} \cite{zan2022cert} \cite{zan2023private} \cite{zhang2023coder} \cite{zhang2024codeagent} \cite{zhou2023codebertscore} \cite{zhuo2023large} \\
 \cline{2-4}
& Prompts & 33 & \cite{cassano2023multipl} \cite{dong2023self} \cite{du2023classeval} \cite{gilbert2023semantic} \cite{jiang2023self} \cite{jiang2023selfevolve} \cite{ke2023discriminating} \cite{khakhar2023pac} \cite{khan2023assessing} \cite{kirinuki2024chatgpt} \cite{li2022generation} \cite{li2023codeie} \cite{li2023novel} \cite{liu2023no} \cite{liu2023your} \cite{luo2023wizardcoder} \cite{nichols2024can} \cite{scao2022bloom} \cite{siddiq2023lightweight} \cite{singla2023evaluating} \cite{tan2023copilot} \cite{tang2023chatgpt} \cite{tarassow2023potential} \cite{tihanyi2023formai} \cite{tu2023llm4cbi} \cite{vikram2023can} \cite{wang2022recode} \cite{wang2023evaluating} \cite{wang2023measuring} \cite{white2023prompt} \cite{xu2023lmpa} \cite{ye2023generating} \cite{zhao2023understanding} \\
 \cline{2-4}
& SO (i.e., Stack Overflow) posts & 12 & \cite{biswas2020achieving} \cite{he2022ptm4tag} \cite{he2023representation} \cite{helmeczi2023few} \cite{huang2023api} \cite{kou2023automated} \cite{luitel2023improving} \cite{mukherjee2023stack} \cite{rahmani2023improving} \cite{shi2023sotana} \cite{wei2022clear} \cite{zhang2020sentiment} \\
 \cline{2-4}
& Bug reports & 11 & \cite{ciborowska2022fast} \cite{ciborowska2023too} \cite{deng2023language} \cite{feng2023prompting} \cite{gomes2023bert} \cite{helmeczi2023few} \cite{huang2024crashtranslator} \cite{isotani2021duplicate} \cite{jin2023inferfix} \cite{kang2022large} \cite{lee2022light} \\
 \cline{2-4}
& Requirements documentation & 9 & \cite{el2023ai} \cite{ezzini2022automated} \cite{helmeczi2023few} \cite{hey2020norbert} \cite{kolthoff2023data} \cite{luo2022prcbert} \cite{moharil2023tabasco} \cite{poudel2023leveraging} \cite{wang2020deep} \\
 \cline{2-4}
& APIs/API documentation & 8 & \cite{deng2023large} \cite{huang2023adaptive} \cite{khan2021automatic} \cite{patil2023gorilla} \cite{wu2024automatic} \cite{yang2022aspect} \cite{yang2023apidocbooster} \cite{zan2022language} \\
 \cline{2-4}
& Q{\&}A pairs & 6 & \cite{kabir2023empirical} \cite{qian2023communicative} \cite{schlag2023large} \cite{von2022validity} \cite{wang2023evaluating} \cite{zhou2023large} \\
 \cline{2-4}
& Vulnerability descriptions & 4 & \cite{pearce2021examining} \cite{sun2024llm4vuln} \cite{thapa2022transformer} \cite{wu2023effective} \\
 \cline{2-4}
& Reviews & 4 & \cite{lu2023llama} \cite{motger2024t} \cite{widyasari2023explaining} \cite{zhang2023revisiting} \\
 \cline{2-4}
& Logs & 3 & \cite{liu2024interpretable} \cite{ma2024knowlog} \cite{yu2023log} \\
 \cline{2-4}
& Methods & 3 & \cite{mastropaolo2021empirical} \cite{mastropaolo2022using} \cite{yuan2023no}\\ 
 \cline{2-4}
& Project issues & 3 & \cite{fu2022gpt2sp} \cite{sun2023gpt} \cite{zhang2023cupid} \\
 \cline{2-4}
& Code comments & 2 & \cite{prenner2021making} \cite{xu2022systematic} \\
 \cline{2-4}
& Theorems & 2 & \cite{first2023baldur} \cite{zhang2024selene} \\
 \cline{2-4}
& Buggy text & 1 & \cite{liu2023testing}\\
 \cline{2-4}
& Dockerfiles & 1 & \cite{henkel2021shipwright}\\
 \cline{2-4}
& Outage descriptions & 1 & \cite{jin2023assess}\\
 \cline{2-4}
& Semantic merge conflicts & 1 & \cite{zhang2022using}\\
 \cline{2-4}
& Site text & 1 & \cite{koide2023detecting} \\
 \cline{2-4}
& Software development tasks & 1 & \cite{zhang2024experimenting} \\
 \cline{2-4}
& User intents & 1 & \cite{jain2022jigsaw} \\
 \cline{2-4}
& Software specifications & 1 & \cite{mandal2023large} \\
 \cline{2-4}
& User reviews & 1 & \cite{wang2022your} \\ 
\hline
Code-based datasets
& Source code & 60 & \cite{agarwal2024structured}
\cite{alam2023gptclonebench} \cite{arakelyan2023exploring} \cite{bui2023codetf} \cite{charalambous2023new} \cite{chen2022transferability} \cite{chochlov2022using} \cite{ciniselli2021empirical} \cite{fan2024rapid} \cite{gao2023constructing} \cite{geng2024large} \cite{hu2023augmenting} \cite{huang2023ai} \cite{izadi2022codefill} \cite{jesse2022learning} \cite{kabir2024zs4c} \cite{kanade2020learning} \cite{li2021toward} \cite{li2023cctest} \cite{li2023ptm} \cite{liu2020multi} \cite{liu2023contrabert} \cite{liu2023refbert} \cite{liu2023repobench} \cite{ma2023scope} \cite{ma2024specgen} \cite{mao2023self} \cite{murali2023codecompose} \cite{nguyen2023snippet} \cite{ochs2023evaluating} \cite{pan2023stelocoder} \cite{prenner2021making} \cite{qi2023sut} \cite{saberi2023multilingual} \cite{shapkin2023entity} \cite{sharma2022exploratory} \cite{shen2022benchmarking} \cite{shi2022cross} \cite{shi2023sotana} \cite{shin2023domain} \cite{shin2023prompt} \cite{siddiq2023exploring} \cite{steenhoek2023reinforcement} \cite{sun2023automatic} \cite{sun2024neural} \cite{tang2023csgvd} \cite{wan2022they} \cite{wang2023codet5+} \cite{wong2023natural} \cite{xie2023chatunitest} \cite{xu2024unilog} \cite{yan2023codetransocean} \cite{yang2023assessing} \cite{yang2023syntax} \cite{yu2023codereval} \cite{zeng2022extensive} \cite{zhang2023neural} \cite{zhang2023toolcoder} \cite{zheng2023outline} \cite{zhu2023automating} \\
 \cline{2-4}
& Bugs/Buggy code & 16 & \cite{cao2023study} \cite{dakhel2023effective} \cite{du2023pre} \cite{du2023resolving} \cite{hao2023v} \cite{kang2023evaluating} \cite{kang2023explainable} \cite{khanfir2023efficient} \cite{paul2023enhancing} \cite{peng2024domain} \cite{silva2023repairllama} \cite{wang2023rap} \cite{wei2023copiloting} \cite{xia2022practical} \cite{xia2023conversational} \cite{zhang2023steam} \\
 \cline{2-4}
& Vulnerable source code & 8 & \cite{chan2023transformer} \cite{chen2023diversevul} \cite{gao2024learning} \cite{grishina2023earlybird} \cite{khare2023understanding} \cite{quan2023xgv} \cite{steenhoek2024dataflow} \cite{zhang2023prompt} \\ 
 \cline{2-4} 
& Patches & 4 & \cite{le2023invalidator} \cite{tian2020evaluating} \cite{tian2023best} \cite{zhang2023boosting} \\
 \cline{2-4} 
& Code changes & 3 & \cite{ghadhab2021augmenting} \cite{li2022auger} \cite{zhang2023multilingual} \\ 
 \cline{2-4}
& Test suites/cases & 3 & \cite{jana2023attention} \cite{xu2022systematic} \cite{zhang2023algo} \\ 
 \cline{2-4}
& Bug-fix pairs & 2 & \cite{fakhoury2023towards} \cite{yuan2022circle} \\ 
 \cline{2-4}
& Error code & 2 & \cite{huang2023chain} \cite{widjojo2023addressing} \\
 \cline{2-4}
& Error-fix pairs & 1 & \cite{chow2024pyty} \\
 \cline{2-4}
& Flaky test cases & 1 & \cite{fatima2022flakify} \\
 \cline{2-4}
\hline
\end{tabular}}
\caption*{\hfill(Continued)}
\label{tab:all_datatypes}
\end{table}

\begin{table}[t]
\captionsetup{labelformat=empty} 
\caption*{Table \ref{tab:all_datatypes}. Continued.}
\resizebox{\linewidth}{!}{
\begin{tabular}{r|p{0.3\linewidth}|c|p{0.55\linewidth}}
\hline
\textbf{Category}& \textbf{Data type} & \textbf{{\#} Studies} & \textbf{References}\\ 
\hline
Code-based datasets
& Identifiers & 1 & \cite{zhang2022beqain} \\ 
 \cline{2-4}
& Labeled clone pairs & 1 & \cite{dou2023towards}\\
 \cline{2-4}
& Packages & 1 & \cite{schafer2023adaptive}\\
\hline
Graph-based datasets & GUI Images & 1 & \cite{kolthoff2023data} \\ \hline
Software repository-
& Code repository & 9  & \cite{bairi2023codeplan} \cite{chen2023effectiveness} \cite{eghbali2024hallucinator} \cite{gupta2023grace} \cite{jimenez2023swe} \cite{mohajer2023skipanalyzer} \cite{tang2023domain} \cite{wang2024teaching} \cite{zhang2024codeagent} \\
based datasets & & & \\
\cline{2-4}
& Android apps & 3 & \cite{liu2023fill} \cite{sun2023dexbert} \cite{yoon2023autonomous} \\
 \cline{2-4}
& Issues and commits & 3 & \cite{alhamed2022evaluation} \cite{lin2021traceability} \cite{zhang2020sentiment} \\
 \cline{2-4}
 & Pull-requests & 2 & \cite{sghaier2023multi} \cite{zhang2020sentiment} \\
 \cline{2-4}
& Industrial projects & 1 & \cite{li2024fine} \\
 \cline{2-4}
& Open-source projects & 1 & \cite{li2024deveval} \\
 \cline{2-4}
& Web applications & 1 & \cite{xu2023guiding} \\
\hline
Combined datasets
& Programming tasks and test suites/cases & 17 & \cite{dong2023codescore} \cite{fan2022automated} \cite{guo2024deepseek} \cite{huang2023agentcoder} \cite{huang2023codecot} \cite{jain2023llm} \cite{li2023novel} \cite{mu2023clarifygpt} \cite{olausson2023demystifying} \cite{ouyang2023llm} \cite{piya2023llm4tdd} \cite{ridnik2024code} \cite{schafer2023empirical} \cite{shirafuji2023exploring} \cite{shypula2023learning} \cite{tian2023chatgpt} \cite{yan2023closer} \\
 \cline{2-4}
& Source code and comments & 12 & \cite{gu2022assemble} \cite{guo2024exploring} \cite{lajko2022towards} \cite{mastropaolo2023towards} \cite{niu2022spt} \cite{prenner2021making} \cite{saberi2023utilization} \cite{tufano2022using} \cite{weyssow2023exploring} \cite{yang2023enhancing} \cite{zhang2022coditt5} \cite{zhang2023self} \\
 \cline{2-4}
& Programming tasks and solutions & 8 & \cite{dibia2022aligning} \cite{doderlein2022piloting} \cite{hajali2023functionconstrained} \cite{khan2023xcodeeval} \cite{kou2023model} \cite{pei2023can} \cite{saieva2023contrastive} \cite{tian2024debugbench} \\
 \cline{2-4} 
& Source code and description & 3 & \cite{liu2023improving} \cite{sun2023clover} \cite{wang2023one} \\ 
 \cline{2-4}
& Code-text pairs & 2 & \cite{manh2023vault} \cite{rao2023ai} \\
 \cline{2-4} 
& Souce code and API usage sequences & 2 & \cite{weyssow2023usage} \cite{zhuo2023pop} \\
 \cline{2-4} 
& Source code and test suites/cases & 2 & \cite{ruiz2024novel} \cite{tufano2023predicting} \\
 \cline{2-4}
& Bug report and test suites/cases & 1 & \cite{plein2023automatic} \\
 \cline{2-4} 
& Buggy code and comments & 1 & \cite{zhao2023right} \\
 \cline{2-4} 
& Buggy code and solutions & 1 & \cite{moon2023coffee} \\
 \cline{2-4} 
& Code files and summaries & 1 & \cite{wang2024sparsecoder} \\
 \cline{2-4} 
& Binary code and related annotations & 1 & \cite{al2023extending} \\
 \cline{2-4} 
& Failing test code and error messages & 1 & \cite{xia2023keep} \\
 \cline{2-4}
& Source code and Q\&A pairs & 1 & \cite{salza2022effectiveness} \\
 \cline{2-4}
& Source code, methods, and logs & 1 & \cite{li2023exploring} \\
 \cline{2-4}
& Vulnerable code and description & 1 & \cite{wu2023deceptprompt} \\
\hline
\end{tabular}}
\label{tab:all_datatypes_continued}
\end{table}
\section{Input Forms}
\label{app:inputs}

In LLM4SE research, data is often transformed into specific formats to be used as input for LLMs. Table ~\ref{tab:all_input-forms} illustrates four input formats, namely token-based input, tree/graph-based input, pixel-based input, and hybrid-based input, along with all the papers that utilize each type.

\begin{table}[h]
\caption{The various input forms of LLMs proposed in prior studies.}
\resizebox{\linewidth}{!}{
\begin{tabular}{r|p{0.35\linewidth}|c|p{0.55\linewidth}}
\hline
\textbf{Category} & \textbf{Input forms} & \textbf{\# Studies} & \textbf{References} \\ \hline
 Token-based input  & Text in tokens & 150 &  \cite{alhamed2022evaluation}  \cite{biswas2020achieving} \cite{buscemi2023comparative} \cite{cassano2023multipl} \cite{chen2021evaluating} \cite{chen2023improving} \cite{chen2023teaching} \cite{ciborowska2022fast} \cite{ciborowska2023too} \cite{deng2023language} \cite{deng2023large} \cite{dong2023codescore} \cite{dong2023self} \cite{du2023classeval} \cite{el2023ai} \cite{ezzini2022automated} \cite{feng2023prompting} \cite{ferreira2024incivility} \cite{first2023baldur} \cite{fu2022gpt2sp} \cite{gilbert2023semantic} \cite{gomes2023bert} \cite{he2022ptm4tag}\cite{he2023representation} \cite{helmeczi2023few} \cite{hendrycks2021measuring} \cite{henkel2021shipwright} \cite{hey2020norbert}  \cite{hong2023metagpt} \cite{hu2024leveraging} \cite{huang2023adaptive} \cite{huang2023agentcoder}  \cite{huang2023anpl} \cite{huang2023api} \cite{huang2024crashtranslator} \cite{isotani2021duplicate} \cite{jain2022jigsaw} \cite{jain2023llm} \cite{ji2023benchmarking} \cite{jiang2023self} \\
\hline
\end{tabular}}
\caption*{\hfill(Continued)}
\label{tab:all_input-forms}
\end{table}

\begin{table}[h]
\captionsetup{labelformat=empty} 
\caption*{Table \ref{tab:all_input-forms}. Continued.}
\resizebox{\linewidth}{!}{
\begin{tabular}{r|p{0.35\linewidth}|c|p{0.55\linewidth}}
\hline
\textbf{Category} & \textbf{Input forms} & \textbf{\# Studies} & \textbf{References} \\ \hline
& Text in tokens & &
\cite{jiang2023selfevolve} \cite{jin2023assess} \cite{jin2023inferfix} \cite{jones2022capturing} \cite{kabir2023empirical} \cite{kang2022large} \cite{ke2023discriminating} \cite{khakhar2023pac} \cite{khan2021automatic} \cite{khan2022automatic} \cite{khan2023assessing} \cite{kirinuki2024chatgpt} \cite{koide2023detecting} \cite{kolthoff2023data} \cite{kou2023automated} \cite{kuznia2022less} \cite{lahiri2022interactive} \cite{le2023codechain} \cite{lee2022light} \cite{li2022generation} \cite{li2023codeie} \cite{li2023enabling} \cite{li2023motcoder} \cite{li2023novel} \cite{li2023structured} \cite{li2023think} \cite{lin2021traceability} \cite{liu2023better} \cite{liu2023no} \cite{liu2023refining} \cite{liu2023testing} \cite{liu2023your} \cite{liu2024interpretable} \cite{lu2023llama} \cite{luitel2023improving} \cite{luo2022prcbert} \cite{luo2023wizardcoder} \cite{ma2024knowlog} \cite{mandal2023large} \cite{mastropaolo2021empirical} \cite{mastropaolo2021studying} \cite{mastropaolo2023robustness} \cite{moharil2022identification} \cite{moharil2023tabasco} \cite{motger2024t} \cite{mukherjee2023stack} \cite{ni2023l2ceval} \cite{nichols2024can} \cite{ouyang2023llm} \cite{patel2023evaluating} \cite{patil2023gorilla} \cite{pearce2021examining} \cite{pegolotti2023qigen} \cite{plein2023automatic} \cite{poudel2023leveraging} \cite{qian2023communicative} \cite{ren2023misuse} \cite{scao2022bloom} \cite{schlag2023large} \cite{sghaier2023multi} \cite{shi2023sotana} \cite{shi2023towards} \cite{shin2023prompt} \cite{siddiq2023lightweight} \cite{singla2023evaluating} \cite{sobania2023analysis} \cite{sun2023gpt} \cite{tan2023copilot} \cite{tang2023chatgpt} \cite{tarassow2023potential} \cite{thapa2022transformer} \cite{tian2023test} \cite{tihanyi2023formai} \cite{tu2023llm4cbi} \cite{vikram2023can} \cite{von2022validity} \cite{wang2020deep} \cite{wang2022recode} \cite{wang2022your} \cite{wang2023chatcoder} \cite{wang2023leti} \cite{wang2023measuring} \cite{wang2023natural} \cite{wang2024oop} \cite{wei2022clear} \cite{wei2023magicoder} \cite{widyasari2023explaining} \cite{wu2023effective} \cite{wu2024automatic} \cite{xie2023impact} \cite{xiong2023program} \cite{xu2023lmpa} \cite{yang2022aspect} \cite{yang2023apidocbooster} \cite{ye2023generating} \cite{yu2023log} \cite{yuan2023no} \cite{zan2022cert} \cite{zan2022language} \cite{zhang2020sentiment} \cite{zhang2022using} \cite{zhang2023coder} \cite{zhang2023cupid} \cite{zhang2023revisiting} \cite{zhang2024experimenting} \cite{zhang2024selene} \cite{zhao2023understanding} \cite{zhou2023codebertscore} \cite{zhou2023large} \cite{zhuo2023large} \\ 
\cline{2-4}
Token-based input
& Code in tokens & 118 & \cite{agarwal2024structured}  \cite{ahmed2024automatic} \cite{alam2023gptclonebench} \cite{arakelyan2023exploring} \cite{bui2023codetf} \cite{chan2023transformer} \cite{charalambous2023new} \cite{chen2022transferability} \cite{chen2023diversevul} \cite{chen2024apigen} \cite{chochlov2022using} \cite{chow2024pyty} \cite{ciniselli2021empirical} \cite{deligiannis2023fixing} \cite{dou2023towards} \cite{du2023resolving} \cite{fakhoury2023towards} \cite{fan2024rapid} \cite{fatima2022flakify} \cite{gao2023constructing} \cite{gao2023far} \cite{gao2024learning} \cite{geng2024large} \cite{ghadhab2021augmenting} \cite{grishina2023earlybird} \cite{guo2023longcoder} \cite{huang2023ai} \cite{huang2023chain} \cite{ibrahimzada2023automated} \cite{islam2024code} \cite{izadi2022codefill} \cite{jana2023attention} \cite{jesse2022learning} \cite{jiang2023impact} \cite{jiang2023nova} 
\cite{kabir2024zs4c} \cite{kanade2020learning} \cite{kang2023evaluating} \cite{kang2023explainable} \cite{kang2023preliminary} \cite{khanfir2023efficient} \cite{khare2023understanding} \cite{lai2023ds}\cite{lajko2022towards} \cite{le2023invalidator} \cite{li2021toward} \cite{li2023codeeditor} \cite{li2023ptm} \cite{li2024rewriting} \cite{liu2020multi} \cite{liu2023contrabert}
\cite{liu2023harnessing} \cite{liu2023refbert} \cite{liu2023repobench} \cite{ma2024specgen} \cite{mao2023self} \cite{mastropaolo2022transfer} \cite{mastropaolo2022using} \cite{nguyen2023snippet} \cite{noever2023can} \cite{pan2023stelocoder} \cite{pan2023understanding} \cite{paul2023enhancing} \cite{peng2024domain} \cite{qi2023sut} \cite{quan2023xgv} \cite{rao2023ai} \cite{ruiz2024novel} \cite{saberi2023multilingual} \cite{schafer2023adaptive} \cite{shapkin2023entity} \cite{sharma2022exploratory} \cite{shen2022benchmarking} \cite{shestov2024finetuning} \cite{shi2022cross} \cite{shi2023sotana} \cite{shin2023domain} \cite{shin2023prompt} \cite{siddiq2023exploring} \cite{silva2023repairllama} \cite{steenhoek2023reinforcement} \cite{steenhoek2024dataflow} \cite{sun2023automatic} \cite{sun2023prompt} \cite{sun2024neural} \cite{tang2023csgvd} \cite{tang2023just} \cite{tian2020evaluating} \cite{tian2023best} \cite{tufano2023predicting} \cite{wadhwa2023frustrated} \cite{wan2022they} \cite{wang2023boosting} \cite{wang2023rap} \cite{wei2023copiloting} \cite{widjojo2023addressing} \cite{wong2023natural} \cite{wu2023large} \cite{xia2022practical} \cite{xia2023conversational} \cite{xie2023chatunitest} \cite{xu2022systematic} \cite{xu2024unilog} \cite{yan2023codetransocean} \cite{yang2023assessing} \cite{yang2023syntax} \cite{yang2024large} \cite{yu2023codereval} \cite{zeng2022extensive} \cite{zhang2023algo} \cite{zhang2023boosting} \cite{zhang2023evaluating} \cite{zhang2023multilingual} \cite{zhang2023steam} \cite{zhang2023toolcoder} \cite{zhang2024appt} \cite{zheng2023outline} \cite{zhu2023automating} \\
\cline{2-4}
& Code and text in tokens & 78 & \cite{bairi2023codeplan} \cite{chen2023effectiveness} \cite{dakhel2023effective} \cite{dibia2022aligning} \cite{dinh2024large} \cite{doderlein2022piloting} \cite{eghbali2024hallucinator} \cite{fan2022automated} \cite{gu2022assemble} \cite{guo2024deepseek} \cite{guo2024exploring} \cite{hajali2023functionconstrained} \cite{hu2023augmenting} \cite{huang2023codecot} \cite{islam2024llm} \cite{jain2023coarse} \cite{jimenez2023swe} \cite{jin2023binary} \cite{khan2023xcodeeval} \cite{kou2023model} \cite{li2022auger} \cite{li2023cctest} \cite{li2023exploring} \cite{li2023nuances} \cite{li2024deveval} \cite{liu2023improving} \cite{manh2023vault} \cite{mastropaolo2023towards} \cite{mohajer2023skipanalyzer} \cite{moon2023coffee} \cite{mu2023clarifygpt} \cite{ni2023lever} \cite{nijkamp2023codegen2} \cite{olausson2023demystifying} \cite{pei2023can} \cite{piya2023llm4tdd} \cite{prenner2021making} \cite{pudari2023copilot} \cite{rahmani2023improving} \cite{ridnik2024code} \cite{saberi2023utilization} \cite{sadik2023analysis} \cite{saieva2023contrastive} \cite{sakib2023extending} \cite{salza2022effectiveness} \cite{schafer2023empirical} \cite{shirafuji2023exploring} \cite{shypula2023learning} \cite{sun2023clover} \cite{sun2023silent} \cite{sun2024llm4vuln} \cite{tang2023domain} \cite{tian2023chatgpt} \cite{tian2024debugbench} \cite{tufano2022using} \cite{wang2023codet5+} \cite{wang2023one} \cite{wang2024sparsecoder} \cite{wang2024teaching} \cite{weyssow2023exploring} \cite{weyssow2023usage} \cite{wu2023deceptprompt} \cite{xia2023keep} \cite{xia2024fuzz4all} \cite{yan2023closer} \cite{yang2023enhancing} \cite{yen2023coladder} \cite{yeticstiren2023evaluating} \cite{yuan2022circle} \cite{zan2023private} \cite{zhang2022beqain} \cite{zhang2022coditt5} \cite{zhang2023prompt} \cite{zhang2023self} \cite{zhang2024codeagent} \cite{zhao2023right} \cite{zheng2023codegeex} \cite{zhuo2023pop}
\\ \hline
Tree/Graph-based input 
& Code in tree structure & 2 &  \cite{ochs2023evaluating} \cite{zhang2023gamma} \\
\cline{2-4}
& Code in graph structure & 3 & \cite{du2023pre} \cite{ma2023scope} \cite{zhang2023neural} \\ \hline
Pixel-based input & Pixel & 1 &\cite{nasir2023llmatic} \\ \hline
Hybrid-based input & Hybrid input forms & 2 & \cite{al2023extending} \cite{niu2022spt} \\ \hline
\end{tabular}}
\label{tab:all_input-forms_continued}
\end{table}

\section{Prompt Engineering}
\label{app:prompting}

Table ~\ref{tab:all_prompting} showcases eight prompt engineering techniques mentioned in 395 studies: Few-shot prompting, Zero-shot prompting, CoT (Chain-of-Thought) prompting, APE (Automatic Prompt Engineer), CoC (Chain of Code) prompting, Auto-CoT (Automatic Chain-of-Thought) prompting, MoT (Modular-of-Thought) prompting, and SCoT (Structured Chain-of-Thought) prompting.

\begin{table}[h]
\caption{Prompt engineering techniques for SE tasks.}
\resizebox{0.95\linewidth}{!}{
\begin{tabular}{p{0.3\linewidth}|c|p{0.69\linewidth}}
\hline
\textbf{Prompt engineering} & \textbf{\# Studies} & \textbf{References} \\ \hline
Few-shot prompting & 88 & \cite{ahmed2024automatic} \cite{alam2023gptclonebench} \cite{arakelyan2023exploring} \cite{bairi2023codeplan} \cite{cao2023study} \cite{chan2023transformer} \cite{charalambous2023new} \cite{chen2021evaluating} \cite{chen2023improving} \cite{dakhel2023effective} \cite{deligiannis2023fixing} \cite{deng2023language} \cite{doderlein2022piloting} \cite{dong2023self} \cite{dou2023towards} \cite{eghbali2024hallucinator} \cite{fakhoury2023towards} \cite{feng2023prompting} \cite{first2023baldur} \cite{gao2023far} \cite{geng2024large} \cite{guo2024deepseek} \cite{hajali2023functionconstrained} \cite{hao2023v} \cite{helmeczi2023few} \cite{hendrycks2021measuring} \cite{hu2024leveraging} \cite{huang2023agentcoder} \cite{huang2023anpl} \cite{huang2023api} \cite{huang2023codecot} \cite{jain2023llm}  \cite{ji2023benchmarking} \cite{jiang2023nova} \cite{jiang2023self} \cite{kang2022large} \cite{kang2023evaluating} \cite{kang2023explainable} \cite{khakhar2023pac} \cite{khare2023understanding} \cite{kuznia2022less} \cite{le2023codechain} \cite{li2023codeie} \cite{li2023exploring} \cite{li2023structured} \cite{liu2023better} \cite{ma2024specgen} \cite{madaan2022language} \cite{mohajer2023skipanalyzer} \cite{moon2023coffee} \cite{mu2023clarifygpt}  \\ 
\hline
\end{tabular}}
\caption*{\hfill(Continued)}
\label{tab:all_prompting}
\end{table}

\begin{table}[h]
\captionsetup{labelformat=empty} 
\caption*{Table \ref{tab:all_prompting}. Continued.}
\resizebox{0.95\linewidth}{!}{
\begin{tabular}{p{0.3\linewidth}|c|p{0.69\linewidth}}
\hline
\textbf{Prompt engineering} & \textbf{\# Studies} & \textbf{References} \\ \hline
Few-shot prompting & & \cite{mukherjee2023stack} \cite{ni2023lever} \cite{nichols2024can} \cite{nijkamp2023codegen2} \cite{olausson2023demystifying} \cite{paranjape2023art} \cite{rao2023ai} \cite{ronanki2023chatgpt} \cite{scao2022bloom} \cite{schlag2023large} \cite{shypula2023learning} \cite{su2022selective} \cite{sun2023automatic} \cite{sun2023clover} \cite{sun2023prompt} \cite{tian2024debugbench} \cite{tufano2023predicting} \cite{wang2023chatcoder} \cite{wang2023codet5+} \cite{wang2023measuring} \cite{wang2024oop} \cite{wei2022clear} \cite{xia2023conversational} \cite{xie2023impact} \cite{xu2023lmpa} \cite{xu2024unilog} \cite{yan2023codetransocean} \cite{yang2023apidocbooster} \cite{yang2023white} \cite{ye2023generating} \cite{zhang2023multilingual} \cite{zhang2023revisiting} \cite{zhang2024experimenting} \cite{zhao2023automatic} \cite{zheng2023codegeex} \cite{zhong2023study} \cite{zhuo2023pop} \\
\hline
Zero-shot prompting	& 79 & \cite{ahmed2024automatic} \cite{bairi2023codeplan} \cite{buscemi2023comparative} \cite{chan2023transformer} \cite{chen2024apigen} \cite{dakhel2023effective} \cite{deng2023language} \cite{deng2023large} \cite{dong2023codescore} \cite{dong2023self} \cite{fakhoury2023towards} \cite{fan2022automated} \cite{fan2024rapid} \cite{gao2023constructing} \cite{gu2022assemble} \cite{guo2024deepseek} \cite{guo2024exploring} \cite{gupta2023grace} \cite{huang2023agentcoder} \cite{ibrahimzada2023automated} \cite{jain2023coarse} \cite{jin2023binary} \cite{kabir2023empirical} \cite{kabir2024zs4c} \cite{kang2023explainable} \cite{kang2023preliminary} \cite{khan2023xcodeeval} \cite{kou2023automated} \cite{li2023codeeditor} \cite{li2023exploring} \cite{li2023novel} \cite{li2023structured} \cite{li2023think} \cite{li2024deveval} \cite{li2024rewriting} \cite{liu2023better} \cite{liu2023refining} \cite{liu2023repobench} \cite{luo2022prcbert} \cite{mastropaolo2023towards} \cite{mohajer2023skipanalyzer} \cite{nijkamp2023codegen2} \cite{ouyang2023llm} \cite{paul2023enhancing} \cite{pearce2021examining} \cite{pegolotti2023qigen} \cite{pei2023can} \cite{ruiz2024novel} \cite{scao2022bloom} \cite{schafer2023empirical} \cite{shen2022benchmarking} \cite{shirafuji2023exploring} \cite{sun2023prompt} \cite{tian2023chatgpt} \cite{tian2024debugbench} \cite{tihanyi2023formai} \cite{tufano2023predicting} \cite{wadhwa2023frustrated} \cite{wang2023chatcoder} \cite{wang2023codet5+} \cite{wang2024oop} \cite{weyssow2023usage} \cite{wu2023deceptprompt} \cite{wu2023effective} \cite{xia2022practical} \cite{yan2023closer} \cite{yang2023apidocbooster} \cite{yang2023assessing} \cite{ye2023generating} \cite{zhang2023algo} \cite{zhang2023coder} \cite{zhang2023cupid} \cite{zhang2023gamma} \cite{zhang2023revisiting} \cite{zhang2023self} \cite{zhang2023steam} \cite{zhao2023right} \cite{zheng2023codegeex} \cite{zhong2023study} \\ 
\hline
CoT (Chain-of-Thought) prompting & 18 & \cite{deng2023language} \cite{feng2023prompting} \cite{huang2023ai} \cite{huang2023chain} \cite{li2023nuances} \cite{li2023structured} \cite{liu2024interpretable} \cite{mu2023clarifygpt} \cite{qian2023communicative} \cite{schlag2023large} \cite{shypula2023learning} \cite{tian2023test} \cite{wang2023leti} \cite{wang2024oop} \cite{yan2023codetransocean} \cite{yang2023syntax} \cite{zhang2023prompt} \cite{zhang2024experimenting} \\ 
\hline
APE (Automatic Prompt Engineer) & 2 & \cite{sun2023prompt} \cite{zhou2023large}  \\ 
\hline
CoC (Chain of Code) prompting & 2 & \cite{huang2023codecot} \cite{le2023codechain} \\ 
\hline
Auto-CoT (Automatic Chain-of-Thought) prompting	& 1 & \cite{paranjape2023art} \\ 
\hline
MoT (Modular-of-Thought) prompting & 1 & \cite{li2023motcoder} \\ 
\hline
SCoT (Structured Chain-of-Thought) prompting & 1 & \cite{li2023structured} \\ 
\hline
Others & 76 & \cite{ahmed2024automatic} \cite{azaria2023chatgpt} \cite{cassano2023multipl} \cite{cheng2023gpt} \cite{dakhel2023effective} \cite{dinh2024large} \cite{du2023classeval} \cite{du2023resolving} \cite{endres2023formalizing} \cite{gandhi2023natural} \cite{gilbert2023semantic} \cite{hong2023metagpt} \cite{hu2023augmenting} \cite{jiang2023impact} \cite{jiang2023selfevolve} \cite{jin2023binary} \cite{jin2023inferfix} \cite{jones2022capturing} \cite{ke2023discriminating} \cite{koide2023detecting} \cite{lai2023ds} \cite{lee2022light} \cite{li2022generation} \cite{li2023cctest} \cite{li2023enabling} \cite{li2023novel} \cite{li2023nuances} \cite{liu2023fill} \cite{liu2023harnessing} \cite{liu2023improving} \cite{liu2023no} \cite{liu2024interpretable} \cite{ma2023scope} \cite{nam2023ide} \cite{nasir2023llmatic} \cite{pan2023understanding} \cite{paul2023enhancing} \cite{pei2023can} \cite{piya2023llm4tdd} \cite{ren2023misuse} \cite{sadik2023analysis} \cite{shi2023sotana} \cite{shin2023prompt} \cite{shypula2023learning} \cite{siddiq2023exploring} \cite{singla2023evaluating} \cite{sridhara2023chatgpt} \cite{sun2023prompt} \cite{sun2024llm4vuln} \cite{tan2023copilot} \cite{tang2023chatgpt} \cite{tang2023just} \cite{tu2023llm4cbi} \cite{wang2023boosting} \cite{wang2023evaluating} \cite{wei2023magicoder} \cite{white2023chatgpt} \cite{white2023prompt} \cite{widjojo2023addressing} \cite{wong2023natural} \cite{wu2023large} \cite{xia2023keep} \cite{xia2024fuzz4all} \cite{xie2023chatunitest} \cite{xiong2023program} \cite{xu2023guiding} \cite{yang2023syntax} \cite{yen2023coladder} \cite{yeticstiren2023evaluating} \cite{yu2023codereval} \cite{yuan2022circle} \cite{yuan2023no} \cite{zan2022language} \cite{zan2023private} \cite{zhang2022using} \cite{zhao2023understanding} \\ 
\hline
\end{tabular}}
\label{tab:all_prompting_continued}
\end{table}

\section{Evaluation Metrics}
\label{app:evaluation}

We categorize the types of tasks that LLMs address in SE into four categories: regression, classification, recommendation, and generation. Each task has commonly used evaluation metrics, as shown in Table ~\ref{tab:all_evaluation}.

\begin{table}[h]
\caption{Evaluation metrics for different types of tasks.}
\resizebox{\linewidth}{!}{
\begin{tabular}{r|p{0.2\linewidth}|c|p{0.75\linewidth}}
\hline
\textbf{Problem Type} & \textbf{Metric} & \textbf{\# Studies} & \textbf{References} \\ \hline
Regression
& MAE (Mean Absolute Error) & 1 & \cite{fu2022gpt2sp} \\ \hline
Classification
& Precision & 35 & \cite{alam2023gptclonebench} \cite{biswas2020achieving} \cite{chan2023transformer} \cite{chen2023diversevul} \cite{chochlov2022using} \cite{dou2023towards} \cite{du2023pre} \cite{ezzini2022automated} \cite{fatima2022flakify} \cite{ferreira2024incivility} \cite{he2022ptm4tag} \cite{hey2020norbert} \cite{huang2023api} \cite{khan2021automatic} \cite{khan2022automatic} \cite{khare2023understanding} \cite{koide2023detecting} \cite{kolthoff2023data} \cite{le2023invalidator} \cite{mukherjee2023stack} \cite{pei2023can} \cite{poudel2023leveraging} \cite{quan2023xgv} \cite{sghaier2023multi} \cite{sharma2022exploratory} \cite{shi2023towards} \cite{steenhoek2024dataflow} \cite{sun2023dexbert} \cite{thapa2022transformer} \cite{tian2023best} \cite{yang2022aspect} \cite{zeng2022extensive} \cite{zhang2020sentiment} \cite{zhang2023prompt} \cite{zhang2024appt} \\
\cline{2-4}
& Recall & 34 & \cite{alam2023gptclonebench} \cite{biswas2020achieving} \cite{chan2023transformer} \cite{chen2023diversevul} \cite{chochlov2022using} \cite{dou2023towards} \cite{ezzini2022automated} \cite{fatima2022flakify} \cite{ferreira2024incivility} \cite{he2022ptm4tag} \cite{hey2020norbert} \cite{huang2023api} \cite{khan2021automatic} \cite{khan2022automatic} \cite{khare2023understanding} \cite{koide2023detecting} \cite{kolthoff2023data} \cite{le2023invalidator} \cite{mukherjee2023stack} \cite{pei2023can} \cite{quan2023xgv} \cite{sghaier2023multi} \cite{sharma2022exploratory} \cite{shi2023towards} \cite{steenhoek2024dataflow} \cite{sun2023dexbert} \cite{tang2023just} \cite{thapa2022transformer} \cite{tian2023best} \cite{zeng2022extensive} \cite{zhang2020sentiment} \cite{zhang2023cupid} \cite{zhang2023prompt} \cite{zhang2024appt} \\
\cline{2-4}
& F1-score & 33 & \cite{biswas2020achieving} \cite{chan2023transformer} \cite{chen2023diversevul} \cite{fatima2022flakify} \cite{ferreira2024incivility} \cite{grishina2023earlybird} \cite{he2022ptm4tag} \cite{hey2020norbert} \cite{huang2023api} \cite{khan2021automatic} \cite{khan2022automatic} \cite{khare2023understanding} \cite{koide2023detecting} \cite{kolthoff2023data} \cite{le2023invalidator} \cite{liu2023harnessing} \cite{luo2022prcbert} \cite{mukherjee2023stack} \cite{pei2023can} \cite{quan2023xgv} \cite{sghaier2023multi} \cite{sharma2022exploratory} \cite{shestov2024finetuning} \cite{shi2023towards} \cite{steenhoek2024dataflow} \cite{tang2023just} \cite{thapa2022transformer} \cite{yang2022aspect} \cite{zeng2022extensive} \cite{zhang2020sentiment} \cite{zhang2023prompt} \cite{zhang2024appt} \cite{zhou2023universalner} \\
\cline{2-4}
& Accuracy & 23 & \cite{chen2023diversevul} \cite{gomes2023bert} \cite{grishina2023earlybird} \cite{jesse2022learning} \cite{kanade2020learning} \cite{kang2023preliminary} \cite{khan2021automatic} \cite{khan2022automatic} \cite{khare2023understanding} \cite{koide2023detecting} \cite{le2023invalidator} \cite{lee2022light} \cite{li2023nuances} \\
\hline

\end{tabular}}
\caption*{\hfill(Continued\_1)}
\label{tab:all_evaluation}
\end{table}

\begin{table}[t]
\captionsetup{labelformat=empty} 
\caption*{Table \ref{tab:all_evaluation}. Continued\_1.}
\resizebox{\linewidth}{!}{
\begin{tabular}{r|p{0.2\linewidth}|c|p{0.75\linewidth}}
\hline
\textbf{Problem Type} & \textbf{Metric} & \textbf{\# Studies} & \textbf{References} \\ \hline
Classification
& Accuracy &  &  \cite{liu2023harnessing} \cite{manh2023vault} \cite{ni2023lever} \cite{pei2023can} \cite{quan2023xgv} \cite{saberi2023utilization} \cite{tian2023best} \cite{zeng2022extensive} \cite{zhang2023prompt} \cite{zhang2024appt} \\
\cline{2-4}
& AUC (Area Under the ROC Curve) & 9 & \cite{alhamed2022evaluation} \cite{el2023ai} \cite{shestov2024finetuning} \cite{tang2023just} \cite{tian2023best} \cite{wang2023evaluating} \cite{yang2022aspect} \cite{yang2024large} \cite{zhang2024appt} \\
\cline{2-4}
& ROC (Receiver Operating Characteristic) & 4 & \cite{alhamed2022evaluation} \cite{du2023pre} \cite{el2023ai} \cite{shestov2024finetuning} \\
\cline{2-4}
& FPR (False Positive Rate) & 4 & \cite{chen2023diversevul} \cite{sun2023gpt} \cite{thapa2022transformer} \cite{wang2023evaluating} \\
\cline{2-4}
& FNR (Falser Negative Rate) & 3 & \cite{sun2023gpt} \cite{thapa2022transformer} \cite{wang2023evaluating} \\ 
\cline{2-4}
& MCC (Matthews Correlation Coefficient) & 2 & \cite{ferreira2024incivility} \cite{yang2022aspect} \\ 
\hline
Recommendation 
& MRR (Mean Reciprocal Rank) & 15 & \cite{chen2024apigen} \cite{ciborowska2022fast} \cite{fan2024rapid} \cite{izadi2022codefill} \cite{li2021toward} \cite{li2023ptm} \cite{li2024rewriting} \cite{lin2021traceability} \cite{luitel2023improving} \cite{mao2023self} \cite{rahmani2023improving} \cite{salza2022effectiveness} \cite{shi2022cross} \cite{wang2023one} \cite{wei2022clear} \\
\cline{2-4}
& Precision/Precision@k & 6 & \cite{chen2024apigen} \cite{he2023representation} \cite{lin2021traceability} \cite{wei2022clear} \cite{wu2024automatic} \cite{zhu2023automating} \\
\cline{2-4}
& MAP/MAP@k & 6 & \cite{chen2024apigen} \cite{ciborowska2022fast} \cite{isotani2021duplicate} \cite{lin2021traceability} \cite{wei2022clear} \cite{zhu2022enhancing} \\
\cline{2-4}
& F-score/F-score@k & 5 & \cite{he2023representation} \cite{lin2021traceability} \cite{wu2024automatic} \cite{zhu2023automating} \cite{zhu2022enhancing}\\
\cline{2-4}
& Recall/Recall@k & 4 & \cite{he2023representation} \cite{wei2022clear} \cite{wu2024automatic} \cite{zhu2023automating} \\
\cline{2-4}
& Accuracy & 3 & \cite{izadi2022codefill} \cite{li2021toward} \cite{salza2022effectiveness} \\ 
\hline
Generation 
& BLEU/BLEU-4/BLEU-DC
& 62 & \cite{agarwal2024structured} \cite{ahmed2024automatic} \cite{al2023extending} \cite{arakelyan2023exploring} \cite{chen2022transferability} \cite{chen2023effectiveness} \cite{ciniselli2021empirical} \cite{dibia2022aligning} \cite{gao2023constructing} \cite{geng2024large} \cite{gu2022assemble} \cite{islam2024code} \cite{islam2024llm} \cite{jana2023attention} \cite{jiang2023self} \cite{jin2023assess} \cite{kuznia2022less} \cite{li2023cctest} \cite{li2023codeeditor} \cite{li2023exploring} \cite{li2023novel} \cite{liu2023contrabert} \cite{liu2023improving} \cite{liu2024reliability} \cite{lu2023llama} \cite{mastropaolo2021empirical} \cite{mastropaolo2021studying} \cite{mastropaolo2022transfer} \cite{mastropaolo2023towards} \cite{murali2023codecompose} \cite{niu2022spt} \cite{paul2023enhancing} \cite{ruiz2024novel} \cite{saberi2023multilingual} \cite{shapkin2023entity} \cite{shi2023sotana} \cite{shi2023towards} \cite{shin2023domain} \cite{shin2023prompt} \cite{shirafuji2023exploring} \cite{sun2023prompt} \cite{tufano2022using} \cite{wang2023codet5+} \cite{wang2023natural} \cite{wang2023one} \cite{wang2023rap} \cite{wang2024sparsecoder} \cite{wang2024teaching} \cite{weyssow2023usage} \cite{yang2023assessing} \cite{yang2023enhancing} \cite{yang2023syntax} \cite{ye2023generating} \cite{zan2023private} \cite{zeng2022extensive} \cite{zhang2022coditt5} \cite{zhang2023multilingual} \cite{zhang2023steam} \cite{zheng2023codegeex} \cite{zhou2023codebertscore} \cite{zhuo2023large} \\ 
\cline{2-4}
& Pass@k & 54 & \cite{agarwal2024structured} \cite{bui2023codetf} \cite{cassano2023multipl} \cite{chen2021evaluating} \cite{chen2023improving} \cite{dinh2024large} \cite{doderlein2022piloting} \cite{dong2023self} \cite{du2023classeval} \cite{fakhoury2023towards} \cite{guo2024deepseek} \cite{hajali2023functionconstrained} \cite{hong2023metagpt} \cite{huang2023agentcoder} \cite{huang2023codecot} \cite{jain2023coarse} \cite{jain2023llm} \cite{jiang2023self} \cite{jiang2023selfevolve} \cite{lahiri2022interactive} \cite{le2023codechain} \cite{li2023enabling} \cite{li2023motcoder} \cite{li2023novel} \cite{li2023structured} \cite{li2023think} \cite{li2024deveval} \cite{liu2023refining} \cite{luo2023wizardcoder} \cite{moon2023coffee} \cite{mu2023clarifygpt} \cite{nichols2024can} \cite{olausson2023demystifying} \cite{ridnik2024code} \cite{sadik2023analysis} \cite{shi2023sotana} \cite{shin2023prompt} \cite{shirafuji2023exploring} \cite{thakur2023verigen} \cite{tian2023test} \cite{wang2023chatcoder} \cite{wang2023leti} \cite{wang2024oop} \cite{wei2023magicoder} \cite{yan2023closer} \cite{yang2023assessing} \cite{yu2023codereval} \cite{zan2022cert} \cite{zan2022language} \cite{zan2023private} \cite{zhang2023algo} \cite{zhang2023self} \cite{zhang2024codeagent} \cite{zheng2023codegeex} \\ 
\cline{2-4}
& Accuracy/Accuracy@k & 38 & \cite{feng2023prompting} \cite{hu2024leveraging} \cite{huang2023ai} \cite{huang2023chain} \cite{ibrahimzada2023automated} \cite{jain2022jigsaw} \cite{jana2023attention} \cite{jin2023inferfix} \cite{kang2022large} \cite{kang2023evaluating} \cite{lahiri2022interactive} \cite{liu2020multi} \cite{liu2023contrabert} \cite{liu2023refbert} \cite{ma2024knowlog} \cite{mastropaolo2021studying} \cite{mastropaolo2022transfer} \cite{mohajer2023skipanalyzer} \cite{ni2023l2ceval} \cite{niu2022spt} \cite{patel2023evaluating} \cite{pegolotti2023qigen} \cite{qi2023sut} \cite{rao2023ai} \cite{ruiz2024novel} \cite{schlag2023large} \cite{shapkin2023entity} \cite{shin2023prompt} \cite{shirafuji2023exploring} \cite{sun2024neural} \cite{tian2020evaluating} \cite{xie2023impact} \cite{xu2024unilog} \cite{ye2023generating} \cite{yu2023log} \cite{zeng2022extensive} \cite{zhang2022beqain} \cite{zhang2024selene} \\ 
\cline{2-4}
& EM (Exact Match) & 36 & \cite{agarwal2024structured} \cite{al2023extending} \cite{chow2024pyty} \cite{eghbali2024hallucinator} \cite{gao2023constructing} \cite{gilbert2023semantic} \cite{guo2023longcoder} \cite{guo2024deepseek} \cite{guo2024exploring} \cite{gupta2023grace} \cite{jana2023attention} \cite{jin2023binary} \cite{li2023codeeditor} \cite{li2023novel} \cite{liu2023refbert} \cite{liu2023repobench} \cite{mastropaolo2023towards} \cite{murali2023codecompose} \cite{paul2023enhancing} \cite{pegolotti2023qigen} \cite{qi2023sut} \cite{ruiz2024novel} \cite{shapkin2023entity} \cite{shi2023towards} \cite{shirafuji2023exploring} \cite{tang2023domain} \cite{wang2023codet5+} \cite{wang2023rap} \cite{weyssow2023exploring} \cite{weyssow2023usage} \cite{widyasari2023explaining} \cite{yang2023assessing} \cite{ye2023generating} \cite{zhang2022coditt5} \cite{zhang2023neural} \cite{zhang2023self} \\ 
\cline{2-4}
& CodeBLEU & 29 & \cite{agarwal2024structured} \cite{arakelyan2023exploring} \cite{chen2023effectiveness} \cite{gao2023constructing} \cite{guo2024exploring} \cite{jana2023attention} \cite{jiang2023self} \cite{li2023novel} \cite{liu2023improving} \cite{mastropaolo2023robustness} \cite{pan2023stelocoder} \cite{paul2023enhancing} \cite{ruiz2024novel} \cite{shapkin2023entity} \cite{shin2023domain} \cite{shin2023prompt} \cite{shirafuji2023exploring} \cite{wang2023codet5+} \cite{wang2023natural} \cite{wang2024teaching} \cite{weyssow2023exploring} \cite{weyssow2023usage} \cite{yang2023assessing} \cite{zan2023private} \cite{zeng2022extensive} \cite{zhang2023multilingual} \cite{zheng2023codegeex} \cite{zhou2023codebertscore} \cite{zhuo2023large} \\ 
\cline{2-4}
& ROUGE/ROUGE-L & 22 & \cite{ahmed2024automatic} \cite{al2023extending} \cite{gao2023constructing} \cite{geng2024large} \cite{gu2022assemble} \cite{islam2024code} \cite{islam2024llm} \cite{jin2023assess} \cite{li2022auger} \cite{li2023exploring} \cite{mastropaolo2021studying} \cite{niu2022spt} \cite{shi2023sotana} \cite{sun2023automatic} \cite{sun2023prompt} \cite{sun2024neural} \cite{wang2024sparsecoder} \cite{yang2023apidocbooster} \cite{yang2023enhancing} \cite{zan2023private} \cite{zhou2023codebertscore} \cite{zhuo2023large} \\ 
\cline{2-4}
& Precision & 18 & \cite{ciniselli2021empirical} \cite{gao2024learning} \cite{ibrahimzada2023automated} \cite{kabir2024zs4c} \cite{kou2023automated} \cite{lu2023llama} \cite{mohajer2023skipanalyzer} \cite{shi2023towards} \cite{sun2024neural} \cite{tian2020evaluating} \cite{wang2022your} \cite{weyssow2023usage} \cite{xie2023impact} \cite{yang2023apidocbooster} \cite{zeng2022extensive} \cite{zhang2022coditt5} \cite{zhang2023revisiting} \cite{zhuo2023pop} \\
\cline{2-4}
& METEOR & 16 & \cite{ahmed2024automatic} \cite{al2023extending} \cite{chen2022transferability} \cite{gao2023constructing} \cite{geng2024large} \cite{gu2022assemble} \cite{jin2023assess} \cite{niu2022spt} \cite{shi2023sotana} \cite{sun2023automatic} \cite{sun2023prompt} \cite{wang2024sparsecoder} \cite{yang2023enhancing} \cite{zhang2022coditt5} \cite{zhou2023codebertscore} \cite{zhuo2023large} \\ 
\cline{2-4}
& Recall & 15 & \cite{gao2024learning} \cite{ibrahimzada2023automated} \cite{kabir2024zs4c} \cite{kou2023automated} \cite{li2024deveval} \cite{lu2023llama} \cite{mohajer2023skipanalyzer} \cite{shi2023towards} \cite{sun2024neural} \cite{tian2020evaluating} \cite{wang2022your} \cite{xie2023impact} \cite{yang2023apidocbooster} \cite{zeng2022extensive} \cite{zhang2023revisiting} \\
\cline{2-4}
& F1-score & 15 & \cite{gao2024learning} \cite{ibrahimzada2023automated} \cite{kabir2024zs4c} \cite{kou2023automated} \cite{li2024deveval} \cite{lu2023llama} \cite{mohajer2023skipanalyzer} \cite{shi2023towards} \cite{sun2024neural} \cite{tian2020evaluating} \cite{wang2022your} \cite{xie2023impact} \cite{yang2023apidocbooster} \cite{zeng2022extensive} \cite{zhang2023revisiting}\\
\cline{2-4}
& MRR (Mean Reciprocal Rank) & 6 & \cite{liu2023contrabert} \cite{niu2022spt} \cite{shi2023towards} \cite{wang2023one} \cite{yang2023enhancing} \cite{zeng2022extensive}\\
\cline{2-4}
& ES (Edit Similarity) & 6 & \cite{eghbali2024hallucinator} \cite{guo2023longcoder} \cite{li2023cctest} \cite{liu2023repobench} \cite{tang2023domain} \cite{wang2024teaching} \\ 
\cline{2-4}
& ED (Edit Distance)  & 5 & \cite{eghbali2024hallucinator} \cite{li2023codeeditor} \cite{liu2023refbert} \cite{ouyang2023llm} \cite{yu2023log} \\
\cline{2-4}
& MAR (Mean Average Ranking) & 4 & \cite{ruiz2024novel} \cite{tu2023llm4cbi} \cite{wang2023one} \cite{zeng2022extensive} \\
\hline
\end{tabular}}
\caption*{\hfill(Continued\_2)}
\label{tab:all_evaluation_continued}
\end{table}

\begin{table}[t]
\captionsetup{labelformat=empty} 
\caption*{Table \ref{tab:all_evaluation}. Continued\_2.}
\resizebox{\linewidth}{!}{
\begin{tabular}{r|p{0.2\linewidth}|c|p{0.75\linewidth}}
\hline
\textbf{Problem Type} & \textbf{Metric} & \textbf{\# Studies} & \textbf{References} \\ \hline
Generation
& ChrF & 3 & \cite{shapkin2023entity} \cite{zhou2023codebertscore} \cite{zhuo2023large} \\
\cline{2-4}
& CrystalBLEU & 3 & \cite{li2023codeeditor} \cite{zhou2023codebertscore} \cite{zhuo2023large} \\
\cline{2-4}
& CodeBERTScore & 2 & \cite{zhou2023codebertscore} \cite{zhuo2023large} \\
\cline{2-4}
& MFR (Mean First Ranking) & 1 & \cite{tu2023llm4cbi} \\
\cline{2-4}
& PP (Perplexity) & 1 & \cite{xu2022systematic} \\
\hline
\end{tabular}}
\label{tab:all_evaluation_continued_2}
\end{table}
\section{SE Tasks}
\label{app:tasks}

According to the software development lifecycle, we have categorized the SE tasks mentioned in 395 studies into six categories: Requirements engineering, Software design, Software development, Software quality assurance, Software maintenance, and Software management. Table ~\ref{tab:all_se_tasks} presents all the papers that apply LLMs to these tasks.

\begin{table}[h]
\caption{Distribution of SE tasks over six activities.}
\resizebox{\linewidth}{!}{
\begin{tabular}{r|p{0.35\linewidth}|c|p{0.5\linewidth}}
\hline
\textbf{SE Activity} & \textbf{SE Task} & \textbf{\# Studies} & \textbf{References} \\ \hline
Requirements engineering
& Anaphoric ambiguity treatment & 4 & \cite{ezzini2022automated} \cite{moharil2022identification} \cite{moharil2023tabasco} \cite{sridhara2023chatgpt} \\
\cline{2-4}
& Requirements classification & 4 & \cite{el2023ai} \cite{helmeczi2023few} \cite{hey2020norbert} \cite{luo2022prcbert} \\
\cline{2-4}
& Requirement analysis and evaluation & 2 & \cite{poudel2023leveraging} \cite{ronanki2023chatgpt} \\
\cline{2-4}
& Specification generation & 2 & \cite{ma2024specgen} \cite{xie2023impact} \\
\cline{2-4}
& Coreference detection & 1 & \cite{wang2020deep} \\
\cline{2-4}
& Requirements elicitation & 1 & \cite{white2023chatgpt} \\
\cline{2-4}
& Specification formalization & 1 & \cite{endres2023formalizing} \\
\cline{2-4}
& Traceability automation & 1 & \cite{lin2021traceability} \\
\cline{2-4}
& Use cases generation & 1 & \cite{zhang2024experimenting} \\ 
\hline
Software design
& GUI retrieval & 1 & \cite{kolthoff2023data}\\
\cline{2-4}
& Rapid prototyping & 1 & \cite{mandal2023large} \\
\cline{2-4}
& Software specification synthesis & 1 & \cite{white2023chatgpt} \\
\cline{2-4}
& System design & 1 & \cite{zhang2024experimenting} \\
\hline
Software development
& Code generation & 118 & \cite{azaria2023chatgpt} \cite{bareiss2022code} \cite{buscemi2023comparative} \cite{cassano2023multipl} \cite{chen2021evaluating} \cite{chen2023effectiveness} \cite{chen2023improving} \cite{chen2023teaching} \cite{dibia2022aligning} \cite{dong2023abilities} \cite{dong2023codescore} \cite{dong2023self} \cite{du2023classeval} \cite{fakhoury2023towards} \cite{gilbert2023semantic} \cite{gong2023intended} \cite{guo2024deepseek} \cite{hendrycks2021measuring} \cite{hong2023metagpt} \cite{hu2024leveraging} \cite{huang2023adaptive} \cite{huang2023agentcoder} \cite{huang2023anpl} \cite{huang2023codecot} \cite{jain2023coarse} \cite{jain2023llm} \cite{ji2023benchmarking} \cite{jiang2023self} \cite{jiang2023selfevolve} \cite{jimenez2023swe} \cite{jones2022capturing} \cite{ke2023discriminating} \cite{khan2023assessing} \cite{khan2023xcodeeval} \cite{kou2023model} \cite{lahiri2022interactive} \cite{lai2023ds} \cite{laskar2023systematic} \cite{le2023codechain} \cite{li2023codeie} \cite{li2023enabling} \cite{li2023motcoder} \cite{li2023novel} \cite{li2023structured} \cite{li2023think} \cite{li2024deveval} \cite{lin2023applications} \cite{liu2023better} \cite{liu2023improving} \cite{liu2023no} \cite{liu2023refining} \cite{liu2023your} \cite{liu2024reliability} \cite{luo2023wizardcoder} \cite{madaan2022language} \cite{mastropaolo2023robustness} \cite{mu2023clarifygpt} \cite{murali2023codecompose} \cite{nascimento2023comparing} \cite{nguyen2023snippet} \cite{ni2023l2ceval} \cite{nichols2024can} \cite{olausson2023demystifying} \cite{ouyang2023llm} \cite{patel2023evaluating} \cite{pegolotti2023qigen} \cite{rao2023ai} \cite{ren2023misuse} \cite{ridnik2024code} \cite{sadik2023analysis} \cite{sakib2023extending} \cite{schlag2023large} \cite{shapkin2023entity} \cite{shi2023sotana} \cite{shi2023towards} \cite{shin2023prompt} \cite{siddiq2023lightweight} \cite{su2022selective} \cite{sun2023clover} \cite{tan2023copilot} \cite{tarassow2023potential} \cite{thakur2023verigen} \cite{tian2023chatgpt} \cite{tian2023test} \cite{tihanyi2023formai} \cite{wang2022recode} \cite{wang2023chatcoder} \cite{wang2023evaluating} \cite{wang2023leti} \cite{wang2023natural} \cite{wang2024oop} \cite{wang2024teaching} \cite{wei2023magicoder} \cite{weyssow2023exploring} \cite{wong2023natural} \cite{wu2023ai} \cite{wu2023deceptprompt} \cite{yan2023closer} \cite{yang2023enhancing} \cite{yang2023syntax} \cite{yen2023coladder} \cite{yeticstiren2023evaluating} \cite{yu2023codereval} \cite{zan2022cert} \cite{zan2022language} \cite{zan2023private} \cite{zelikman2023self} \cite{zeng2022extensive} \cite{zhang2023coder} \cite{zhang2023multilingual} \cite{zhang2023self} \cite{zhang2024codeagent} \cite{zhang2024experimenting} \cite{zheng2023codegeex} \cite{zheng2023outline} \cite{zhong2023study} \cite{zhou2023codebertscore} \cite{zhuo2023large} \\
\cline{2-4}
& Code completion & 22 & \cite{ciniselli2021empirical} \cite{ding2023static} \cite{dinh2024large} \cite{doderlein2022piloting} \cite{eghbali2024hallucinator} \cite{guo2023longcoder} \cite{izadi2022codefill} \cite{kabir2024zs4c} \cite{khan2022automatic} \cite{li2021toward} \cite{li2023cctest} \cite{liu2020multi} \cite{liu2023repobench} \cite{ochs2023evaluating} \cite{pearce2021examining} \cite{prenner2021making} \cite{pudari2023copilot} \cite{shi2023towards} \cite{sun2024neural} \cite{tang2023domain} \cite{wong2023natural} \cite{xu2022systematic} \\
\cline{2-4}
& Code summarization & 21 & \cite{ahmed2024automatic} \cite{al2023extending} \cite{arakelyan2023exploring} \cite{chen2022transferability} \cite{gao2023constructing} \cite{gao2024learning} \cite{gu2022assemble} \cite{jin2023binary} \cite{mastropaolo2021studying} \cite{mastropaolo2022transfer} \cite{saberi2023multilingual} \cite{sadik2023analysis} \cite{shi2023sotana} \cite{shi2023towards} \cite{shin2023prompt} \cite{sun2023automatic} \cite{sun2023prompt} \cite{tian2023chatgpt} \cite{wang2023one} \cite{wang2024sparsecoder} \cite{yang2023enhancing} \\
\cline{2-4}
& Code search & 12 & \cite{fan2024rapid} \cite{li2022generation} \cite{li2022pre} \cite{li2024rewriting} \cite{liu2023contrabert} \cite{mao2023self} \cite{saieva2023contrastive} \cite{salza2022effectiveness} \cite{shi2022cross} \cite{shi2023towards} \cite{wang2023natural} \cite{wang2023one} \\
\hline
\end{tabular}}
\caption*{\hfill(Continued\_1)}
\label{tab:all_se_tasks}
\end{table}

\begin{table}[h]
\captionsetup{labelformat=empty} 
\caption*{Table \ref{tab:all_se_tasks}. Continued\_1.}
\resizebox{\linewidth}{!}{
\begin{tabular}{r|p{0.35\linewidth}|c|p{0.5\linewidth}}
\hline
\textbf{SE Activity} & \textbf{SE Task} & \textbf{\# Studies} & \textbf{References} \\ \hline
Software development
& Code translation & 12 & \cite{jana2023attention} \cite{jiang2023nova} \cite{liu2023contrabert} \cite{liu2024reliability} \cite{pan2023stelocoder} \cite{pan2023understanding} \cite{qi2023sut} \cite{shin2023prompt} \cite{wong2023natural} \cite{yan2023codetransocean} \cite{yang2023assessing} \cite{yang2023enhancing} \\
\cline{2-4}
& Code understanding & 8 & \cite{kanade2020learning} \cite{ma2023scope} \cite{manh2023vault} \cite{nam2023ide} \cite{shen2022benchmarking} \cite{wang2023codet5+} \cite{wong2023natural} \cite{zhao2023understanding} \\
\cline{2-4}
& API inference & 5 & \cite{huang2023adaptive} \cite{patil2023gorilla} \cite{wang2023measuring} \cite{weyssow2023usage} \cite{zhuo2023pop} \\
\cline{2-4}
& Program synthesis & 6 & \cite{gandhi2023natural} \cite{hajali2023functionconstrained} \cite{jain2022jigsaw} \cite{kuznia2022less} \cite{shirafuji2023exploring} \cite{singla2023evaluating} \\
\cline{2-4}
& API recommendation & 5 & \cite{chen2024apigen} \cite{li2023ptm} \cite{wei2022clear} \cite{wu2024automatic} \cite{zhang2023toolcoder} \\
\cline{2-4}
& Code editing & 5 & \cite{bairi2023codeplan} \cite{gupta2023grace} \cite{li2023codeeditor} \cite{moon2023coffee} \cite{shypula2023learning} \\
\cline{2-4}
& Code representation & 3 & \cite{agarwal2024structured} \cite{niu2022spt} \cite{wan2022they} \\
\cline{2-4}
& Code comment generation & 2 & \cite{geng2024large} \cite{mastropaolo2021empirical} \\
\cline{2-4}
& Method name generation & 2 & \cite{sridhara2023chatgpt} \cite{zhu2023automating} \\
\cline{2-4}
& Code recommendation & 2 & \cite{luitel2023improving} \cite{rahmani2023improving} \\
\cline{2-4}
& Agile story point estimation & 1 & \cite{fu2022gpt2sp} \\
\cline{2-4}
& API documentation augment & 1 & \cite{yang2023apidocbooster} \\
\cline{2-4}
& API documentation smells & 1 & \cite{khan2021automatic} \\
\cline{2-4}
& API entity and relation extraction & 1 & \cite{huang2023api} \\
\cline{2-4}
& Data analysis & 1 & \cite{cheng2023gpt} \\
\cline{2-4}
& Fuzz driver generation & 1 & \cite{zhang2023understanding} \\
\cline{2-4}
& Control flow graph generation & 1 & \cite{huang2023ai} \\
\cline{2-4}
& Identifier normalization & 1 & \cite{zhang2022beqain} \\
\cline{2-4}
& Instruction generation & 1 & \cite{zhou2023large} \\
\cline{2-4}
& Type inference & 1 & \cite{jesse2022learning} \\
\cline{2-4}
& Others & 14 & \cite{bui2023codetf} \cite{he2023representation} \cite{khakhar2023pac} \cite{mukherjee2023stack} \cite{nasir2023llmatic} \cite{nijkamp2023codegen2} \cite{paranjape2023art} \cite{piya2023llm4tdd} \cite{qian2023communicative} \cite{scao2022bloom} \cite{sheng2023flexgen} \cite{ye2023generating} \cite{zhao2023automatic} \cite{zhou2023universalner} \\
\hline
Software quality assurance
& Vulnerability detection & 18 & \cite{chan2023transformer} \cite{chen2023diversevul} \cite{gao2023far} \cite{gao2024learning} \cite{grishina2023earlybird} \cite{khare2023understanding} \cite{koide2023detecting} \cite{liu2023contrabert} \cite{noever2023can} \cite{shestov2024finetuning} \cite{steenhoek2024dataflow} \cite{sun2023gpt} \cite{sun2023silent} \cite{sun2024llm4vuln} \cite{tang2023csgvd} \cite{thapa2022transformer} \cite{yang2023white} \cite{zhang2023prompt} \\
\cline{2-4}
& Test generation & 17 & \cite{azaria2023chatgpt} \cite{dakhel2023effective} \cite{gao2024learning} \cite{kirinuki2024chatgpt} \cite{plein2023automatic} \cite{schafer2023adaptive} \cite{schafer2023empirical} \cite{shin2023domain} \cite{siddiq2023exploring} \cite{steenhoek2023reinforcement} \cite{tang2023chatgpt} \cite{vikram2023can} \cite{xia2024fuzz4all} \cite{xie2023chatunitest} \cite{xiong2023program} \cite{yuan2023no} \cite{zhang2023algo} \\
\cline{2-4}
& Bug localization & 5 & \cite{ciborowska2022fast} \cite{ciborowska2023too} \cite{du2023pre} \cite{feng2023prompting} \cite{kang2023preliminary} \\
\cline{2-4}
& Verification & 5 & \cite{charalambous2023new} \cite{first2023baldur} \cite{ni2023lever} \cite{tihanyi2023formai} \cite{zhang2024selene} \\
\cline{2-4}
& Testing automation & 4 & \cite{deng2023language} \cite{deng2023large} \cite{hu2023augmenting} \cite{khanfir2023efficient} \\
\cline{2-4}
& Fault localization & 3 & \cite{li2023nuances} \cite{wu2023large} \cite{yang2024large} \\
\cline{2-4}
& Defect detection & 2 & \cite{sun2023dexbert} \cite{wong2023natural} \\
\cline{2-4}
& GUI testing & 2 & \cite{liu2023fill} \cite{yoon2023autonomous} \\
\cline{2-4}
& Static analysis & 2 & \cite{hao2023v} \cite{mohajer2023skipanalyzer} \\
\cline{2-4}
& Binary taint analysis & 1 & \cite{liu2023harnessing} \\
\cline{2-4}
& Compiler fuzzing & 1 & \cite{quan2023xgv} \\
\cline{2-4}
& Decompilation & 1 & \cite{xu2023lmpa} \\
\cline{2-4}
& Invariant prediction & 1 & \cite{pei2023can} \\
\cline{2-4}
& Malicious code localization & 1 & \cite{sun2023dexbert} \\
\cline{2-4}
& Mobile app crash detection & 1 & \cite{liu2023testing} \\
\cline{2-4}
& Resource leak detection & 1 & \cite{wang2023boosting} \\
\cline{2-4}
& Test prediction & 1 & \cite{fatima2022flakify} \\
\hline
Software maintenance
& Program repair & 35 & \cite{cao2023study} \cite{charalambous2023new} \cite{deligiannis2023fixing} \cite{fan2022automated} \cite{gao2023constructing} \cite{huang2023chain} \cite{ibrahimzada2023automated} \cite{islam2024code} \cite{jiang2023impact} \cite{jin2023inferfix} \cite{lajko2022towards} \cite{le2023invalidator} \cite{liu2024reliability} \cite{mastropaolo2022transfer} \cite{paul2023enhancing} \cite{peng2024domain} \cite{ruiz2024novel} \cite{silva2023repairllama} \cite{sobania2023analysis} \cite{tian2020evaluating} \cite{tian2023best} \cite{tian2023chatgpt} \cite{wang2023rap} \cite{wei2023copiloting} \cite{white2023prompt} \cite{widjojo2023addressing} \cite{wu2023effective} \cite{xia2022practical} \cite{xia2023conversational} \cite{xia2023keep} \cite{yuan2022circle} \cite{zhang2023boosting} \cite{zhang2023gamma} \cite{zhang2023neural} \cite{zhang2023steam} \\
\cline{2-4}
& Code clone detection & 8 & \cite{alam2023gptclonebench} \cite{chochlov2022using} \cite{dou2023towards} \cite{jiang2023nova} \cite{liu2023contrabert} \cite{saberi2023utilization} \cite{sharma2022exploratory} \cite{shi2023towards} \\
\cline{2-4}
& Code review & 7 & \cite{guo2024exploring} \cite{li2022auger} \cite{liu2024reliability} \cite{lu2023llama} \cite{sghaier2023multi} \cite{tufano2022using} \cite{zhang2022coditt5} \\
\cline{2-4}
& Debugging & 4 & \cite{kang2023explainable} \cite{sakib2023extending} \cite{tian2024debugbench} \cite{tu2023llm4cbi} \\
\cline{2-4}
& Bug reproduction & 3 & \cite{huang2024crashtranslator} \cite{kang2022large} \cite{kang2023evaluating} \\
\cline{2-4}
& Review/commit/code classification & 3 & \cite{ghadhab2021augmenting} \cite{kou2023automated} \cite{yang2022aspect} \\
\cline{2-4}
& Duplicate bug report detection & 3 & \cite{helmeczi2023few} \cite{isotani2021duplicate} \cite{zhang2023cupid} \\
\hline
\end{tabular}}
\caption*{\hfill(Continued\_2)}
\label{tab:all_se_tasks_continued}
\end{table}

\begin{table}[h]
\captionsetup{labelformat=empty} 
\caption*{Table \ref{tab:all_se_tasks}. Continued\_2.}
\resizebox{\linewidth}{!}{
\begin{tabular}{r|p{0.35\linewidth}|c|p{0.5\linewidth}}
\hline
\textbf{SE Activity} & \textbf{SE Task} & \textbf{\# Studies} & \textbf{References} \\ \hline
Software maintenance
& Logging & 3 & \cite{li2023exploring} \cite{mastropaolo2022using} \cite{xu2024unilog} \\
\cline{2-4}
& Log parsing & 3 & \cite{liu2024interpretable} \cite{ma2024knowlog} \cite{yu2023log} \\
\cline{2-4}
& Sentiment analysis & 3 & \cite{biswas2020achieving} \cite{zhang2020sentiment} \cite{zhang2023revisiting} \\
\cline{2-4}
& Code revision & 2 & \cite{kabir2023empirical} \cite{wadhwa2023frustrated} \\
\cline{2-4}
& Vulnerability repair & 2 & \cite{islam2024llm} \cite{pearce2021examining} \\
\cline{2-4}
& API misuses repair & 1 & \cite{zhang2023evaluating} \\
\cline{2-4}
& Bug prediction & 1 & \cite{gomes2023bert} \\
\cline{2-4}
& Bug triage & 1 & \cite{lee2022light} \\
\cline{2-4}
& Code coverage prediction & 1 & \cite{tufano2023predicting} \\
\cline{2-4}
& Code review explained & 1 & \cite{widyasari2023explaining} \\
\cline{2-4}
& Code-Review defects repair & 1 & \cite{zhao2023right} \\
\cline{2-4}
& Crash bug repair & 1 & \cite{du2023resolving} \\
\cline{2-4}
& Crash bug repair & 1 & \cite{du2023resolving} \\
\cline{2-4}
& Dockerfile Repair & 1 & \cite{henkel2021shipwright} \\
\cline{2-4}
& Patch correctness prediction & 1 & \cite{zhang2024appt} \\
\cline{2-4}
& Patch detection & 1 & \cite{tang2023just} \\
\cline{2-4}
& Program merge conflicts repair & 1 & \cite{zhang2022using} \\
\cline{2-4}
& Rename Refactoring & 1 & \cite{liu2023refbert} \\
\cline{2-4}
& Tag recommendation & 1 & \cite{he2022ptm4tag} \\
\cline{2-4}
& Technical debt payback & 1 & \cite{mastropaolo2023towards} \\
\cline{2-4}
& Traceability recovery & 1 & \cite{zhu2022enhancing} \\
\cline{2-4}
& Web test repair & 1 & \cite{xu2023guiding} \\
\cline{2-4}
& Type error repair & 1 & \cite{chow2024pyty} \\
\cline{2-4}
& Others & 5 & \cite{jin2023assess} \cite{motger2024t} \cite{schroder2023autoscrum} \cite{von2022validity} \cite{wang2022your} \\
\hline
Software management
& Effort estimation & 2 & \cite{alhamed2022evaluation} \cite{li2024fine} \\
\cline{2-4}
& Software tool configuration & 1 & \cite{kannan2023can} \\
\hline
\end{tabular}}
\label{tab:all_se_tasks_continued_2}
\end{table}

\end{document}